\documentclass[%
 reprint,
 amsmath,amssymb,
 aps,
prd,
]{revtex4-2}

\usepackage{graphicx}
\usepackage{dcolumn}
\usepackage{bm}

\newcommand{\Msun}{${\rm M}_{\odot}$\ }

\newcommand{\mnras}{Monthly Notices of the Royal Astronomical Society}
\newcommand{\apjl}{The Astrophysical Journal Letters}

\newcommand{\aap}{Astronomy and Astrophysics}
\newcommand{\araa}{Annual Review of Astronomy and Astrophysics}
\newcommand{\jcap}{Journal of Cosmology and Astroparticle Physics}

\begin{document}

\preprint{APS/123-QED}

\title{Constraining the abundance of primordial black holes with gravitational lensing of gravitational waves at LIGO frequencies}

\author{Jose M. Diego}
 \email{jdiego@ifca.unican.es}
\affiliation{%
 Instituto de F\'isica de Cantabria (CSIC-UC)\\
 Edificio Juan Jord\'a. Avda Los Castros s/n. 39005 Santander, Spain 
}%




\date{\today}

\begin{abstract}
Gravitational waves from binary black holes that are gravitationally lensed can be distorted by small microlenses along the line of sight. Microlenses with masses of a few tens of solar masses, and that are close to a critical curve in the lens plane, can introduce a time delay of a few millisecond. Such time delay would result in distinctive interference patterns in the gravitational wave that can be measured with current experiments such as LIGO/Virgo.  We consider the particular case of primordial black holes with masses between 5 and 50 solar masses acting as microlenses. We study the effect of a population of primordial black holes constituting a fraction of the dark matter, and contained in a macrolens (galaxy or cluster), over gravitational waves that are being lensed by the combined effect of the macrolens plus microlenses. We find that at the typical magnifications expected for observed GW events, the fraction of dark matter in the form of compact microlenses, such as primordial black holes, can be constrained to percent level. Similarly, if a small percentage of the dark matter is in the form of microlenses with a few tens of solar masses, at sufficiently large magnification factors, all gravitational waves will show interference effects. These effects could have an impact on the inferred parameters. The effect is more important for macroimages with negative parity, which usually arrive after the macroimages with positive parity.   
\end{abstract}

\maketitle


\section{\label{sec_Intro}Introduction}
Gravitational waves (GW hereafter) open new opportunities to study the universe. Due to the all-sky sensitivity of an array of GW detectors, an event with an amplitude above the sensitivity of the detector can be observed independently of its location in the sky (although in reality a detector will see the signal modulated by the geometric factor that does depend on the sky location). Rare events can then be observed provided their amplitude is large enough. Strong gravitational lensing of distant background objects can be considered as a rare event since the probability that a distant source ($z>1$) is strongly lensed is less than 0.1 \%. Observations of transients (like supernovae) that are strongly lensed are extremely rare since detecting  transients has traditionally required to be observing that particular region of the sky at the time of the transient and only a small fraction of these transients are expected to be strongly lensed. In the case of GWs, one does not need to be pointing a telescope to a particular region in the sky. Then, if a distant GW is being strongly lensed, and its magnified amplitude surpasses the detector threshold sensitivity, it will be detected independently of its sky position. If the rate of GWs is sufficiently high at $z>1$, the number of GWs per redshift interval at large redshifts (that scales to first order as $z^3$) may be large  enough to compensate for the small probability of lensing. Then, some of the events that could not be observed at $z>1$ because they are too distant, could be promoted beyond the sensitivity threshold of the detector if they are magnified by some factor. Strong lensing of GWs is different than standard lensing of galaxies in different ways. Owing to the very low frequency of the GWs, and depending on the mass of the lens, diffraction effects may be important. This will be studied in more detail in the sections below. On the other hand, due to the very small size of the volume emitting GWs, magnification factors of many thousands are possible. This is not true in the case of strongly lensed galaxies, where magnification factors above a few tens are rare. This is due to the fact that the maximum magnification is proportional to $\sqrt{R}^{-1}$, where $R$ is the radius of the object being lensed. Allowing for magnification factors larger than a few hundred has important implications on the detectability of distant objects. GWs that are emitted at a redshift $z\approx3 $, if magnified by a factor $\approx 300$ can compensate the increase in the luminosity distance and appear as luminous as a similar event at $z \approx 0.3$. On the contrary, the galaxy hosting this GW would have its total flux magnified only by a factor of a few tens at most. 

A high rate of events at high redshift may require the existence of an exotic candidate for dark matter; the primordial black holes (or PBHs hereafter). PBHs have been hypothesized as a possible candidate for dark matter \citep{Carr1974,Carr2016,GarciaBellido1996,Bellido2017,Clesse2015}. Rates of events (with masses $\approx 30$ \Msun) as high as $10^4$ events per Gpc$^3$ and year (at $z \approx$ 1--2) are possible if the fraction of dark matter in the form of PBH is of order 10\% \citep{Kavanagh2018}. If a fraction of the dark matter is composed of PBHs, they will not only have an impact on the rate of merger events. PBHs can also act as microlenses, and split an incoming GW into multiple images with a  time delay between them. If the mass of the microlens is in a certain range, diffraction effects may take place on the GW \citep{Deguchi1986,Nakamura1999}. The GW may interfere with itself if the time delay between the lensed microimages is of the order of the period of the GW.
Lensing of GWs by compact astrophysical bodies has been considered in the past, although normally earlier work considers masses above 100 \Msun \citep{Wang1996,Nakamura1998,Sereno2010,Cao2014,Ng2018}. 
 In \cite{Jung2019}, the authors consider lensing by PBHs with masses between $\sim10$ and $\sim 10^5$ \Msun. In that work, the lensing probability is computed assuming a simple model in which all microlenses are isolated, that is, the halo (galaxy, group or cluster) hosting the microlens is being ignored. As shown by \cite{Diego2019b}, the role of the macrolens can not be ignored since it can modify significantly the time delays, and consequently the detectability of the microlensing event. The calculation of the optical depth of lensing is also affected by the macrolens. This is particularly important for the most distant detected GWs, most of which will be strongly lensed. At large magnification factors, the probability of microlensing grows as the magnification of the macrolens \citep{Diego2018}. As discussed in \cite{Diego2019b}, at sufficiently large magnifications, all strongly lensed GWs will be affected by microlensing. This happens when the effective optical depth is $\approx 1$

In this work we focus on lensed BBH, although the results presented in this paper are also relevant for other GWs at higher frequencies like BNS or NSBH. However, since these alternative candidates have a smaller chirp mass, the probabilities of observing these events at higher redshifts is smaller. As demonstrated in earlier work, and also shown in the sections below, the probability of strong lensing is sufficiently high only at redshifts above 0.5. BBH are luminous enough that, after being magnified by some factor, can be observed even if they are produced at $z>1$. 

Under the hypothesis that dark matter is composed of PBHs with masses around 30 solar masses, GWs that are being strongly lensed by massive haloes are useful probes of dark matter. Events that are strongly lensed necessarily must travel through areas with a relatively large surface mass density. As strong lensing events must take place near the critical curves of halos, the surface mass density near critical curves is close to the critical surface mass density for lensing, which typically is in the range of a few thousand solar masses per parsec square. This means that, in projection, a parsec square must contain dozens of PBHs in the intervening macrolens alone (and more when one accounts for additional matter along the line of sight)

We adopt a cosmological model with $\Omega_m=0.3$, $\Lambda=0.7$, and $h=0.7$. A few terms will be used throughout the paper. We refer to the massive haloes (galaxies, groups and clusters) as macrolenses. The stars, remnants and PBHs in these macrolenses that perturb the macrolens on the small scale are referred to as microlenses. The images produced by the macrolens are referred to as macroimages. In the presence of microlenses in the lens plane, macroimages of small sources can break up into smaller microimages around the positions of the microlenses. The source frame is the reference frame where the GW is originated while the observer frame is Earth. 
The structure of this paper is as follows. In section~\ref{sec_LensingI} we give a brief introduction to lensing of GWs in the regime of geometric optics. This regime is the appropriate one to address the probability of lensing of a GW by a macrolens. Section~\ref{sec_PDFmass} introduces the chirp mass function of GWs. Section~\ref{sec_Tau} discusses the probability of lensing by macrolenses. In section~\ref{sec_Rate} we describe two models for the evolution of the rate of GWs as a function of time. Section~\ref{sec_dNdz} estimates the expected number of strongly lensed events for the two models assumed. 
In section~\ref{sect_microlensing} we discuss the effects that microlenses have on GWs, including wave effects that, at LIGO frequencies, become relevant for microlenses with masses below a few hundred ${\rm M}_{\odot}$. In section \ref{sec_Results} we describe the simulation used to test the sensitivity of microlensing of GWs to the abundance of PBHs. Finally, in section \ref{sec_discuss} we present the main results of this work and discuss its implications. 

\section{\label{sec_LensingI}Lensing of Gravitational Waves}
Lensing of GWs has been studied in detail in earlier work  \citep{Wang1996,Nakamura1998,Sereno2010} and more recently in \citep{Dai2017,Jung2019,Broadhurst2018,Lai2018,Dai2018,Christian2018,Oguri2018b,Li2018}. In general, wave optics is the appropriate regime for studying the lensing of GWs when the Schwarzschild radius of the lens is comparable to the wavelength of the GW \citep{Nakamura1998,Nakamura1999,takahashi2003gravitational}. For the frequencies probed by experiments such as LIGO/Virgo, wave optics is relevant when the lens has a mass smaller than a few thousand solar masses. For lenses with masses above $10^4$ solar masses, one can rely on geometric optics. We discuss the regime of wave optics in section~\ref{sect_microlensing}.This section presents only a brief discussion of lensing of GWs in the regime of geometric optics. 

The most basic quantity that can be measured from a GW is the observed (redshifted) chirp mass, 
\begin{equation}
 \mathcal{M} = (1+z) \mathcal{M}_{\rm chirp} = (1+z) \frac{(m_1 m_2)^{3/5} }{ (m_1 +m_2)^{1/5}},
 \label{Eq_1}
\end{equation}
where $m_1$, and $m_2$ are masses of the two objects before coalescence (in the source frame), and $\mathcal{M}_{\rm chirp}$ is the chirp mass in the source frame. 

Only GWs that are above the detection threshold of the detector can be observed. This threshold is determined by the signal-to-noise-ratio, $\rho$, of the GW. To first order,  the value of $\rho$ depends on the luminosity distance of the GW, $D_l(z)$, and the observed chirp mass, $\mathcal{M}$. More specifically, $\rho$ is obtained after integrating the square of the Fourier transform of the observed signal over a frequency range, and weighted by the inverse of the power spectral density of the detector (see for instance \cite{Cutler1994,Finn1996}). 
\begin{equation}
\rho = \sqrt{\mu}\frac{\mathcal{M}^{5/6}}{D_l(z)}\sqrt{\mathcal{F}(\Theta,s,p,\theta)\zeta(f_{max})}, 
\label{Eq_2}
\end{equation}
The term $\mathcal{F}(\Theta,s,p,\theta)$ accounts for the geometric configuration, including the angular position in the sky, orientation of the detector, spin, polarization and orbital inclination of the binary. The term $\zeta$ encapsulates the detector response through the noise spectral density and the maximum frequency of the GW, $f_{max}$. 
This maximum frequency (twice the frequency of the last orbit) is determined by the innermost stable circular orbit (ISCO). The ISCO frequency scales as the inverse of the chirp mass.  
Larger chirp masses correspond to smaller $f_{max}$, and consequently to smaller $\zeta$, partially compensating the dependency of $\rho$ with $\mathcal{M}^{5/6}$. More specifically, the term $\zeta$ scales as  $\int_{f_{min}}^{f_{max}} df f^{-7/3}/\mathcal{S}(f)$ \citep{Finn1996}, where $S(f)$ is the power spectral density of the detector. For the range of frequencies and sensitivity in LIGO, one can use the published $\mathcal{S}(f)$ for the observed LIGO events to estimate $\zeta$. For the typical BBH events detected by LIGO, $\zeta$ has a mild dependency with $f_{max}$. In particular, $\zeta$ scales roughly as $f_{max}^{0.1}$ (or as $\mathcal{M}^{-0.1}$) for $f_{max}$ between 300 Hz and 700 Hz, which is the typical range of the observed BBH in LIGO. This modest dependency with $f_{max}$ is due to the fact that most of the power of the signal is contained in the lower frequencies, so above 200 Hz, the contribution to $\rho$ is relatively small.
The focus of this paper is in the often neglected first term in equation~\ref{Eq_2}. This is the square root of the magnification, $\mu$, which is usually assumed to be $\mu=1$, and hence ignored\footnote{One should note that the most likely value for $\mu$ is slightly below 1.0 (i.e a subtle demagnification) which, when ignored, leads to an overprediction of the distance and hence a systematic bias in the estimation of $H_o$ if not properly accounted for}. If a GW is being amplified by strong lensing, by ignoring $\mu$, the inferred luminosity distance will be smaller than the true distance by a factor $\sqrt{\mu}$, and the inferred chirp mass (in the rest frame) larger in order to maintain $\mathcal{M}$ constant. This mechanism has been invoked in the past to explain the unusually high chirp masses observed for some of the events detected in LIGO/Virgo \cite{Broadhurst2018}. 
\section{Chirp mass distribution of the background population}\label{sec_PDFmass}
%
\begin{figure}[ht]
\includegraphics[width=9cm]{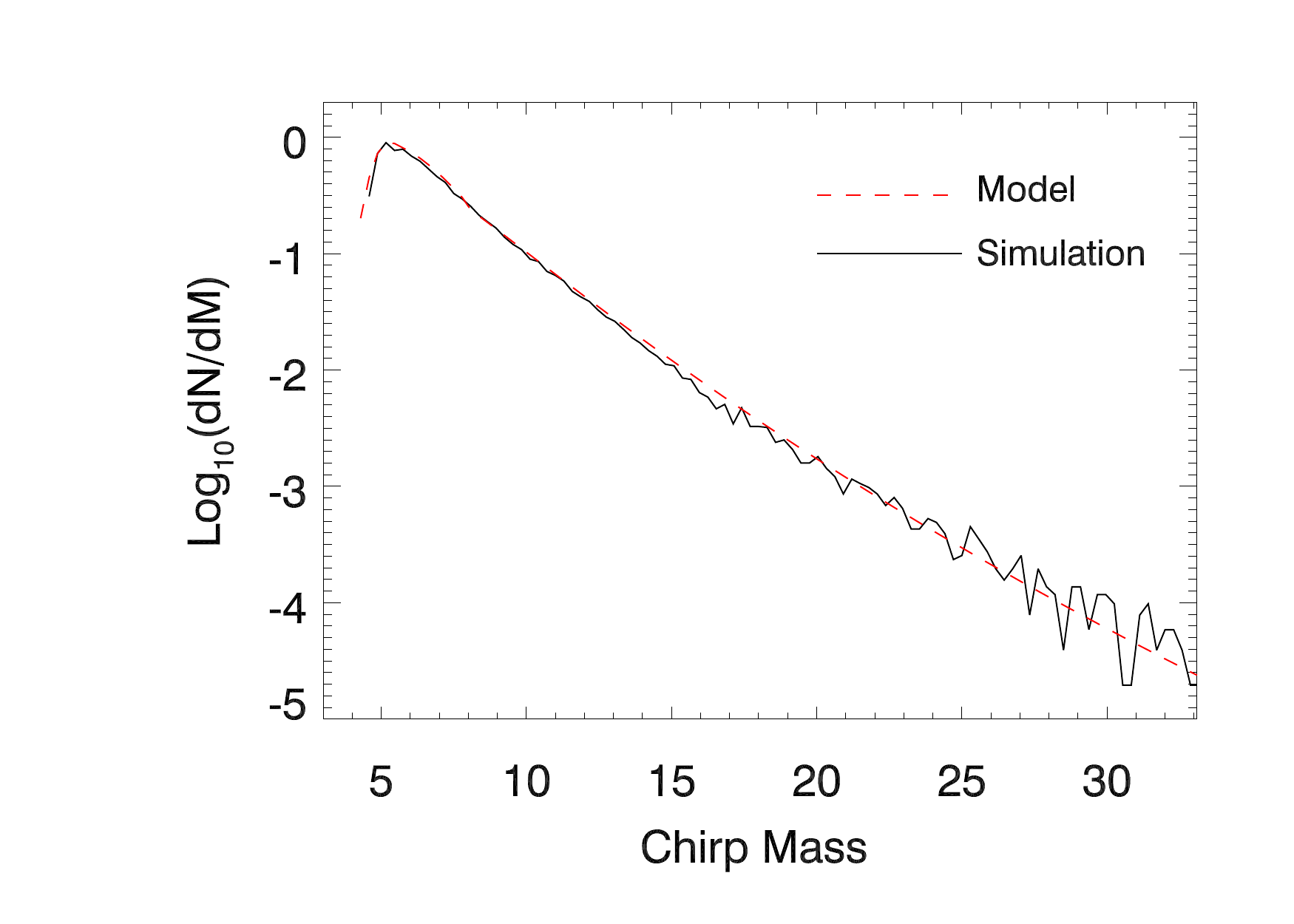}
\caption{\label{fig:FigChirpMassFunction}Mass function for the chirp mass. The solid line shows the result of a simulation where the chirp mass is computed from individual masses following the standard power law model with exponent 2.3. The dashed line is a model that reproduces well the simulation.}
\end{figure}
A fundamental ingredient in this work is the distribution of chirp masses. Little is known about this distribution. The observed events by LIGO/Virgo are consistent with standard power law models ($M^{-2.3}$) but also with a bimodal function having a peak at high masses (at $\approx 40$ \Msun) \citep{LigoCatalog2019II}. 
In \cite{Broadhurst2018}, the authors make the interesting suggestion that the peak in the bimodal mass function could be the consequence of lensing of the GWs. Since a lensed GW can be misinterpreted (if lensing is ignored) as a closer GW with a larger chirp mass, lensed GWs could naturally produce observed shallow, or even bimodal mass functions. 
In this work we assume a simple model for the Chirp mass function. We start from a standard stellar model where heavy stars follow the standard power law $M^{-2.3}$, appropriate for masses above 1 ${\rm M}_{\odot}$ \citep{Kroupa2001}, and assume that the remnants (BHs) left by these massive stars have a mass that is  proportional to the parent star. Individual masses ($m_1$ and $m_2$ in equation~\ref{Eq_1}) are drawn from this distribution (with $m_1$ and $m_2$ in the range $5 {\rm M}_{\odot} < m_1,m_2 < 50 {\rm M}_{\odot}$) and the Chirp mass is computed from the randomly generated $m_1$ and $m_2$. This approach ignores possible correlations between $m_1$ and $m_2$. The resulting distribution is shown in Fig.~\ref{fig:FigChirpMassFunction} as a solid line. The dashed line is the model we use in this work which is built from a lognormal function, $f(\mathcal{M})\propto \mathcal{M}^{-1}exp((ln(\mathcal{M})-\eta)^2/(2\sigma^2))$ where $(\eta=1.68,\sigma=0.12)$ for masses below 5.5 ${\rm M}_{\odot}$,  $(\eta=1.7,\sigma=0.3)$ for masses between 5.5 and 8  ${\rm M}_{\odot}$, and  $(\eta=1.68,\sigma=0.12)$ for masses above 8 ${\rm M}_{\odot}$. Since $\mathcal{M}$ is dominated by the lightest component (between $m_1$ and $m_2$), the resulting distribution peaks at low chirp masses and falls faster with mass than the original mass function. The minimum chirp mass is 4.35 ${\rm M}_{\odot}$ (when $m_1=m_2=5 {\rm M}_{\odot}$).
The resulting mass function resembles the low mass peak of Model C considered in  \cite{LigoCatalog2019II}, although with a lower mean mass (see Fig.1 in that reference). It is important to note that, as demonstrated below (and originally proposed by \cite{Broadhurst2018}), the high mass peak of model C in \cite{LigoCatalog2019II} can be naturally produced by lensing, provided the rate of events at high redshift is sufficiently high.  
\section{Probability of observing a lensed GW}\label{sec_Tau}
 %
\begin{figure}[ht]
\includegraphics[width=9cm]{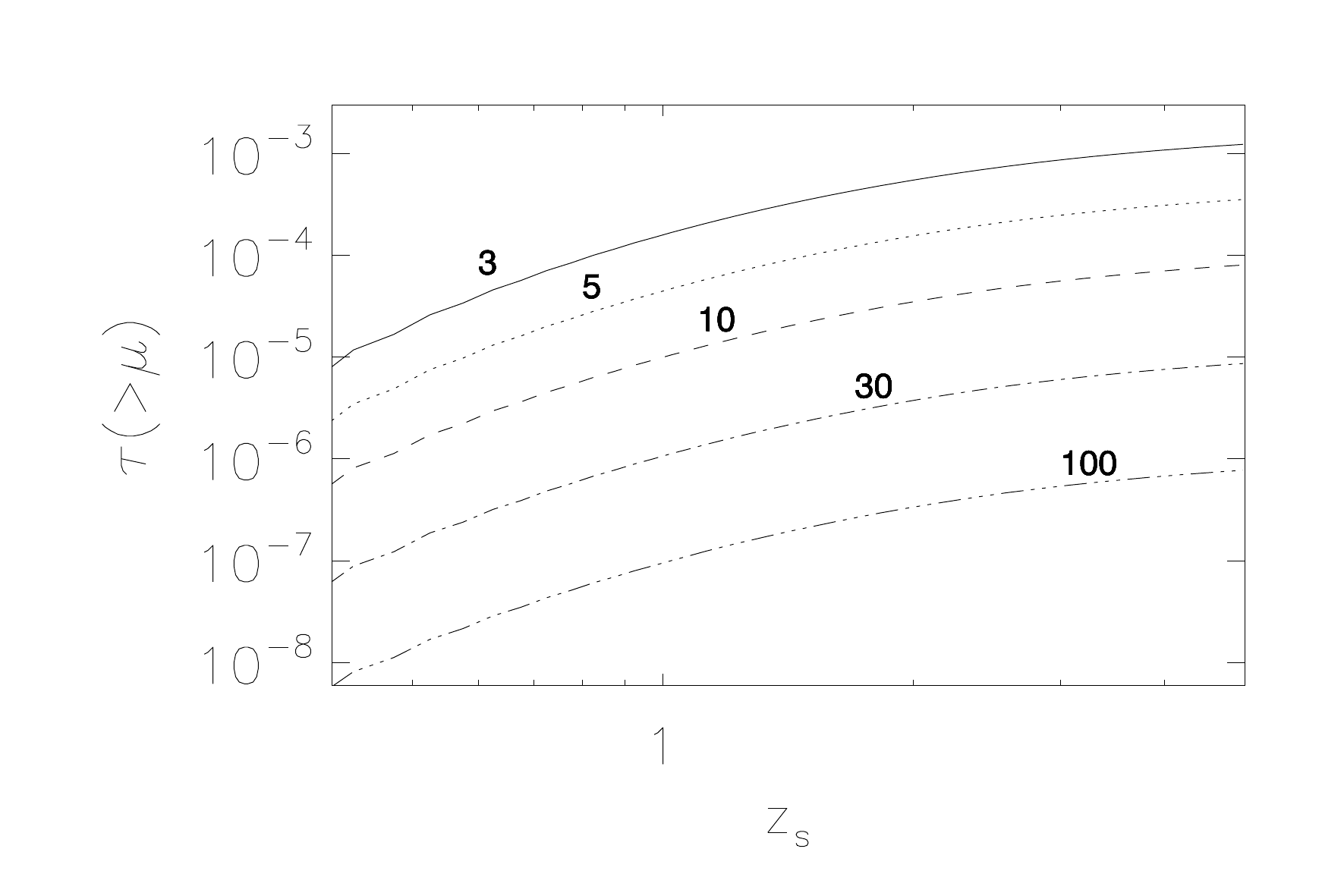}
\caption{\label{fig:FigOpticalDepth} Optical depth of lensing as a function of the redshift of the background source for different values of the  magnification. }
\end{figure}
The probability of lensing is referred to as the optical depth of lensing, $\tau$, and in its most basic form depends on the magnification, $\mu$ and the redshift of the source, $z_s$, i.e $\tau(\mu,z_s)$. Different authors have estimated $\tau(\mu,z_s)$ adopting different assumptions. Most estimates of $\tau$ are based on analytic models which have the advantage of flexibility, speed and achieving high spatial resolution (needed to resolve small areas with larger magnifications). On the negative side, analytic calculations do not account for corrections such as projection effects or substructure that can increase the optical depth. N-body simulations can account for both projection effects and substructure but at the expense of resolution. In this work we follow \cite{Diego2019} and use an analytic model based on the mass function of \cite{Watson2014} and an elliptical halo model to compute $\tau(\mu,z_s)$. We improve on the work of \cite{Diego2019} and modify the elliptical profile by adding a steeper component in the central region. This component accounts for the baryonic contribution that is important in smaller haloes and was neglected in \cite{Diego2019}. We also extend the mass range down to $10^{11} M_{\odot}$ from the minimum mass of  $10^{12} M_{\odot}$ considered in \cite{Diego2019}. The baryonic contribution makes smaller haloes more relevant, and even though their contribution to $\tau$ is still subdominant, they can not be ignored. 
 An additional improvement involves computing the optical depth in the image plane rather than in the source plane. In \cite{Diego2019}, the author used the optical depth computed in the source plane and later corrected for the multiplicity of lensed images. The optical depth computed in the source plane accounts for the total magnification and it is the best choice when estimating rates of unresolved images. Since GWs can be resolved in time (two counterimages from the same event will arrive separated by a time interval that can range between a few hours to a  few days), the optical depth computed in the source plane has the advantage that no corrections are needed to account for multiple images. It also has the advantage of achieving higher spatial resolution (since there is no need to map the image plane into the smaller source plane). The improvement in spatial resolution translates into an improvement into the maximum magnification that can be computed before limited resolution affects the computation. Our calculation can reach magnification factors of 100 before being affected by resolution effects (even for the smallest haloes). Above magnification 100, one can safely extrapolate the optical depth with the standard $\mu^{-3}$ law as $\tau(\mu_{>100},z_s)=\tau(\mu_{100},z_s)(100/\mu)^3$. 
In Fig.~\ref{fig:FigOpticalDepth} we show examples of the optical depth as a function of $z_s$ and for different values of the magnification factor. Note how between $\mu=10$ and $\mu=100$ the optical depth has decreased by two orders of magnitude, as expected from the scaling $P(\mu) \propto \mu^{-3}$  in strong lensing. This result is similar to the one the one derived from  N-body simulations and ray tracing \citep{Hilbert2008,Takahashi2011}. 

For magnification values $\mu>10 10$, and for redshifts $z_s>1$, the probability of lensing is $\approx 10^{-4}$. This number can be confronted with the number of observed strongly lensed QSOs (few dozens) out of the total number of known QSOs ($>10^5$). Between $z=1$ and $z=2$, one expects that 1 in $\approx 4\times10^5$ events will be magnified with factors $\mu>30$. Since the volume in this redshift interval is $\approx 450$ Gpc$^3$, if the rate of GWs above $z=1$ is above 1000 events per Gpc$^3$ and year, one would expect to observe at least 1 strongly lensed event above $z=1$ per year and with $\mu>30$. The rate of events as a function of redshift is discussed in the next section. 

\section{Rate evolution of BBH}\label{sec_Rate}
\begin{figure}[b]
\includegraphics[width=9cm]{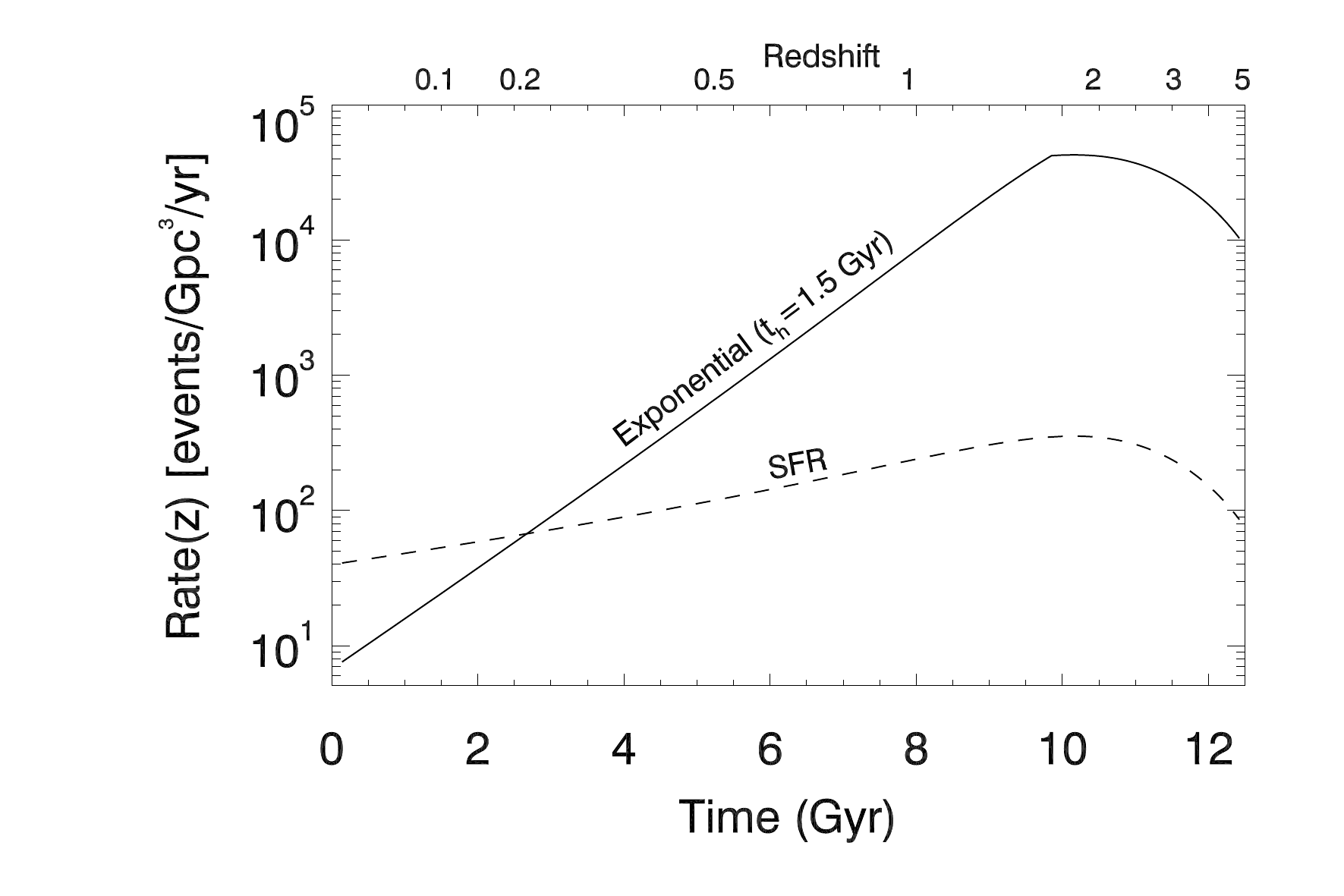}
\caption{\label{fig:FigRate} Rate of events as a function of (lookback) time. The solid line represents a model that grows as the star formation rate at high redshift but below redshift $z\approx 1.8)$ it falls of exponentially with a half-life time of 1 Gyr (in the main text, we refer to this model as the BDS model). The dashed line is a model that traces the star formation rate history. For comparison, the volumetric rate of SNe (of all types) at $z\approx 2$ is approximately $10^6$ per year and Gpc$^3$ }
\end{figure}

The rate of GWs as a function of redshift is unknown and depends on the formation mechanism of BBH. The simplest models rely on the assumption that the rate of GW events trace the star formation rate. We adopt the model of \cite{Madau2014} as a conservative one (and we denote it as the SFR case). This model predicts that the evolution of the rate of GWs is mild in redshift and peaks at $z \approx 2$. We normalize this model at redshift $z=0$ to a rate of a few tens of events per year and Gpc$^3$, consistent with the inferred rate from the LIGO collaboration (\cite{LigoCatalog2019II}, under the assumption of no lensing). This model is shown as a dashed line in Fig.~\ref{fig:FigRate}. 
As an alternative case \citep[inspired by][and that we refer to as the BDS model]{Broadhurst2018}, we consider also a model in which lensed GWs occur more frequently at high redshift, and has the attractive feature of explaining the apparent bi-modality of the mass function of BHs, as well as the unexpectedly high number density of very massive events (with chirp masses of 30 ${\rm M}_{\odot}$ or above). In this model, the rate of events at high redshift (where the lensing probability is highest) is increased by $\approx 2$ orders of magnitude (but still comfortably below the rate of SNe at all redshifts) in order to compensate for the low probability of lensing. To keep the observed rate of events comparable to the ones predicted by the conservative model, the rate at the lowest redshifts must be smaller. We attain this by allowing the rate to decay exponentially between the maximum of the star formation rate at $z\approx 2$ and $z=0$. Using a half-life time of 1.5 Gyr results in similar predicted number of observed events for both models so we adopt this value. The fast decline in the rate can be justified by the rapid evolution of BBH if these are formed in dense environments like globular clusters \cite[see e.g][]{Clesse2017}. In this case, the BHs are formed after the SN explosion of the parent star (typically a short-lived massive star), and through mass segregation they sink towards the center of the cluster (since after the massive stars undergo SN, the most massive objects in the globular cluster are still their remnants). Typical relaxation times are 1 Gyr or less \citep{Zwart2000,Banerjee2010,Rodriguez2015}. The duration of the phase between the BBH formation and the final coalescence is more uncertain and depends on mechanisms  like dynamical friction and three-body interactions. Most models agree that, once the binary is formed, only a small fraction of the binaries will merge within a Hubble time due to radiation of GWs \citep[see for instance]{Ziosi2014}. Other mechanisms may need to be invoked to explain a rapid evolution in the merger rate between the peak of star formation and redshift $z=0$. 

Also, it is important to consider the fact that massive BHs are typically produced from massive stars with low metallicity \citep{Postnov2019}. Below $z\approx 2$, massive stars with low metallicity are expected to be exceedingly rare. Hence, one may expect that BBH with elevated chirp masses are mostly produced at high redshifts and, if the time of BBH formation and coalescence is relatively short, a rapid evolution between redshift 2 and redshift 0 is expected. Above $z\approx 2$, the rate decays with redshift in a similar fashion as in the SFR model. 

The BDS model is shown as a solid line in Fig.~\ref{fig:FigRate}. Note that we choose to express the rate as a function of lookback time in order to accentuate the extended period up to the epoch of maximum star formation rate. Interestingly, the LIGO collaboration finds evidence for a rapid evolution in the rate ($\mathcal{R} \propto (1+z)^{6.5}$, \cite{LigoCatalog2019II}) when GW170729 is included in their analysis. On the contrary, if GW170729 is excluded, the rate is consistent with a non-evolving law, reflecting the large degree of uncertainty on the redshift dependency of the merger rate.
Note that the BDS model is consistent with the model C in the above reference when the 10 BBHs from O1 and O2 are included in the analysis \citep[See the redshift evolution model in Figure 6 in ][]{LigoCatalog2019II}. On the other hand, it is worth noticing that the relatively high rate of events at high redshift of this model may be in tension with upper limits on the stochastic background of GWs \citep{Clesse2017B}. However, the sensitivity of LIGO to the stochastic background is still $\approx 2$ orders of magnitude above the current estimate of this background \citep{LIGO2019} so a model such a BDS is still marginally consistent with the lack of detection of this background. Moreover, since GWs at low redshift (with a lower rate in the BDS model at $z<2$) contribute more to the stochastic background, a model such as the BDS model is expected to produce a background less than 2 orders of magnitude above the model used in \cite{LIGO2019}
 
\section{Expected number of strongly lensed events}\label{sec_dNdz}
%
\begin{figure}[ht]
\includegraphics[width=9cm]{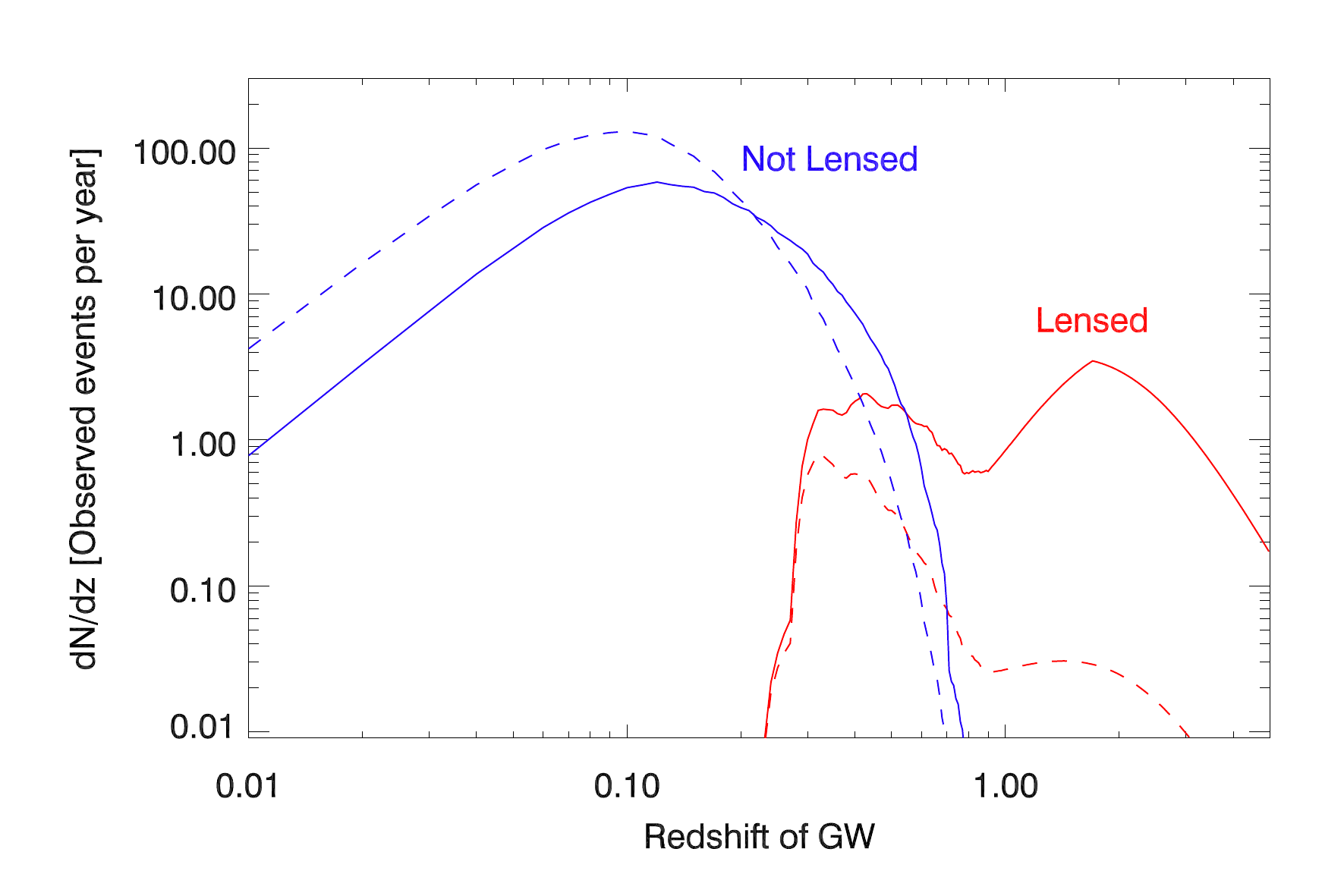}
\caption{\label{fig:FigdNdZ} Number of events above the detection threshold. The dashed line is fir the SFR model and the solid line for the BDS model. Lensed events are given by the red curves. In the BDS model, most of the lensed events are all originating at redshifts $z_s>1$. }
\end{figure}
\begin{figure}[ht]
\includegraphics[width=9cm]{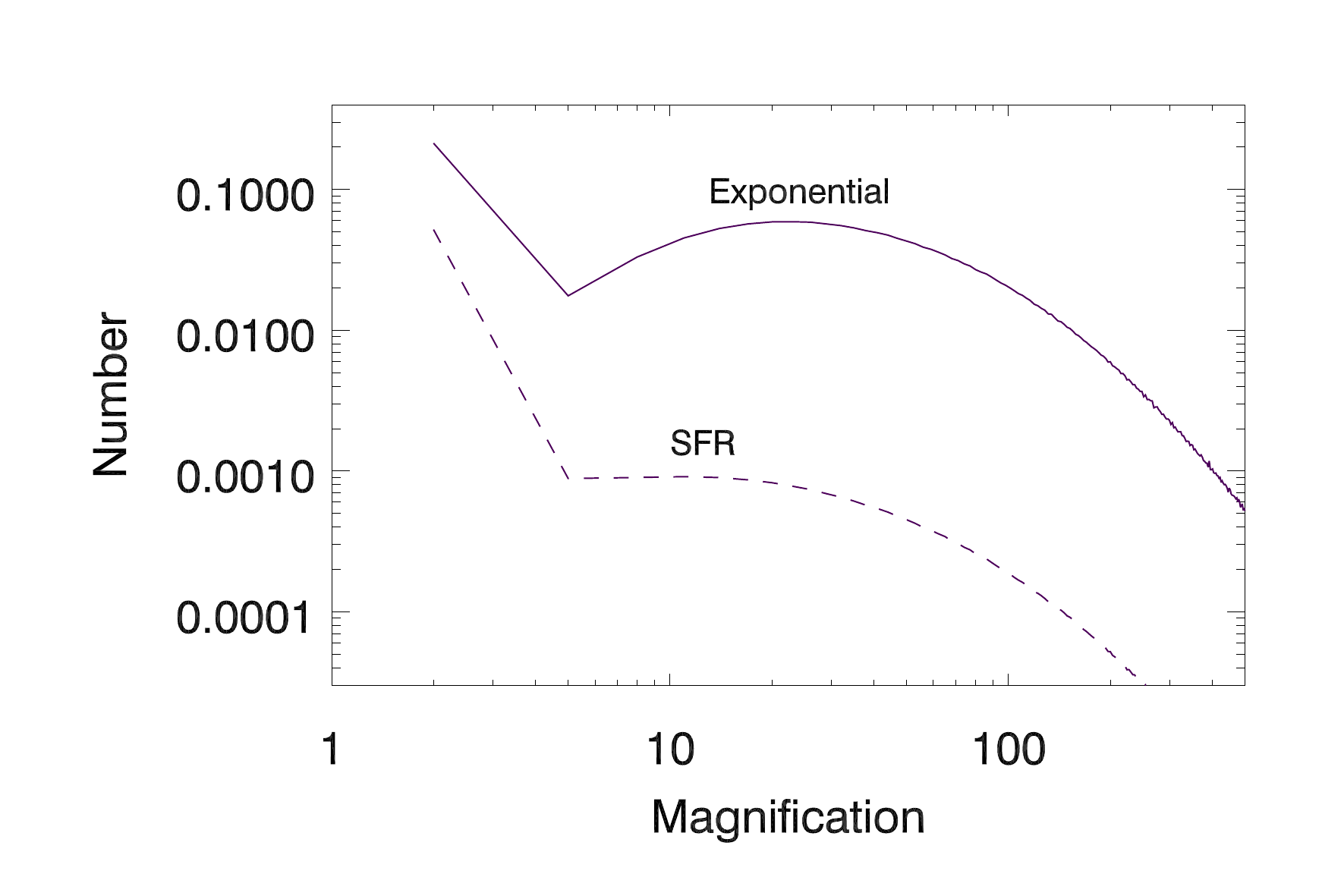}
\caption{\label{fig:FigMagnification} Distribution of observed events as a function of magnification. }
\end{figure}

With the ingredients presented in the previous sections, we can now compute the expected number of strongly lensed events. This number is basically given by the integral,
\begin{equation}
\frac{dN}{dVdz} = \mathcal{R}(z)\int_{M_a}^{M_b}\frac{d\mathcal{N}}{d\mathcal{M}}d\mathcal{M} \int P(\Theta) \tau(>\mu_{min},z)d\Theta 
\end{equation}
where $\mathcal{R}(z)$ is the rate of events discussed in the previous section, $d\mathcal{N}/d\mathcal{M}$ is the chirp mass function discussed in section \ref{sec_PDFmass} (that to first order we assume it does not evolve with redshift). The integration mass limits in the integral are fixed to $M_a=5$ \Msun and $M_b=50$\Msun. $ P(\Theta)$ is the probability distribution for the geometric factor, encoding all possible orientations of the detector and GW. For this term we follow \cite{Finn1996}. Finally, $\mu_{min}$ is the minimum magnification required for a GW at redshift $z$, with chirp mass $\mathcal{M}$ and geometric factor $\Theta$ to be above the detection threshold of the experiment (see Eq.~\ref{Eq_2}). We set a threshold in the signal-to-noise ratio of $\rho=8$ for noise properties similar to those in O1 and O2 runs in LIGO/Virgo (see section \ref{sec_LensingI}). 
By setting $\mu_{min}=1$ we compute also the number of events that are not being lensed. 
The resulting number of observed events it is shown in Fig.~\ref{fig:FigdNdZ} as a function of the redshift of the GW and for the two rate models discussed in section \ref{sec_Rate}. 
The dashed line (SFR model) shows results consistent with earlier estimates that predict that a small probability of lensed events is expected. The peak in lensed events at $z\approx 0.3$ corresponds to events with a mild magnification of $\approx 2$ for which the optical depth of lensing is higher (see Fig.~\ref{fig:FigMagnification}).  
On the contrary, the BDS model (solid line) predicts similar number of lensed and not lensed events. The lensed events originate mostly at $z>1$. As in the SFR model, a peak of low magnification can be appreciated also at $z\approx 0.3$
Fig.~\ref{fig:FigMagnification} shows the distribution of magnifications that result in observed events. For the SFR model, most of the events are magnified by moderate factors which makes them difficult to be recognized as lensed events. However, for the BDS model, most of the lensed events have magnifications of a few tens, and a non-negligible fraction can have magnifications above 100. 

\begin{figure}[b]
\includegraphics[width=9cm]{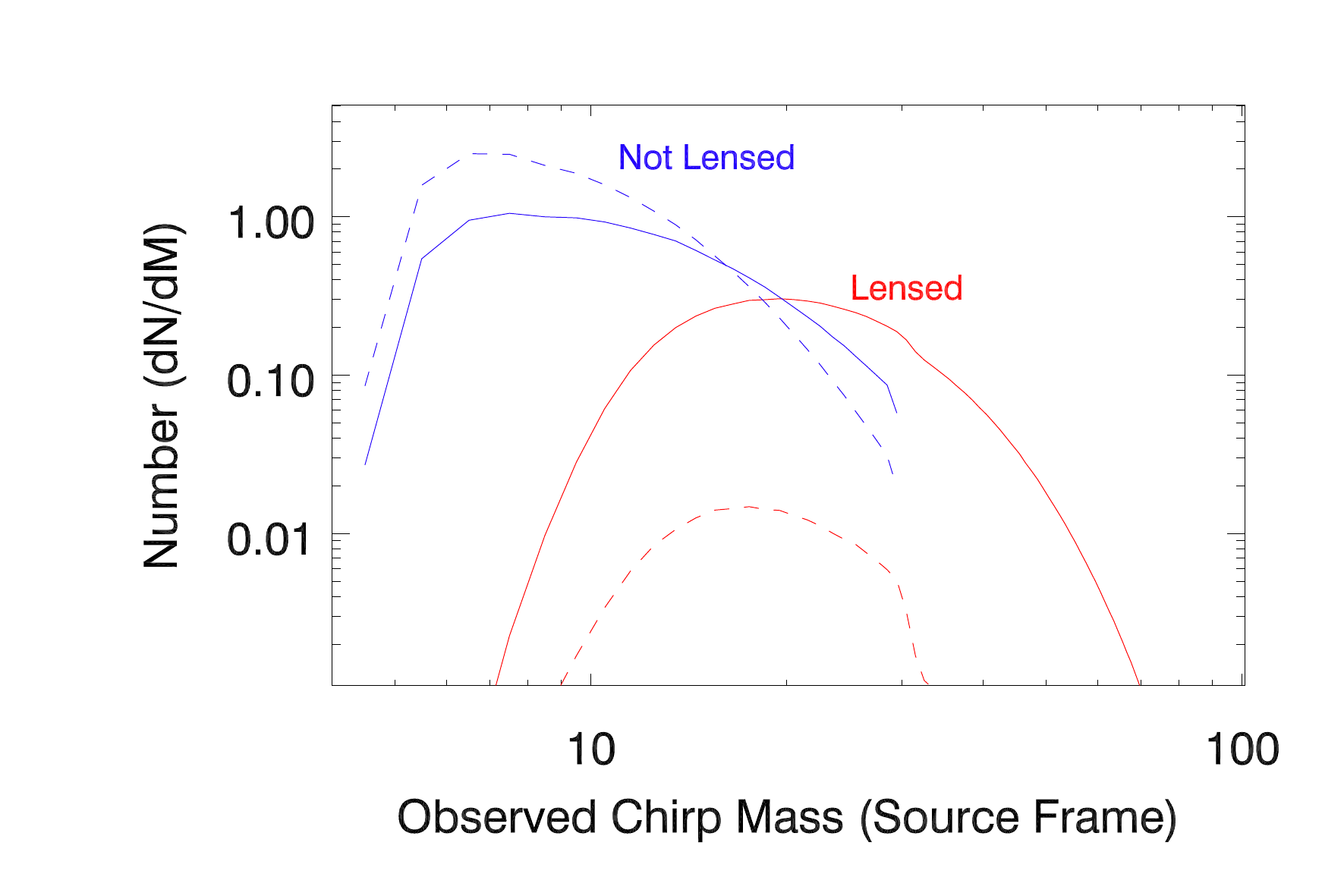}
\caption{\label{fig:FigObservedChirp} Inferred observed chirp masses in the source frame assuming there is no magnification. Note how the lensed events appear to have a higher chirp mass and the apparent bi-modality of the observed mass function.}
\end{figure}

Also interesting is the distribution of inferred chirp masses. Figure~\ref{fig:FigObservedChirp} shows the chirp masses inferred for the not lensed and lensed events. As expected the inferred masses for the not lensed events resembles the underlying chirp mass (with a bias towards higher masses because lower masses are less likely to exceed the detection threshold). However, for the lensed events (for which the mass has been inferred assuming the wrong magnification, $\mu=1$), the peak of the distribution is around masses of 20 or 30 solar masses, in striking resemblance to the current observed LIGO masses. \cite{Broadhurst2018} originally suggested that this mechanism was responsible for the apparently high chirp masses of most of the LIGO events. 
\section{Lensing by a macrolens plus microlenses}\label{sect_microlensing}
This section briefly reviews the lensing formalism for microlenses in a macrolens. This topic has been extensively covered in the literature \citep{Chang1979,Chang1984,Kayser1986,Paczynski1986}. Our model involves a macrolens and a population of microlenses. The microlenses include stellar microlenses (stars and remnants) and a population of PBHs. The PBHs account for a fraction $f$ of the total dark matter where $f=1$ would imply that all dark matter is made of PBHs. Since in the range of $\approx 5-50 {\rm M}_{\odot}$, the fraction $f=1$ is already excluded by other observations, we consider only values consistent with current constraints. The surface mass density of dark matter is given by the convergence, $\kappa$ of the macrolens. Then, the surface mass density of PBHs is simply $f\kappa$. 

For the macrolens, we follow \cite{Diego2019b} and define the macrolens with just two parameters, the magnification factors in the radial and tangential directions, or $\mu_r$ and $\mu_t$ respectively. This is a valid approach for the macrolens since we are dealing with very small regions of the sky. 
Without loss of generality, we assume that $\mu_t \gg \mu_r$ and that the main direction of the shear, $\gamma$, is oriented in the horizontal direction, that is $\gamma_2=0$ and $\gamma=\sqrt{\gamma_1^2 + \gamma_2^2}=\gamma_1$. 
Given $\mu_t$ and $\mu_r$, the corresponding values of $\kappa$ (convergence) and $\gamma$ (shear) can be found from the relation between $\kappa$, $\gamma$, $\mu_r$ and $\mu_t$. 
For a given choice of $\kappa$, and $\gamma$, the lens equation ($\vec{\beta} = \vec{\theta} - \vec{\alpha}(\vec{\theta})$) of the macrolens can be expressed as 
 \begin{equation}
 \vec{\beta}=  \vec{\theta} - \vec{\alpha}(\vec{\theta}) =
\begin{pmatrix} 1-\kappa-\gamma_1 & -\gamma_2 \\ -\gamma_2 & 1-\kappa+\gamma_1 \end{pmatrix} \vec{\theta},
\end{equation}
where the positions in the source plane are given by the coordinates $\beta=(\beta_x,\beta_y)$ and the positions in the image plane are given by the coordinates $\theta=(\theta_x,\theta_y)$.

The lensing potential of the macrolens, $\psi$, is given by 
\begin{equation}
\psi=\frac{\kappa}{2}(\theta_x^2 + \theta_y^2) + \frac{\gamma_1}{2}(\theta_x^2 - \theta_y^2) - \gamma_2\theta_x\theta_y 
\label{Eq_Psi}
\end{equation}
where we remind the reader that we adopt a reference system where $\gamma_2=0$, $\theta_x$ and $\theta_y$ are given in radians, and we ignore a constant additive term (i.e., the potential is identically zero at the origin of coordinates of $\theta$). 

Since both the deflection field and lensing potential are linear with the addition of new masses, if a population of $N$ point masses are present, the deflection, $\vec{\alpha}_{PS}(\vec{\theta})$,  and potential, $\psi_{PS}(\vec{\theta})$, from the distribution of point masses  can be simply added to the above equations with;  
\begin{equation}
\vec{\alpha}_{PS}(\vec{\theta})=\sum_i^N\frac{4GM_iD_i(z_l,z_s)}{c^2}\frac{\delta\vec{\theta}_i}{|\delta\vec{\theta}_i|^2},
\end{equation}
and,
\begin{equation}
\psi_{PS}(\vec{\theta})=\sum_i^N\frac{4GM_iD_i(z_l,z_s)}{c^2}ln(|\delta\vec{\theta_i}|),
\label{Eq_Psi_ps}
\end{equation}
where $\delta\vec{\theta}_i=\vec{\theta}-\vec{\theta}_i$ is the distance to the point mass $i$ at $\vec{\theta}_i$ and with mass $M_i$, $D_i(z_l,z_s)$ is the geometric factor $D_i(z_l,z_s)=D_{ls}(z_l,z_s)/(D_l(z_l)D_s(z_s))$ with $D_{ls}(z_l,z_s)$, $D_l(z_l)$ and $D_s(z_s)$ the angular diameter distances between the lens and the source, between the observer and the lens, and between the observer and the source respectively. 

A quantity of interest, that will be relevant in sections \ref{sec_Results} and \ref{sec_discuss}, is the effective optical depth, $\tau_{eff}$ introduced by \cite{Diego2018},
\begin{equation}
\tau_{eff}=(4.2\times 10^{-4})\frac{\Sigma}{{\rm M}_{\odot} {\rm pc}^{-2}}\frac{\mu}{\mu_r}
\label{Eq_Tau}
\end{equation} 
where the total magnification ($\mu$) is the product of the tangential and radial magnifications (i.e., $\mu=\mu_t\times\mu_r$), and $\Sigma$ (expressed in units of $M_{\odot}/pc^2$ in the expression above) is the microlens surface mass density. When $\tau_{eff}\approx 1$, the saturation regime is reached. In this regime, caustics always overlap in the source plane, and any source moving across a field with $\tau_{eff} > 1$ will be constantly experiencing microlensing \citep{Diego2018,Diego2019}. Since typical values for $\Sigma(M_{\odot}/pc^2)$ range between a few to a few tens, and assuming a typical value for $\mu_r \sim 1$, it is clear from the expression above, that the saturation regime is reached when the macrolens magnification is in the range of a few hundred to a few thousand. Similar values for the macrolens magnification have been observed already at the position of the microlensing events of the Icarus and Warhol high redshift stars \citep{Kelly2018,Chen2019}. Given the fact that GW experiments have access to the entire sky at any given moment (as opposed to the aforementioned examples of Icarus and Warhol where a significant luck-factor had to be involved) events with similar, or even more extreme, macrolens magnifications are expected \citep[see][for a detailed estimation of the probability of these events]{Diego2019}. In those scenarios, we expect microlensing to play a significant role. 

Finally, the time delay is given by
\begin{equation}
\Delta T = \frac{1+z_l}{cD(z_l,z_s)}\left[ \frac{1}{2}|\vec{\theta}-\vec{\beta}|^2 - \psi \right] 
\label{Eq_TimeDelay}
\end{equation}
Where we have assumed that all point masses are at the same redshift in the lens plane so the factors  $D_i(z_l,z_s)$ are the same for all of them. 

The time delay can be expressed in dimensionless units by re-scaling both the angular positions and potential by the Einstein radius, $\theta_E^2=(4GM/c^2)D(z_l,z_s)$. This redefinition of the time delay expression makes most sense when one is dealing with a single microlens since in this case the Einstein radius can be defined without ambiguity, but in general one can still set the Einstein radius to any arbitrary mass, or scale, and still redefine the time delay equation. 
\begin{equation}
\Delta T = \frac{2R_s(z_l)}{c}\left[ \frac{1}{2}|\vec{x}-\vec{y}|^2 - \tilde{\psi} \right],
\end{equation}
where $\vec{x}=\vec{\theta}/\theta_E$, $\vec{y}=\vec{\beta}/\theta_E$, $\tilde{\psi}=\psi/\theta_E^2$, and $R_s(z_l)$ is the redshifted Schwarzschild radius of the lens. 
The first term sets the scale of the time delay for a given lens mass,
$2R_s(z_l)/c = 1.97\times 10^{-5} (1+z_l)(M/{\rm M}_{\odot})$ seconds. This simple scaling, shows that for GW with frequencies $\nu\sim 100$ Hz, interference and diffraction effects are expected for isolated microlenses with masses $> 500/(1+z_l)$ \Msun \citep{Deguchi1986,Nakamura1998}. Below this mass, the Schwarzschild radius of the lens is smaller than the wavelength of the GW (with $\nu\sim 100$ Hz) and diffraction effects are not important resulting in the GW not seeing the microlens. However, as discussed in the section below, this minimum mass can be lowered substantially when the microlens is near a macrolens critical curve.

\subsection{The impact of microlenses on gravitational waves}\label{sec_WaveEffects}
 In earlier work, \cite{Diego2018} discussed how a microlens with mass $M$ embedded in a region of a macrolens where the macrolens magnification is $\mu$, behaves as a microlens with an effective mass $\mu M$.  Also, at larger magnifications, an area $A$ in the lens plane, maps into a smaller area $A/\mu$ in the source plane. Hence the density of microcaustics increases by a factor $\mu$. Since the surface mas density of microlenses is roughly a constant fraction of the convergence (which is of order 1 near the critical curves of macrolenses), it is then easy to realize that a GW that is being magnified by a macrolens with a factor $\mu$, will inevitably be affected by microlensing (and its associated wave effects), if the magnification factor of the macrolens is large enough. 

 When the wavelength of a GW is comparable to the Schwarzschild radius of the microlens, one needs to consider wave optics instead of geometric optics. The magnification factor in wave optics is given by the diffraction integral \citep[see for instance][]{SchneiderBook1992}. 
\begin{equation}
    F(w, \beta) =  A_o \frac{\nu}{2\pi i} \int d^2\theta\, e^{i2\pi\nu\Delta T(\theta,\beta)}
\label{Eq_Fw}
\end{equation}
where $\nu$ is the frequency of the GW in Hz and the normalization $A_o$ guarantees that at very high frequencies one recovers the  geometric optics magnification when averaged over a frequency range. $\Delta T$ is the time delay between lensed images expressed in seconds. 
The total magnification, $\mu$, and phase shift,  $\phi$, are given by \citep[see for instance][]{takahashi2003gravitational} 
\begin{equation}
\begin{split}
\mu &= |F(w,\beta)|^2,
\end{split}
\label{Eq_Fw2}
\end{equation}
\begin{equation}
\begin{split}
\phi &= -i \ln\left( \frac{F(w,\beta)}{|F(w,\beta)|}\right).
\end{split}
\label{Eq_Fw_phase}
\end{equation}
For isolated microlenses and simple macrolenses, analytic solutions can be found for the magnification. The magnification depends on the frequency and that at low frequencies the magnification tends to 1 (i.e the microlens has a negligible effect over the GW). In realistic situations, where a GW is strongly lensed, it may intersect not only one microlens but different microlenses. In this case, the integral in Eq.~\ref{Eq_Fw} can be solved numerically as described in \cite{UlmerGoodman1995,Nakamura1999,Diego2019b}. The case of a pair of overlapping microlenses was studied in \cite{Diego2019}. In this paper we study the more realistic situation where a distribution of microlenses intersects the GW. The probability that a GW intersects a microlens and suffers a distortion in its waveform depends on the density of microlenses and the  size of the caustic region. As discussed at the beginning of this section, this probability increases with $\mu$. Hence GWs with large magnification factors are more likely to suffer interference  than GWs with modest magnification factors. We consider two scenarios, the first one where the macrolens magnification is 30 and a second scenario where the macrolens magnification is 5 times larger (that is, 150). As shown in section~\ref{sec_dNdz}, the most likely magnification factor for models such as the BDS one is $\mu \approx 30$. Despite the rapid decline of the optical depth with magnification, observed events with $\mu=150$ are only $\approx 5$ times less likely than observed events with $\mu=30$ (for the BDS model) since the reduction in lensing probability is partially compensated by the increase in the accessible volume. For the SFR model, the rate of probabilities is similar but the probability of observing events with $\mu=30$ is two orders of magnitude smaller, so events like the ones described in this work would be observed only after the sensitivity of the detectors improve.   

\section{Primordial Black holes as dark matter}\label{sec_Results}
%
\begin{figure}[ht]
\includegraphics[width=9cm]{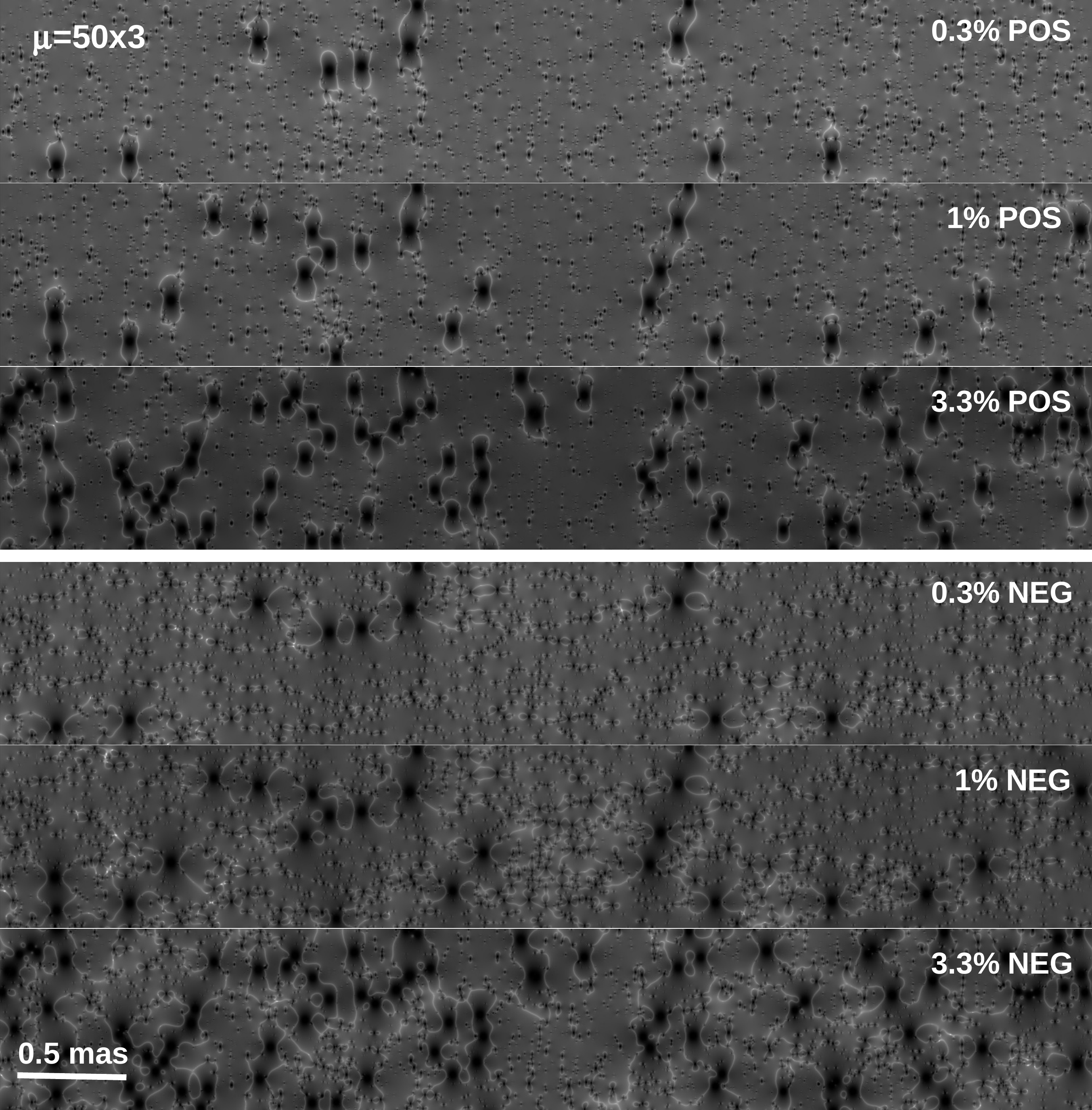}
\caption{\label{fig:Fig_Magnif} Magnification in the image plane showing the critical curves (white lines) for a macrolens magnification $\mu=\mu_t \mu_r = 50\times 3 = 150$. The top three panels correspond to three simulations (with different contribution to the dark matter content) on the side of the image plane with positive parity. The bottom panels are the corresponding simulations at the same macrolens magnification in the side with negative parity.}
\end{figure}
In the previous sections we discussed the lensing probability for the SFR and the BDS models. In both cases, lensing events are expected to take place although in the case of the BDS model, these events are $\approx$ two orders of magnitude more likely, and should have already been observed by LIGO. As mentioned above, the BDS model is appealing because it predicts a bi-modal  mass function in the observed chirp mass, consistent with the observed distribution of chirp masses. On the contrary, if the BDS model is correct, lensed events with inferred large chirp masses should come in pairs with time delays between a few hours to several days, and from the same location in the sky. To this date, no clear multiply lensed GW has been observed \citep{hannuksela2019search}\footnote{see however the discussion above about the reduced probability of observing the counterimage with negative parity}, although several candidates have been proposed \citep{Broadhurst2019}.  
More recently, two events during the 03 run were observed on August 28th 2019, which are separated in time by $\approx$ 19 minutes, and seem to originate from similar luminosity distances and positions in the sky. These positions are separated by only $\approx 10$ degrees. To first order, the probability of having two events so close in time and space is comparable to the probability of having a strongly lensed event. However, based on the sky location of these two events, \cite{Singer2019} concluded that the two events can not be interpreted as a pair of strongly lensed images. 
As shown below, if GWs are strongly lensed and microlenses are present, the observed strains can be perturbed, specially at the largest frequencies. The impact of this perturbation on the estimation of the parameters of the GW  is something that needs to be addressed carefully. 

In this section we assume that strongly lensed GWs have been observed already (as suggested by the BDS model), or that they will be observed in the near future, and investigate the possibility of constraining the abundance of PBHs. We take advantage of the magnifying power of the macrolens hosting the PBHs, which increase the concentration of caustics in the source plane, resulting in complex time delay distributions and interference patterns of the lensed GWs.    
We create a set of simulations where we vary the fraction of dark matter contained in PBHs, and compute the distortion in the magnification as a function of frequency for different positions of the GWs in the source plane. Sampling the source plane allows us to perform a statistical analysis, where we estimate the probability that a GW is distorted by some fraction at a given frequency. The simulations involve three components, a macrolens (galaxy or cluster), stellar microlenses from the macrolens, and PBHs that constitute a fraction, $f$, of the dark matter of the macrolens. By construction $f<1$, and we consider fractions, $f$, that are consistent with current upper limits.
We neglect projection effects and only consider deflectors that are in the macrolens plane. Projection effects are expected to contribute at the percent level since the adopted value for the convergence of the macrolens is of order 1. For the macrolens, we follow \cite{Diego2019b} and model the macrolens with just two parameters, the convergence $\kappa$, and the shear $\gamma$. The specific values of $\kappa$ and $\gamma$ are derived from predetermined values of the magnification in the tangential, $\mu_t$, and radial directions, $\mu_r$. We consider two scenarios for the magnification, $\mu$, of the macrolens. In the first scenario, we adopt a relatively modest magnification of 30, resulting from the product of a tangential magnification of 10 and a radial magnification of 3, that is $\mu = \mu_t\mu_r=10\times3=30$. In the second scenario, we consider a more extreme value of the magnification $\mu = \mu_t\mu_r=50\times3=150$. The two scenarios can be considered as two nearby positions in the image (or source) plane where the second scenario corresponds to the position that is closest to the critical curve (or caustic). For typical macrolenses with the mass of a massive galaxy, the magnification can change from 30 to 150 in image plane distances smaller than one arcsecond.  
In addition to the two considered values for the macrolens magnification, we also consider the two possible parities of the macroimages. A source located near a caustic produces typically two bright macroimages (in general additional macroimages are also formed, but with significantly less magnification and time delays much larger than the duration of a GW event, so they can be neglected). One of the two macroimages has positive parity (i.e, it has the same orientation of the orbital momentum as the source) and the other has negative parity (i.e it resembles a mirror image of the source). Other than the change in symmetry, in the presence of microlenses in the lens plane, the parity of the macroimage is important when dealing with very small sources. As shown in earlier work (see for instance \cite{Diego2019b} for a discussion in the context of GW), macroimages of very small sources with negative parity are significantly more likely to have smaller magnifications than  the corresponding macroimage with positive parity, if microlenses intersect the line of sight. The effect is even more dramatic for GWs since the microlenses introduce a time delay that can be more pronounced between microimages forming around a macroimage with negative parity. The time delay between microimages results in constructive and destructive interference patterns affecting the GW. This effect depends on the frequency of the GW and can resemble (or can be confused with) spin misalignment. Since the results depend on the parity of the macroimage, this allows to confirm that a GW is being strongly lensed, since one would expect (on average) that the image with positive parity arrives first and a few hours (or days) later the second image with negative parity would arrive. This second image is expected to show stronger distortions with frequency, on average, if microlensing is involved.  If $\mu$ is sufficiently large ($\mu>$ larger than a few hundred), this type of behavior is always expected since, even in the absence of PBHs, microlenses from the intergalactic or intracluster medium can saturate the source plane with microcaustics. Events with extreme magnification can be recognized because they may originate at very high redshifts ($z>2$) and be misinterpreted (if lensing is ignored) as events with unusually high masses $m_1$ and $m_2$, that could even fall within the mass gap above 50 \Msun.  
\begin{figure}[ht]
\includegraphics[width=9cm]{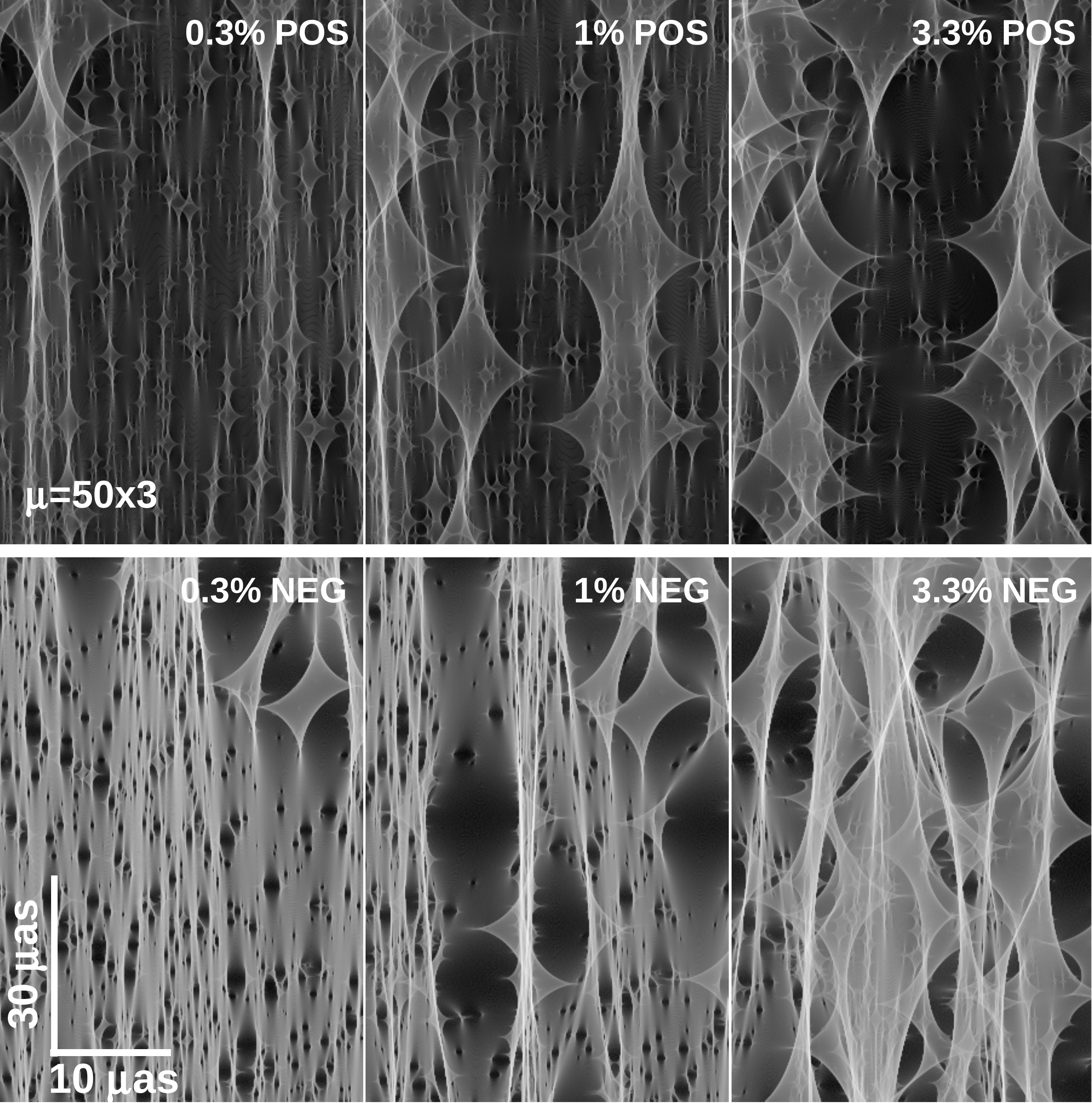}
\caption{\label{fig:Fig_Caust} Corresponding caustics in the sub-region used to compute the time delay statistics. The images have been compressed by a factor $\approx 2$ in the vertical direction in order to show a larger area.}
\end{figure}

In all simulations we include the stellar component using the same model described in \cite{Diego2019b}. For the two macrolens magnifications considered (30 and 150), we simulate a circular area in the image plane of $3$ mas in diameter, but consider only the central horizontal rectangular area of $3\times 0.5$ mas$^2$. The resolution is 100 nanoarcsec per pixel, sufficient to properly resolve the caustic regions around microlenses with stellar masses. The area in the source plane is reduced by the corresponding magnification factor of the macrolens. The contribution of the stellar component is modest and only the (rare) most massive stars and remnants are expected to produce significant effect on the GWs. 
The contribution to $\kappa$ of the stars and remnants is typically at the $\sim 1\%$ level, and most of the mass is contained in objects with masses $\sim 1$ \Msun or less. 
At masses above $\sim 5$ \Msun, where the interference effects are more prominent, stars and remnants contribute to the total mass a fraction much smaller than the contribution of the PBHs considered in this work. 
Because of this, the impact of the particular modeling of the stellar component is expected to be minor in our conclusions. Nevertheless, they are included assuming a surface mass density of 11.7 \Msun pc$^{-2}$ which is the same value used in \cite{Diego2019b}, and a typical value found around the critical curves of macrolenses for background sources with $z>1$ \citep{Morishita2017,Kelly2018}. 

For the mass function of the PBHs, we assume that their mass function is not necessarily the same as the one assumed for the sources (see section~\ref{sec_PDFmass}). Instead, we assume two different models. First, we consider a Gaussian distribution. This is a simple extension to the monochromatic model with a fixed mass, that allows for a broadening of the mass function. A log-normal mass function could be used instead, but we expect negligible differences in our conclusions, provided both Gaussian and log-normal have comparable dispersion. Similar models have been considered in the past to describe a possible population of PBHs with masses comparable to those found by LIGO. This type of model is already constrained by current observations which rule out fractions larger than $f\approx 0.1$. In this work we consider fractions that do not violate these limits. In particular we consider values of $f=0.033,0.01,0.003$, that is, $3.3\%, 1\%$, and $0.3\%$ the amount of dark matter respectively. We refer to these three models as Gaussian models and more specific details above these models are provided below. Interestingly, a fraction $\approx 1$\% may be sufficient to explain the rate of mergers observed by LIGO \citep{Liu2018,Kavanagh2018}.

As a second model for the population of PBH, we consider a much broader mass function, which is still consistent with current constraints \citep{Lehman2018}, but can allow for a larger fraction of the dark matter to be explained by PBHs. This is possible as long as the PBH span a wide range of masses, and the fraction of mass contained in each range does not violate current limits \citep{Lehman2018,Niikura2019,Smyth2019}. We model this second model as a power law between $M_{min}$ and $M_{max}$ and refer to this type of model as power-law model.
\begin{equation}
    f = \frac{\int_{M_{min}}^{M_{max}} dM M \Psi(M)}{\kappa*\Sigma_{crit}}
\end{equation}
where $\kappa$ is the convergence, $\Sigma_{crit}$ is the critical surface mass density, and $\Psi(M)$ is the mass function of the PBH per unit area. We assume that it follows a power law, $\Psi(M)= dN/dM\propto M^{\gamma_p-1}$. The index, $\gamma_p$, is a free parameter and we consider two values, $\gamma_p=0$ and $\gamma_p=1$.  When $\gamma_p=0$, $f$ is a constant as a function of mass. When $\gamma_p=1$, the number of PBHs per mass interval is constant and hence more massive PBHs contribute more to $f$ than lower mass PBHs. 

This mass function was also considered in \cite{Carr2017} that argues that "a mass function of this form arises naturally if the PBHs form from scale-invariant density fluctuations or from the collapse of cosmic strings".
For the limits of the integral, we consider $M_{min}=5$ \Msun and  $M_{max}=50$ \Msun. This range is justified because wave effects are not very important for masses below 5 \Msun. On the other extreme, above 50 \Msun the fraction of PBH is heavily constrained. Outside this mass range, there may still be PBHs that contribute to $f$. To account for this possibility, we consider values of $f$ smaller than 1, but still larger than the values of $f$ adopted for the Gaussian model. 
Note however that in \cite{Carr2019} the authors propose a mass function that, they claim, is consistent with current microlensing observations and can explain all dark matter as PBHs. In particular, in this model the mass function of PBHs is multi-modal, with the peaks of the mass function concentrated around $10^{-6}$, $1$, $30$, and $10^{6}$\,\Msun. 

For a given choice of the mass function, the number of PBHs per unit area inserted in the simulation is determined by the specific values of $f$ and $\kappa$. The convergence is determined by the values of $\mu_t$ and $\mu_r$ as described in \cite{Diego2018}. 

\subsection{Gaussian Model}\label{sec_Gauss}

%
\begin{figure}[b]
\includegraphics[width=9cm]{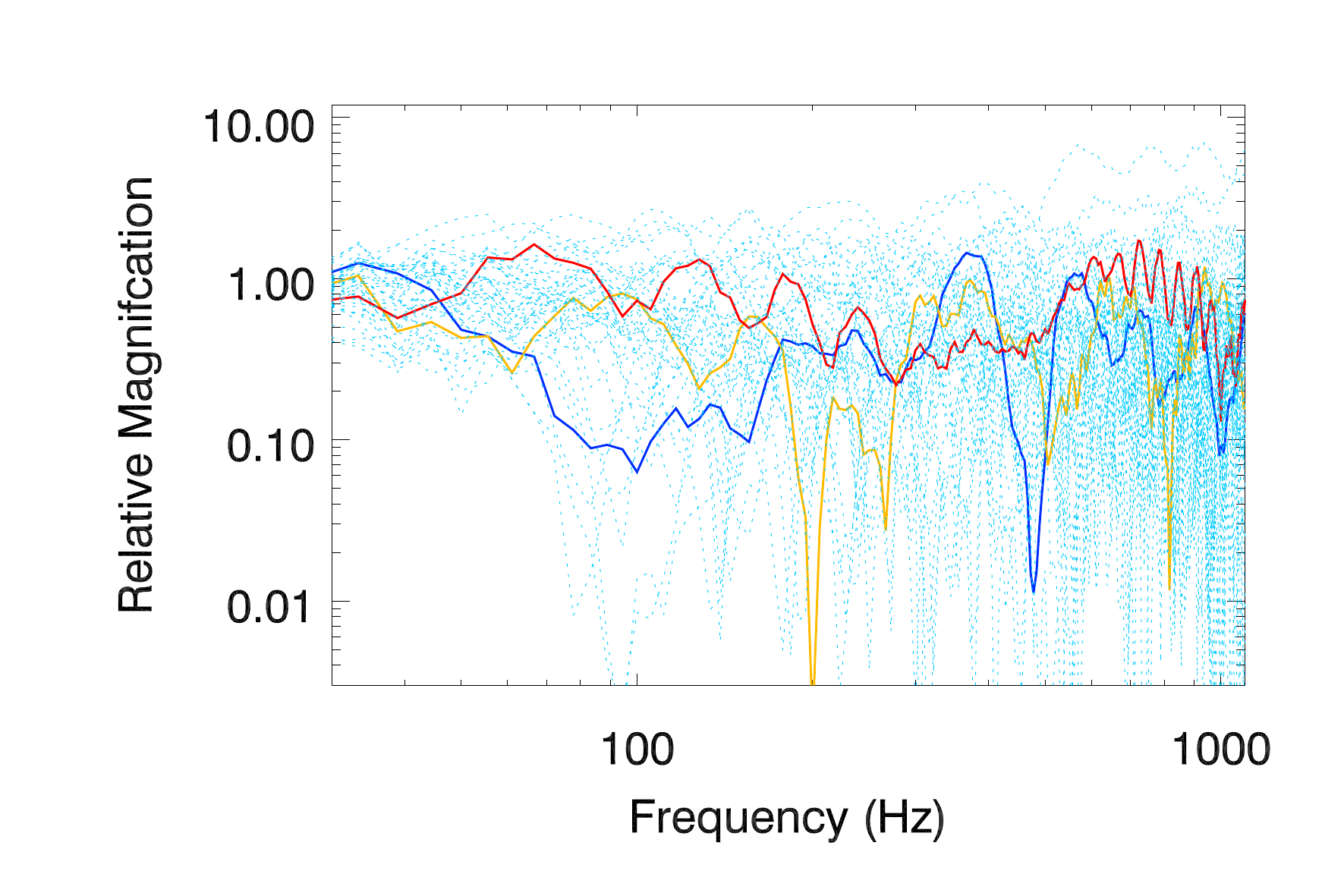}
\caption{\label{fig:FwExamples1} Examples of the relative magnification induced by microlensing as a function of GW frequency for the case of negative parity and the power law model with $\gamma=1$. The colored lines are the first three curves from the randomly selected positions. The light-blue dotted lines show the first 50 curves from the random positions.}
\end{figure}

First we consider the case of the Gaussian model. We assume a Gaussian mass function for the PBHs with mean 30 \Msun and dispersion 5 \Msun. This model is inspired by the high-mass peak found in the bi-modal mass function in \cite{LigoCatalog2019II}. Since the dispersion is relatively small, the results presented in this section would be close to those obtained with a monochromatic mass function with masses equal to the mean value of the Gaussian. 
In Figure~\ref{fig:Fig_Magnif} we show the magnification in the image plane for the simulations in the Gaussian model case and for the case with $\mu=150$. The magnification diverges at the critical curves (white lines). The top 3 panels show the magnification on the side of the image plane where the parity is positive and for the three values of $f$ that are consistent with current constraints \citep{Raidal2019}. Form top to bottom $f=0.3\%$,$f=1\%$, and $f=3.3\%$. The bottom three panels are the corresponding magnification maps in the image plane for the side of the image plane with negative parity and the same values of $f$. In both cases (positive and negative parities), the dark-gray areas have magnification factors of a few.  The largest critical curves are around the PBHs. Smaller critical curves correspond to stars and remnants in the macrolens. Note how the orientation (and shape) of the critical curves depend on the sign of the parity. The case with $\mu=30$ is not shown but would look similar with the exception that the critical curves around the microlenses would be smaller by a factor $\sqrt{150/30}$. 
By using the lens equation, we map the magnification in the image plane into the magnification in the source plane, where the critical curves map into caustics. Since the image plane gets compressed by a factor $\mu$ in the source plane, it is often the case that, at large macrolens magnifications, the caustics overlap. In this scenario, images can form around critical curves that are separated by relatively large distances in the image plane. For these images, the geometric time delay can be larger than the time delay that would be observed if the caustics did not overlap, resulting in interference effects at even lower frequencies. 
This overlapping effect in the source plane is shown in Figure~\ref{fig:Fig_Caust} where the three top panels are for the case with positive parity and the three panels in the bottom are for the case with negative parity. In this case, the caustics look very different depending on the sign of the parity. On the side with positive parity, the minimum magnification (dark gray areas) is of the order of the macrolens magnification (i.e, $\mu_{min}\approx 150$) while in the side with negative parity the dark-gray areas have magnification of just a few. For both parities, at fractions $f$ above a few percent and at this macrolens magnification, caustics start to overlap across the source plane. At even larger macrolens magnification factors, the overlap start to take place at smaller values of $f$. 

We use the simulated source planes to estimate the probability that a GW is distorted by some amount at a given frequency. We place 300 GWs at random positions in the source plane shown in Fig.~\ref{fig:Fig_Caust} and compute the magnification of the GW (relative to the macrolens magnification that does not depend on frequency) as a function of frequency following \cite{Diego2019b} for each position. Each of the 300 random positions ($\beta$ in Eq.~\ref{Eq_Fw}) results in a distortion curve of relative magnification as a function of frequency ($w$ in Eq.~\ref{Eq_Fw}). In Figure~\ref{fig:FwExamples1} we show the first 50 curves, out of the 300 examples for a particular choice of the PBH (that is described below in subsection~\ref{subsect_PL}). Let us remind the reader that these curves represent the relative magnification with respect to the macromodel and that the macromodel magnification is expected to be independent of frequency. For clarity purposes, we highlight the first three curves as colored solid lines (blue, orange, and red). As shown in the figure, some of the curves show significant deviations from unity at frequencies as low as 100 Hz. For instance, the dark blue solid curve shows a large decrement between $\approx 80$ Hz and $\approx 150 Hz$. The same curve, shows an increase in the magnification (relative to the macrolens) at $\approx 400$ Hz, followed by a large reduction in the relative magnification at $\approx 500$ Hz. All curves tend to 1 at lower frequencies, as expected, since at these frequencies the GWs become insensitive to microlenses with the masses considered in this work. 

\begin{figure}[ht]
\includegraphics[width=9cm]{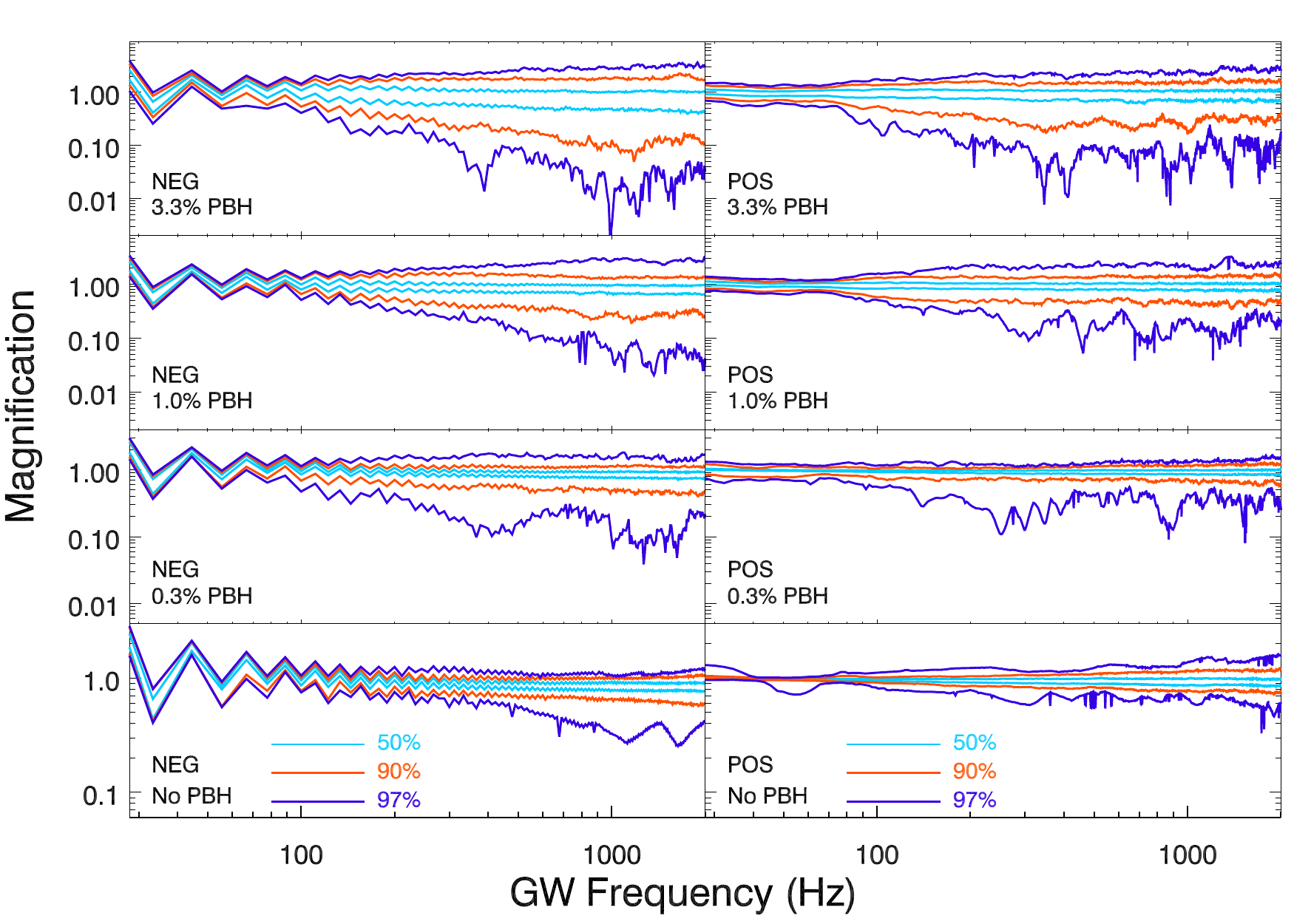}
\caption{\label{fig:FigStatMu1} Probability of magnification as a function of GW frequency for a macrolens with magnification $\mu=\mu_t \mu_r = 10\times 3 = 30$. }
\end{figure}

With the 300 magnification curves, we compute the probability that a the microlensing distortion of GW is larger than some amount as a function of frequency. For this goal we compute the envelopes of the curves at different percentage levels (50\%, 90\% and 97\%), and as a function of frequency. The envelopes at the 50\% level indicate the minimum distortion (at a given frequency) in the magnification of 50\% of the curves, that is 1 in 2 GWs will have a distortion at least as large as the one shown by the envelope at 50 \% at that frequency.  The envelopes at the 90\% level indicate the minimum distortion in the magnification of 10\% of the cases. That is, 1 in 10 GWs, will have a distortion in the magnification at a given frequency that is at least as large as the one shown by the envelope. Similarly, for the envelope at 97\%, approximately 1 in 30 GWs, will have a distortion at a given frequency at least as large as the one shown by the envelope.  Figure~\ref{fig:FigStatMu1} shows the envelopes (or probabilities) for the case with macrolens  magnification $\mu=30$. The cases where the macrolens parity is negative are shown in the left column while the cases where the parity is positive are on the right column. From top to bottom, we show the cases with $f=0.033, 0.01, 0.003$ and $f=0$. The case with $f=0$ has only the stellar microlenses (stars and remnants with a surface mass density of 11.7 \Msun$/pc^2$ following the model of \cite{Fryer2012}), and no PBHs. The serrated pattern observed in the envelopes with negative parity at low frequencies is an artifact due to numerical limitations (and related to the limited size of the simulation) when solving numerically the integral in equation~\ref{Eq_Fw}. The envelope at this low frequencies is supposed to be a smooth version of the curves. As expected, the distortions are larger at higher frequencies. If one considers the case with $f=0.01$ (or 1\% of the dark matter in the form of PBHs) and positive parity (i.e., right column, second panel starting from the top) at 500 Hz, then 10\% (90\% envelope) of the strongly lensed GWs with $\mu=30$ are expected to show a strong distortion in the magnification where destructive interference demagnifies the GW by a factor $\approx 3$ (relative magnification $\approx 0.3$). Since the strain goes as the square root of the magnification, this translates into a reduction of a factor $\approx 2$ in the amplitude of the strain at 500 Hz.
\begin{figure}[ht]
\includegraphics[width=9cm]{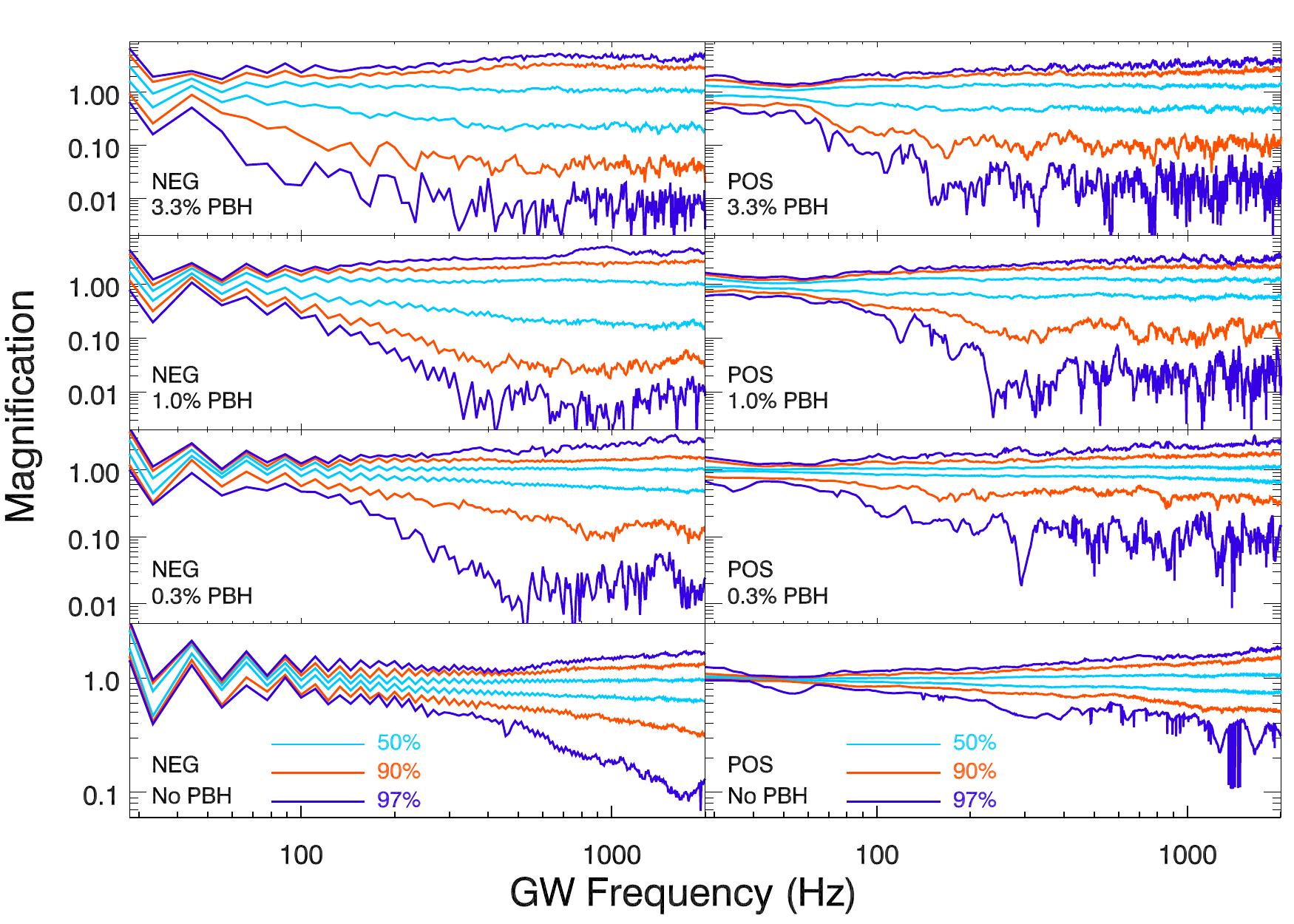}
\caption{\label{fig:FigStatMu2} As in Figure~\ref{fig:FigStatMu1} but for a position in the image plane where the macrolens has magnification $\mu=\mu_t \mu_r = 50\times 3 = 150$.}
\end{figure}

If the fraction of dark matter in PBHs increases, at $f=3.3\%$, destructive interference with even greater intensity is evident at frequencies as low as $\approx 200 Hz$, well within range of detectors such as LIGO. The distortion is even more apparent for the counterimages with negative parity (that we remind the reader should arrive between a few hours to a few days after the GW with positive parity) as shown by the left column. In this case, at least 1 in 10 GWs with magnification above $\mu=30$ should show a reduction in the strain by a factor $\approx 2$ at frequencies of 200 Hz. We should note that the envelope accounts only for the most extreme distortion at that frequency. This means, that the event mentioned above with a reduction of a factor $\approx 2$ at frequencies of 200 Hz, at higher frequencies may show a smaller distortion (or even no distortion at all in a given frequency range). 

The above result shows the potential of strongly lensed GWs to probe the abundance of PBHs but it requires observing several events in order for one of them to show significant distortions in the observed strain. However, if we focus on events with even larger macrolens magnification factors, the caustics will overlap even more  and individual microlenses will behave as microlenses with larger masses (with an effective mass that scales as $\mu M$). In Figure~\ref{fig:FigStatMu2} we show the case for a macrolens magnification $\mu=150$. Note that although the probability of an event with $\mu>150$ is $5^2$ times smaller than the probability with $\mu>30$, the probability of observing such event is actually only a few times smaller since at $\mu>150$ one is  probing a larger volume than at $\mu>30$ (see for instance Fig.~\ref{fig:FigMagnification}). 
For lensed events with $\mu>150$, and if one considers the case where 1\% of the dark matter is in the form of PBHs, the 50\% envelope is already at magnification factors 0.3 at frequencies $\approx 300$ Hz. That is, a reduction in the strain of a factor $\approx 2$ at 300 Hz is expected in 50\% of the lensed events with $\mu>150$. One in ten of these events (90\% envelope) would have a similar reduction of a factor 2 in the strain at frequencies as low as $\approx 130$ Hz. The same fraction would show reductions of a factor 5  in the strain at 400 Hz. 

\subsection{Power Law Model}\label{subsect_PL}
The second scenario we consider is the case where the mass function of PBHs between 5 \Msun and 50 \Msun is described by a  power law. Given the current constraints, PBH can make up to $\approx 10\%$ of the dark matter \citep{Lehman2018}.  If one relaxes the constraints at very low masses (asteroids) from PBH evaporation into gamma rays and at masses above 50 \Msun from the CMB (Planck), PBHs can make up to 100\% of the dark matter. In \cite{Carr2017}, the authors explicitly consider extended mass functions with power laws similar to the ones considered in this work and find that in the mass regime we are interested in, a fraction as high as 10\% is consistent with current constraints. 
We consider this value of $f=0.1$, in the case of the power law and make a set of simulations similar to the ones in the previous section ($\mu=30$,  $\mu=150$, positive, and negative parities) for the two power laws considered, $\gamma_p=0$, and $\gamma_p=1$.

\begin{figure}[ht]
\includegraphics[width=9cm]{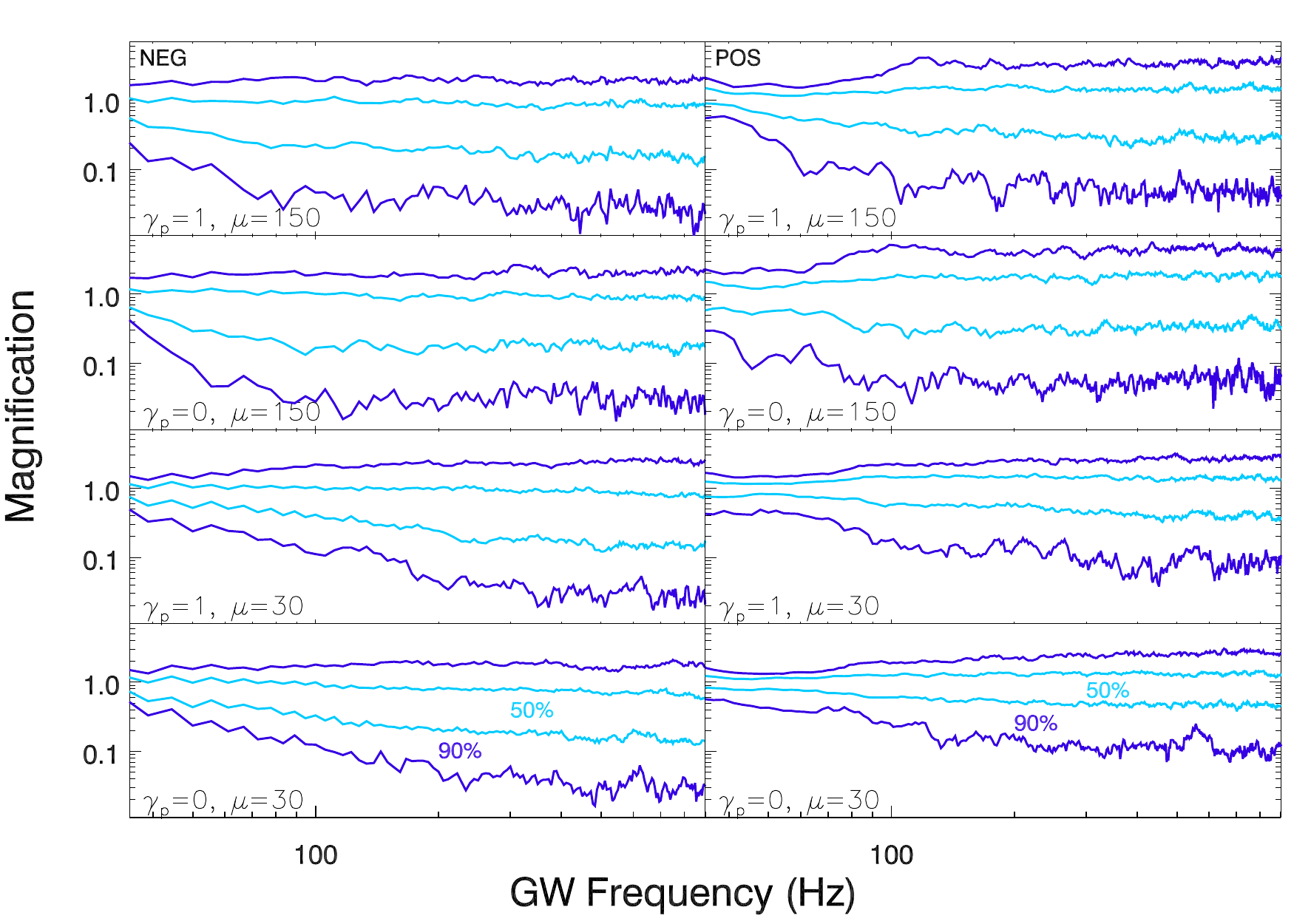}
\caption{\label{fig:FigPowLaw1} Envelopes at 50\% and 90\% showing the probability of distortion as a function of frequency for the two power law models, the two magnifications considered and $f=10\%$.}
\end{figure}

\begin{figure}[ht]
\includegraphics[width=9cm]{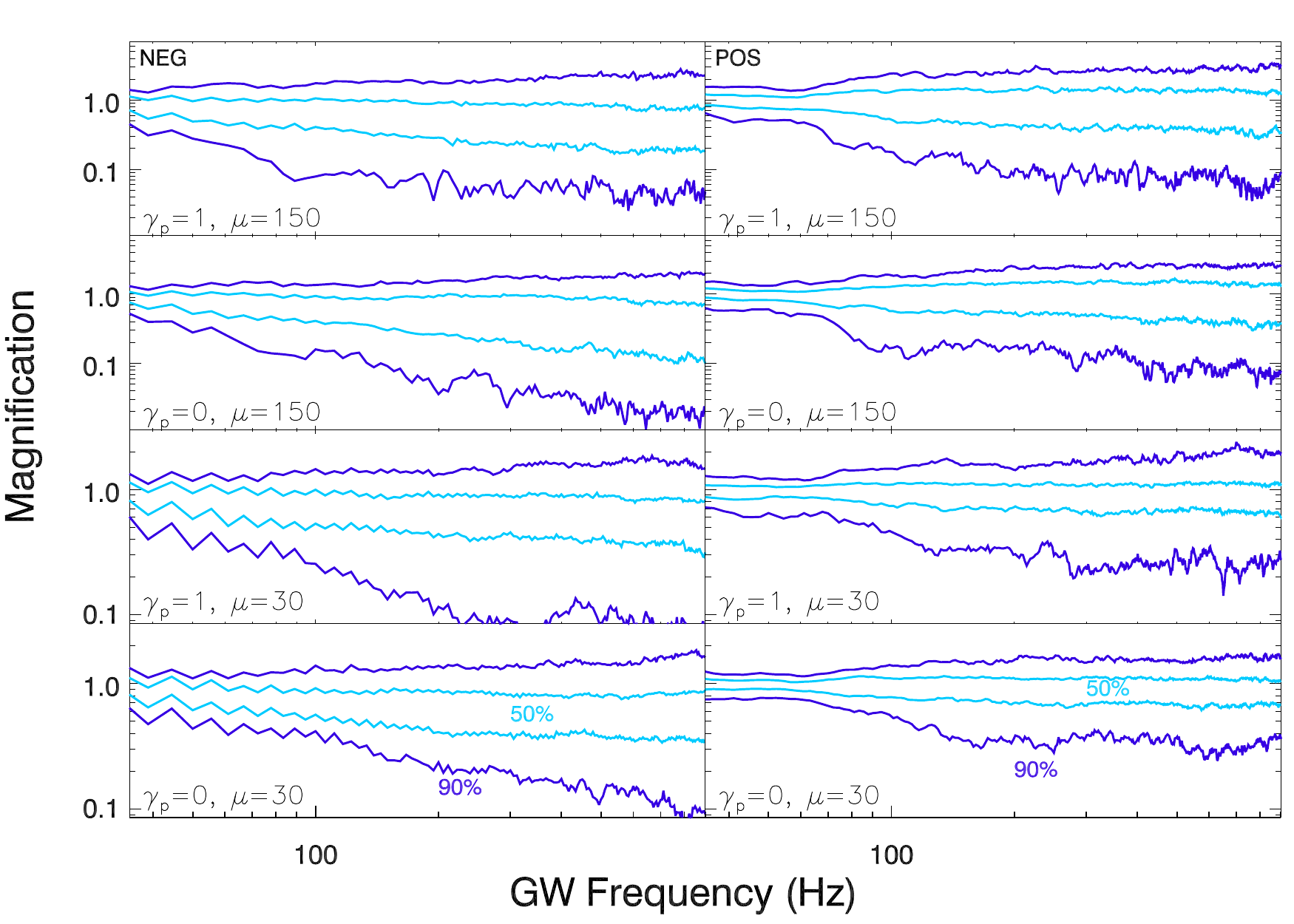}
\caption{\label{fig:FigPowLaw2} As in Fig.~\ref{fig:FigPowLaw1} but for a smaller fraction $f=3.3\%$. Note that the scale in the y-axis is different than in Fig.~\ref{fig:FigPowLaw1}.}
\end{figure}

The result is shown in Figure~\ref{fig:FigPowLaw1}. In this case, and for simplicity, we consider only two envelopes at 50\% and 90\%. Not surprisingly, since there are more PBHs per unit area, the effect is larger than for the Gaussian model described in the previous subsection. As a general trend, we observe again that images with negative parity are more likely to show interference effects. The differences between the two power law models is relatively small. The difference is also small when comparing the $\mu=30$ and $\mu=150$ cases, except at low frequencies, where interference effects in the case with $\mu=150$ are more evident. 
This can be explained because at fractions $f=0.1$, and for magnifications $\mu>30$, the saturation level is reached in the source plane \citep[see][for a discussion of the saturation level]{Diego2018}, where microcaustics overlap over the entire source plane,  and the statistical properties of the magnification are very similar at $\mu=30$ and $mu=150$. Only at smaller fractions, $f$, the probability of interference becomes sensitive to the macrolens magnification for magnification factors $\mu>30$, as shown in Fig.~\ref{fig:FigPowLaw2}. 
For smaller magnification factors, the saturation regime is not reached and the probability of intersecting a  caustic becomes sensitive to the masses of the microlenses.  

An interesting trend is observed in all cases with negative parity (left column in Figs.~\ref{fig:FigPowLaw1} and~\ref{fig:FigPowLaw2}), where in more than  50\% of the cases the magnification decreases with increasing frequency. In the 90\% envelope, both for the images with positive and negative parity, GWs are heavily suppressed above 100 Hz. For the images with negative parity, this happens at even lower frequencies. In particular, 10\% of the GWs with magnification $\mu=150$ would be heavily suppressed below 60 Hz. Since most of the signal to noise in the events observed by LIGO is concentrated below 100 Hz, this means that these events would basically be unobserved. In 50\% of the cases, at $\mu=150$ the suppression factor is typically a factor $>2$ at frequencies $>100$ Hz for images with negative parity and a bit smaller than 2 for the images with positive parity, and at the same frequencies.  
\section{Discussion and conclusions}\label{sec_discuss}
The results in the previous sections show how, if a fraction of the dark matter is composed of microlenses with masses above a few \Msun, a strongly lensed GW may reveal this population of microlenses through interference effects. In general, the distortion is larger the larger is the fraction of dark matter in the form of compact microlenses, but at sufficiently high magnifications, the saturation level is reached (when $\tau_{eff}\approx 1$, see equation~\ref{Eq_Tau}) and the microcaustics overlap in the source plane. Once this saturation level is reached, the results are less sensitive to the number and masses of the microlenses but the point where the saturation level is reached depends on the surface mass density of microlenses.  Since the critical surface mass density for typical redshifts of lenses and background sources is $\Sigma_{crit}\approx 2000$ \Msun pc$^{-2}$ and near the critical curves $\kappa\approx 1$, if the fraction of dark matter in microlenses is $\approx 3\%$, then $\tau_{eff}\approx 1$ when the magnification of the macrolens is $\mu \sim 100$ (assuming the radial magnification factor is $\approx 2$). At smaller magnification factors, such as $\mu \approx 30$, one expects to be more sensitive to the particular mass function of microlenses but at the expense of having fewer GWs with a significant distortion in the magnification, since the area in the source plane which is not close to a microcaustic is larger in this case. 

\begin{figure}[ht]
\includegraphics[width=9cm]{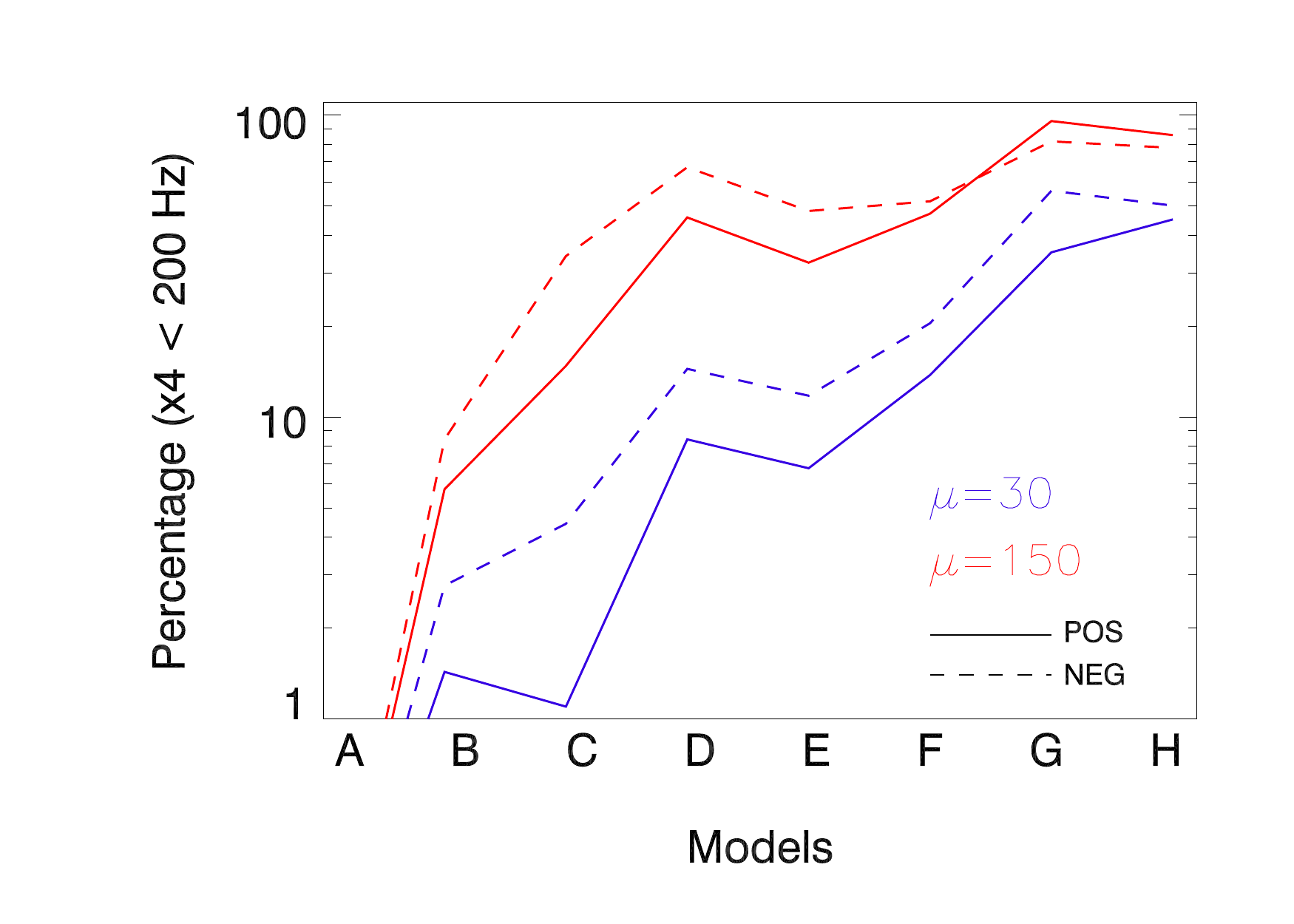}
\caption{\label{fig:Fig_ProbAll} Probability of having a distortion in the magnification larger than a factor 4 at frequencies below 200 Hz. The solid lines are for images with positive parity while the dashed lines are for images with negative parity. Blue lines are for macroimages with magnification 30 and red lines are for macroimages with magnification 150. Model A is the one with no PBHs (i.e only stars and remnants with $\Sigma=11.7$ \Msun pc$^{-2}$). Models B, C and D are the three Gaussian models with $f$=0.3\%, 1\%, and 3.3\% respectively. Models E and F are power law models with $f$=3.3\% and with $\gamma=0$ (E) and $\gamma=1$ (F). Models G and H  are power law models with $f$=10\% and with $\gamma=0$ (G) and $\gamma=1$ (H).}
\end{figure}

In the previous section, the envelope curves shown in Figures~{fig:FigStatMu1}--\ref{fig:FigPowLaw1} account for the probability of distortion of certain magnitude at a given frequency. Alternatively, it is interesting to ask the question of what is the probability of having a magnification of at least a factor $\mu^*$ (a strain distortion of a factor $\sqrt{\mu^*}$) below some frequency $\nu^*$. In Figure~\ref{fig:Fig_ProbAll} we show the case for $\mu^*=4$ and $\nu^*=200$ Hz. Below this frequency, one can expect that a distortion in the strain of a factor $>2$ at some frequency $\nu<200$ Hz could be picked by optimally filtering the raw data. Each curve represents all the models discussed in the previous section.  The Gaussian models are one the left side of the x-axis while the power law models are on the right side. Dashed lines are for macroimages with negative curvature and solid lines are for macroimages with positive curvature. In general, and as expected, it can be appreciated how models with a larger fraction $f$ are  more likely to produce distortions in the strain. The exception is model C in the blue solid line which has fewer cases than model B in the same curve but this is purely a fluctuation due to small number statistics. 
The power law models are also more likely to produce distortions than the Gaussian models except when comparing models D (Gaussian with 3.3\% dark matter) and model E (power law with  3.3\% dark matter and $\gamma=0$) which predict almost identical results, although with the Gaussian model predicting a few more cases with this magnitude in the distortion. Note how models G and H are close to the saturation regime and produce virtually identical results. The plot also shows the gain in probability of distortion when one considers events magnified by larger factors (red curves). If one considers for instance the Gaussian model with 1\% dark matter at $\mu=150$ (red curves, model C), between $\approx 20$\% and $\approx 30$\% of the cases show distortions in the magnification of at least a factor 4 below 200 Hz. At even larger magnification factors, PBH models with smaller fractions of dark matter would show similar results.
Note how the Gaussian model with 0.3\% dark matter as PBHs (model B) is clearly more sensitive than model A (only stars and remnants) despite the fact that model A has a larger surface mass density than model B. However, most of the mass in model A is contained in low-mass stars, for which diffraction effects are small below 200 Hz.  
It is important to stress out that the mass function of the PBHs assumed in this work is not the same as the one assumed for the two components producing the GWs (see section~\ref{sec_PDFmass}). In other words, we assume that the merger rate of BBHs is fixed and independent of the PBHs. For the GWs we assumed a chirp mass that results from drawing $m_1$ and $m_2$ from a power law which in our notation would have an exponent $\gamma_p=-1.3$. Such a steep mass function would produce, on average,  diffraction effects which are smaller than the ones presented in this work for $\gamma_p=0,1$, since most of the mass would concentrate in smaller objects.

\begin{figure}[b]
\includegraphics[width=9cm]{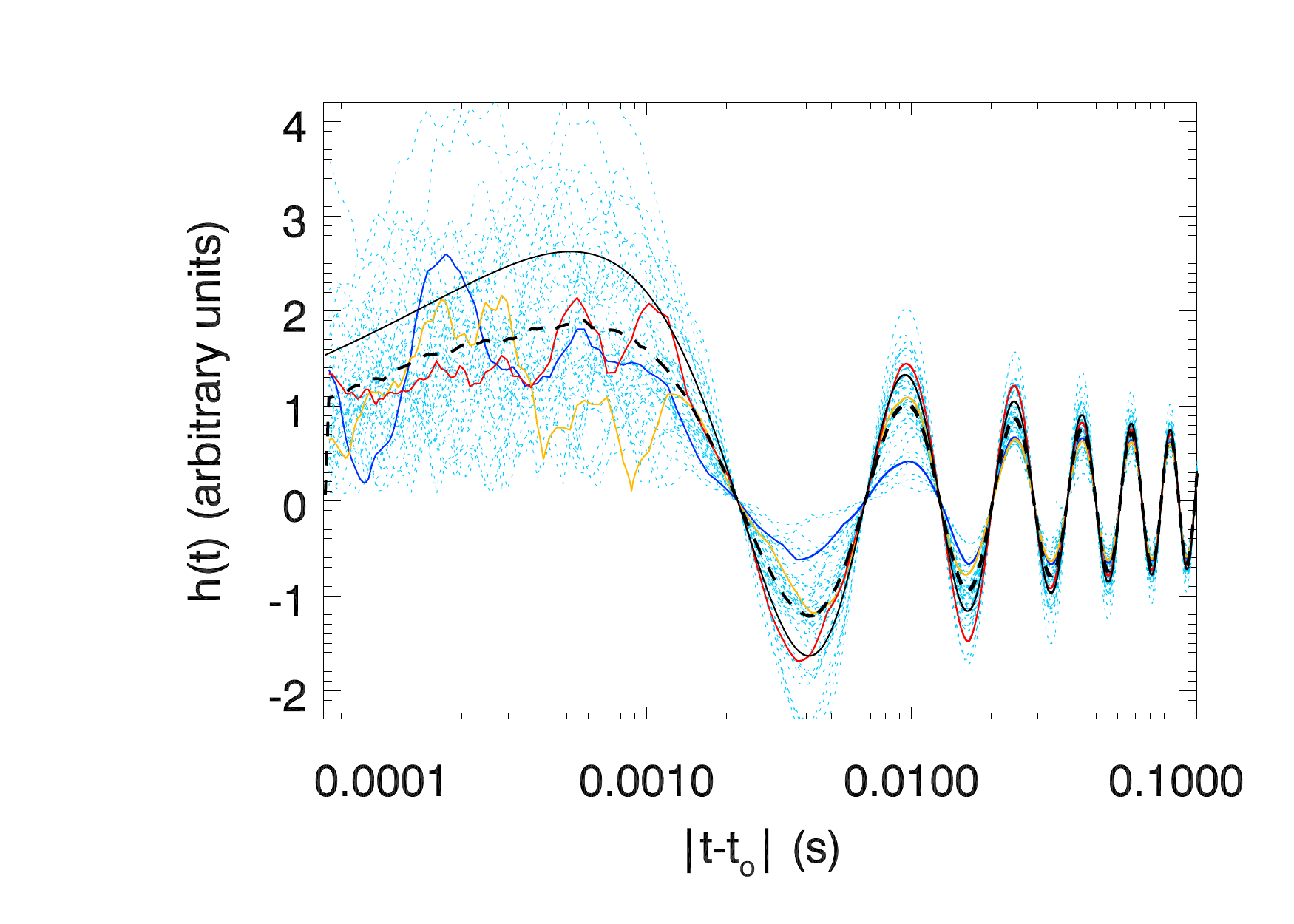}
\caption{\label{fig:StrainExamples1} Distortion of the strain for a lensed GW with $m_1=m_2=8$ \Msun at $z=1.5$ (black solid line). The colored curves correspond to the same cases shown in Figure~\ref{fig:FwExamples1}. The time or merger is at $t_o$.  The dashed black line shows the average over all random positions, where a systematic damping of the strain is clearly observed as the frequency of the GW increases.}
\end{figure}

Although all the information of the distortion due to microlensing is encapsulated in the relative magnification as a function of frequency shown in the previous section, it is interesting to show the effect over the observed strain. In Fig.~\ref{fig:StrainExamples1} we show examples for the case with magnification $\mu=150$, negative parity, and for the power law mass function model with $\gamma=1$. The plot shows the original signal of an undisturbed strain derived in the post-Newtonian approximation as a black solid line. For this strain we have assumed that $m_1=m_2=8$ \Msun and that the GW originated at z=1.5. The dotted lines show 50 examples of 50 GWs placed at random positions in the source plane. The first three of these 50 examples are shown as colored solid lines (blue, orange and red). These 50 curves (and the colored lines) are the same cases shown in Fig.~\ref{fig:FwExamples1} in section~\ref{sec_Gauss}. The time of coalescence takes place at $t=t_o$. Note how in some cases, the amplitude in the last cycles is affected substantially or even suppressed almost completely. Also, it is interesting to see how on average (black-dashed line), the amplitude in the last cycles is reduced by some factor that increases with frequency. This trend was already observed in Figs.~\ref{fig:FigPowLaw1} and~\ref{fig:FigPowLaw2} but it is observed more clearly in this plot. The suppression of the last cycles is more prominent in the macroimages with negative parity than in macroimages with positive parity \citep[see][ where it is also shown how the mean of the magnification remains unchanged when microlenses are present]{Diego2019}. This follows the expected behavior of lensing in the wave optics regime, where at low frequencies one is expected to be insensitive to the presence of microlenses (i.e, one recovers the mean of the magnification) while at high frequencies, the GWs can see the microlenses (i.e, the typical magnification tends towards the median of the magnification). 
If microlensing is predominant in strongly lensed GWs, and as shown by Figure~\ref{fig:StrainExamples1}, the last cycle before the ringdown is more likely to be damped (specially in macroimages with negative parity), it is possible that the estimated parameters of the GW (specially the one with negative parity) are biased if microlensing effects are ignored. 

If one assumes that the BDS model is correct (i.e, all the low-frequency events detected by LIGO are strongly lensed with magnification factors up to several hundreds), this would imply that the fraction of dark matter in the form of PBHs between 5 \Msun and 50 \Msun must be below the 10\% level. Otherwise, about half of the observed lensed GWs would show clear signs of microlensing in the strains. Similarly, if the fraction of dark matter in the form of PBHs in the same mass regime is of the order, or above, 10\%, this would imply that the BDS model can not be correct, and that the observed low frequency (or high chirp mass) GWs are not strongly lensed. More data, and with higher sensitivity (specially at high frequencies) is needed to settle this question. 

Future detectors of GWs, such as the Einstein telescope, will soon increase the sensitivity and extend the range in frequencies where events can be observed. Extending the range towards lower frequencies is interesting, not only for early warning reasons (events can be detected hundreds of seconds before the merger takes place), but also because at these low frequencies, microlensing effects are expected to be very small, and the parameters estimated at low frequencies can be later used to unveil signs of microlensing at high frequencies. Increasing the sensitivity at high frequencies would  allow to measure, with higher significance, the ring-down (very sensitive to microlensing) and also detect events with smaller chirp masses at cosmological distances. Binary neutron stars (or BNS) are expected to be more common than BBHs (the inferred volumetric rate of BNS at $z=0$ is $\approx 10^3$ per year and Gpc$^3$ \cite{LigoCatalog2019I}). However, since the chirp mass is smaller, the signal-to-noise ratio is also smaller than for BBHs and hence they can only be observed at smaller distances (where the optical depth of lensing is small). Once the sensitivity of future GW detectors increase, it will allow to observe BNS at larger cosmological distances. If magnified by factors of a few tens to a few hundreds, future observations can reveal lensed GWs from distant BNS at $z>1$. At higher redshifts, this rate could be even higher, offering a chance of observing these events through lensing. Interestingly, these events, when redshifted (assuming $z \sim 1-2$), would appear to be falling within the forbidden mass gap at low mass. 
Also, the relatively smaller chirp mass results in a larger $f_{max}$ and it is at these higher frequencies where wave effects are more prominent. 

\begin{acknowledgments}
The author thanks Tom Broadhurst, George Smoot, Bradley Kavanagh, Leo Singer, and Juan Garcia-Bellido for useful comments and discussions, and Patrick Kelly for gently providing the stellar+remnant mass function. 
J.M.D. acknowledges the support of project PGC2018-101814-B-100 (MCIU/AEI/MINECO/FEDER, UE) Ministerio de Ciencia, investigaci\'on y Universidades. 
\end{acknowledgments}

\nocite{*}

\bibliography{MyBib}

\begin{thebibliography}{76}%
\makeatletter
\providecommand \@ifxundefined [1]{%
 \@ifx{#1\undefined}
}%
\providecommand \@ifnum [1]{%
 \ifnum #1\expandafter \@firstoftwo
 \else \expandafter \@secondoftwo
 \fi
}%
\providecommand \@ifx [1]{%
 \ifx #1\expandafter \@firstoftwo
 \else \expandafter \@secondoftwo
 \fi
}%
\providecommand \natexlab [1]{#1}%
\providecommand \enquote  [1]{``#1''}%
\providecommand \bibnamefont  [1]{#1}%
\providecommand \bibfnamefont [1]{#1}%
\providecommand \citenamefont [1]{#1}%
\providecommand \href@noop [0]{\@secondoftwo}%
\providecommand \href [0]{\begingroup \@sanitize@url \@href}%
\providecommand \@href[1]{\@@startlink{#1}\@@href}%
\providecommand \@@href[1]{\endgroup#1\@@endlink}%
\providecommand \@sanitize@url [0]{\catcode `\\12\catcode `\$12\catcode
  `\&12\catcode `\#12\catcode `\^12\catcode `\_12\catcode `\%12\relax}%
\providecommand \@@startlink[1]{}%
\providecommand \@@endlink[0]{}%
\providecommand \url  [0]{\begingroup\@sanitize@url \@url }%
\providecommand \@url [1]{\endgroup\@href {#1}{\urlprefix }}%
\providecommand \urlprefix  [0]{URL }%
\providecommand \Eprint [0]{\href }%
\providecommand \doibase [0]{https://doi.org/}%
\providecommand \selectlanguage [0]{\@gobble}%
\providecommand \bibinfo  [0]{\@secondoftwo}%
\providecommand \bibfield  [0]{\@secondoftwo}%
\providecommand \translation [1]{[#1]}%
\providecommand \BibitemOpen [0]{}%
\providecommand \bibitemStop [0]{}%
\providecommand \bibitemNoStop [0]{.\EOS\space}%
\providecommand \EOS [0]{\spacefactor3000\relax}%
\providecommand \BibitemShut  [1]{\csname bibitem#1\endcsname}%
\let\auto@bib@innerbib\@empty
\bibitem [{\citenamefont {{Carr}}\ and\ \citenamefont
  {{Hawking}}(1974)}]{Carr1974}%
  \BibitemOpen
  \bibfield  {author} {\bibinfo {author} {\bibfnamefont {B.~J.}\ \bibnamefont
  {{Carr}}}\ and\ \bibinfo {author} {\bibfnamefont {S.~W.}\ \bibnamefont
  {{Hawking}}},\ }\bibfield  {title} {\bibinfo {title} {{Black holes in the
  early Universe}},\ }\href {https://doi.org/10.1093/mnras/168.2.399}
  {\bibfield  {journal} {\bibinfo  {journal} {\mnras}\ }\textbf {\bibinfo
  {volume} {168}},\ \bibinfo {pages} {399} (\bibinfo {year}
  {1974})}\BibitemShut {NoStop}%
\bibitem [{\citenamefont {{Carr}}\ \emph {et~al.}(2016)\citenamefont {{Carr}},
  \citenamefont {{K{\"u}hnel}},\ and\ \citenamefont {{Sandstad}}}]{Carr2016}%
  \BibitemOpen
  \bibfield  {author} {\bibinfo {author} {\bibfnamefont {B.}~\bibnamefont
  {{Carr}}}, \bibinfo {author} {\bibfnamefont {F.}~\bibnamefont
  {{K{\"u}hnel}}},\ and\ \bibinfo {author} {\bibfnamefont {M.}~\bibnamefont
  {{Sandstad}}},\ }\bibfield  {title} {\bibinfo {title} {{Primordial black
  holes as dark matter}},\ }\href {https://doi.org/10.1103/PhysRevD.94.083504}
  {\bibfield  {journal} {\bibinfo  {journal} {\prd}\ }\textbf {\bibinfo
  {volume} {94}},\ \bibinfo {eid} {083504} (\bibinfo {year} {2016})},\ \Eprint
  {https://arxiv.org/abs/1607.06077} {arXiv:1607.06077 [astro-ph.CO]}
  \BibitemShut {NoStop}%
\bibitem [{\citenamefont {{Garc{\'\i}a-Bellido}}\ \emph
  {et~al.}(1996)\citenamefont {{Garc{\'\i}a-Bellido}}, \citenamefont
  {{Linde}},\ and\ \citenamefont {{Wands}}}]{GarciaBellido1996}%
  \BibitemOpen
  \bibfield  {author} {\bibinfo {author} {\bibfnamefont {J.}~\bibnamefont
  {{Garc{\'\i}a-Bellido}}}, \bibinfo {author} {\bibfnamefont {A.}~\bibnamefont
  {{Linde}}},\ and\ \bibinfo {author} {\bibfnamefont {D.}~\bibnamefont
  {{Wands}}},\ }\bibfield  {title} {\bibinfo {title} {{Density perturbations
  and black hole formation in hybrid inflation}},\ }\href
  {https://doi.org/10.1103/PhysRevD.54.6040} {\bibfield  {journal} {\bibinfo
  {journal} {\prd}\ }\textbf {\bibinfo {volume} {54}},\ \bibinfo {pages} {6040}
  (\bibinfo {year} {1996})},\ \Eprint {https://arxiv.org/abs/astro-ph/9605094}
  {arXiv:astro-ph/9605094 [astro-ph]} \BibitemShut {NoStop}%
\bibitem [{\citenamefont {{Garc{\'\i}a-Bellido}}(2017)}]{Bellido2017}%
  \BibitemOpen
  \bibfield  {author} {\bibinfo {author} {\bibfnamefont {J.}~\bibnamefont
  {{Garc{\'\i}a-Bellido}}},\ }\bibfield  {title} {\bibinfo {title} {{Massive
  Primordial Black Holes as Dark Matter and their detection with Gravitational
  Waves}},\ }in\ \href {https://doi.org/10.1088/1742-6596/840/1/012032} {\emph
  {\bibinfo {booktitle} {Journal of Physics Conference Series}}},\ \bibinfo
  {series} {Journal of Physics Conference Series}, Vol.\ \bibinfo {volume}
  {840}\ (\bibinfo {year} {2017})\ p.\ \bibinfo {pages} {012032},\ \Eprint
  {https://arxiv.org/abs/1702.08275} {arXiv:1702.08275 [astro-ph.CO]}
  \BibitemShut {NoStop}%
\bibitem [{\citenamefont {{Clesse}}\ and\ \citenamefont
  {{Garc{\'\i}a-Bellido}}(2015)}]{Clesse2015}%
  \BibitemOpen
  \bibfield  {author} {\bibinfo {author} {\bibfnamefont {S.}~\bibnamefont
  {{Clesse}}}\ and\ \bibinfo {author} {\bibfnamefont {J.}~\bibnamefont
  {{Garc{\'\i}a-Bellido}}},\ }\bibfield  {title} {\bibinfo {title} {{Massive
  primordial black holes from hybrid inflation as dark matter and the seeds of
  galaxies}},\ }\href {https://doi.org/10.1103/PhysRevD.92.023524} {\bibfield
  {journal} {\bibinfo  {journal} {\prd}\ }\textbf {\bibinfo {volume} {92}},\
  \bibinfo {eid} {023524} (\bibinfo {year} {2015})},\ \Eprint
  {https://arxiv.org/abs/1501.07565} {arXiv:1501.07565 [astro-ph.CO]}
  \BibitemShut {NoStop}%
\bibitem [{\citenamefont {{Kavanagh}}\ \emph {et~al.}(2018)\citenamefont
  {{Kavanagh}}, \citenamefont {{Gaggero}},\ and\ \citenamefont
  {{Bertone}}}]{Kavanagh2018}%
  \BibitemOpen
  \bibfield  {author} {\bibinfo {author} {\bibfnamefont {B.~J.}\ \bibnamefont
  {{Kavanagh}}}, \bibinfo {author} {\bibfnamefont {D.}~\bibnamefont
  {{Gaggero}}},\ and\ \bibinfo {author} {\bibfnamefont {G.}~\bibnamefont
  {{Bertone}}},\ }\bibfield  {title} {\bibinfo {title} {{Merger rate of a
  subdominant population of primordial black holes}},\ }\href
  {https://doi.org/10.1103/PhysRevD.98.023536} {\bibfield  {journal} {\bibinfo
  {journal} {\prd}\ }\textbf {\bibinfo {volume} {98}},\ \bibinfo {eid} {023536}
  (\bibinfo {year} {2018})},\ \Eprint {https://arxiv.org/abs/1805.09034}
  {arXiv:1805.09034 [astro-ph.CO]} \BibitemShut {NoStop}%
\bibitem [{\citenamefont {{Deguchi}}\ and\ \citenamefont
  {{Watson}}(1986)}]{Deguchi1986}%
  \BibitemOpen
  \bibfield  {author} {\bibinfo {author} {\bibfnamefont {S.}~\bibnamefont
  {{Deguchi}}}\ and\ \bibinfo {author} {\bibfnamefont {W.~D.}\ \bibnamefont
  {{Watson}}},\ }\bibfield  {title} {\bibinfo {title} {{Diffraction in
  gravitational lensing for compact objects of low mass}},\ }\href
  {https://doi.org/10.1086/164389} {\bibfield  {journal} {\bibinfo  {journal}
  {\apj}\ }\textbf {\bibinfo {volume} {307}},\ \bibinfo {pages} {30} (\bibinfo
  {year} {1986})}\BibitemShut {NoStop}%
\bibitem [{\citenamefont {{Nakamura}}\ and\ \citenamefont
  {{Deguchi}}(1999)}]{Nakamura1999}%
  \BibitemOpen
  \bibfield  {author} {\bibinfo {author} {\bibfnamefont {T.~T.}\ \bibnamefont
  {{Nakamura}}}\ and\ \bibinfo {author} {\bibfnamefont {S.}~\bibnamefont
  {{Deguchi}}},\ }\bibfield  {title} {\bibinfo {title} {{Wave Optics in
  Gravitational Lensing}},\ }\href {https://doi.org/10.1143/PTPS.133.137}
  {\bibfield  {journal} {\bibinfo  {journal} {Progress of Theoretical Physics
  Supplement}\ }\textbf {\bibinfo {volume} {133}},\ \bibinfo {pages} {137}
  (\bibinfo {year} {1999})}\BibitemShut {NoStop}%
\bibitem [{\citenamefont {{Wang}}\ \emph {et~al.}(1996)\citenamefont {{Wang}},
  \citenamefont {{Stebbins}},\ and\ \citenamefont {{Turner}}}]{Wang1996}%
  \BibitemOpen
  \bibfield  {author} {\bibinfo {author} {\bibfnamefont {Y.}~\bibnamefont
  {{Wang}}}, \bibinfo {author} {\bibfnamefont {A.}~\bibnamefont {{Stebbins}}},\
  and\ \bibinfo {author} {\bibfnamefont {E.~L.}\ \bibnamefont {{Turner}}},\
  }\bibfield  {title} {\bibinfo {title} {{Gravitational Lensing of
  Gravitational Waves from Merging Neutron Star Binaries}},\ }\href
  {https://doi.org/10.1103/PhysRevLett.77.2875} {\bibfield  {journal} {\bibinfo
   {journal} {Physical Review Letters}\ }\textbf {\bibinfo {volume} {77}},\
  \bibinfo {pages} {2875} (\bibinfo {year} {1996})},\ \Eprint
  {https://arxiv.org/abs/astro-ph/9605140} {astro-ph/9605140} \BibitemShut
  {NoStop}%
\bibitem [{\citenamefont {{Nakamura}}(1998)}]{Nakamura1998}%
  \BibitemOpen
  \bibfield  {author} {\bibinfo {author} {\bibfnamefont {T.~T.}\ \bibnamefont
  {{Nakamura}}},\ }\bibfield  {title} {\bibinfo {title} {{Gravitational Lensing
  of Gravitational Waves from Inspiraling Binaries by a Point Mass Lens}},\
  }\href {https://doi.org/10.1103/PhysRevLett.80.1138} {\bibfield  {journal}
  {\bibinfo  {journal} {Physical Review Letters}\ }\textbf {\bibinfo {volume}
  {80}},\ \bibinfo {pages} {1138} (\bibinfo {year} {1998})}\BibitemShut
  {NoStop}%
\bibitem [{\citenamefont {{Sereno}}\ \emph {et~al.}(2010)\citenamefont
  {{Sereno}}, \citenamefont {{Sesana}}, \citenamefont {{Bleuler}},
  \citenamefont {{Jetzer}}, \citenamefont {{Volonteri}},\ and\ \citenamefont
  {{Begelman}}}]{Sereno2010}%
  \BibitemOpen
  \bibfield  {author} {\bibinfo {author} {\bibfnamefont {M.}~\bibnamefont
  {{Sereno}}}, \bibinfo {author} {\bibfnamefont {A.}~\bibnamefont {{Sesana}}},
  \bibinfo {author} {\bibfnamefont {A.}~\bibnamefont {{Bleuler}}}, \bibinfo
  {author} {\bibfnamefont {P.}~\bibnamefont {{Jetzer}}}, \bibinfo {author}
  {\bibfnamefont {M.}~\bibnamefont {{Volonteri}}},\ and\ \bibinfo {author}
  {\bibfnamefont {M.~C.}\ \bibnamefont {{Begelman}}},\ }\bibfield  {title}
  {\bibinfo {title} {{Strong Lensing of Gravitational Waves as Seen by LISA}},\
  }\href {https://doi.org/10.1103/PhysRevLett.105.251101} {\bibfield  {journal}
  {\bibinfo  {journal} {Physical Review Letters}\ }\textbf {\bibinfo {volume}
  {105}},\ \bibinfo {eid} {251101} (\bibinfo {year} {2010})},\ \Eprint
  {https://arxiv.org/abs/1011.5238} {arXiv:1011.5238 [astro-ph.CO]}
  \BibitemShut {NoStop}%
\bibitem [{\citenamefont {Cao}\ \emph {et~al.}(2014)\citenamefont {Cao},
  \citenamefont {Li},\ and\ \citenamefont {Wang}}]{Cao2014}%
  \BibitemOpen
  \bibfield  {author} {\bibinfo {author} {\bibfnamefont {Z.}~\bibnamefont
  {Cao}}, \bibinfo {author} {\bibfnamefont {L.-F.}\ \bibnamefont {Li}},\ and\
  \bibinfo {author} {\bibfnamefont {Y.}~\bibnamefont {Wang}},\ }\bibfield
  {title} {\bibinfo {title} {Gravitational lensing effects on parameter
  estimation in gravitational wave detection with advanced detectors},\ }\href
  {https://doi.org/10.1103/PhysRevD.90.062003} {\bibfield  {journal} {\bibinfo
  {journal} {Phys. Rev. D}\ }\textbf {\bibinfo {volume} {90}},\ \bibinfo
  {pages} {062003} (\bibinfo {year} {2014})}\BibitemShut {NoStop}%
\bibitem [{\citenamefont {{Ng}}\ \emph {et~al.}(2018)\citenamefont {{Ng}},
  \citenamefont {{Wong}}, \citenamefont {{Broadhurst}},\ and\ \citenamefont
  {{Li}}}]{Ng2018}%
  \BibitemOpen
  \bibfield  {author} {\bibinfo {author} {\bibfnamefont {K.~K.~Y.}\
  \bibnamefont {{Ng}}}, \bibinfo {author} {\bibfnamefont {K.~W.~K.}\
  \bibnamefont {{Wong}}}, \bibinfo {author} {\bibfnamefont {T.}~\bibnamefont
  {{Broadhurst}}},\ and\ \bibinfo {author} {\bibfnamefont {T.~G.~F.}\
  \bibnamefont {{Li}}},\ }\bibfield  {title} {\bibinfo {title} {{Precise LIGO
  lensing rate predictions for binary black holes}},\ }\href
  {https://doi.org/10.1103/PhysRevD.97.023012} {\bibfield  {journal} {\bibinfo
  {journal} {\prd}\ }\textbf {\bibinfo {volume} {97}},\ \bibinfo {eid} {023012}
  (\bibinfo {year} {2018})},\ \Eprint {https://arxiv.org/abs/1703.06319}
  {arXiv:1703.06319} \BibitemShut {NoStop}%
\bibitem [{\citenamefont {{Jung}}\ and\ \citenamefont
  {{Shin}}(2019)}]{Jung2019}%
  \BibitemOpen
  \bibfield  {author} {\bibinfo {author} {\bibfnamefont {S.}~\bibnamefont
  {{Jung}}}\ and\ \bibinfo {author} {\bibfnamefont {C.~S.}\ \bibnamefont
  {{Shin}}},\ }\bibfield  {title} {\bibinfo {title} {{Gravitational-Wave
  Fringes at LIGO: Detecting Compact Dark Matter by Gravitational Lensing}},\
  }\href {https://doi.org/10.1103/PhysRevLett.122.041103} {\bibfield  {journal}
  {\bibinfo  {journal} {Physical Review Letters}\ }\textbf {\bibinfo {volume}
  {122}},\ \bibinfo {eid} {041103} (\bibinfo {year} {2019})},\ \Eprint
  {https://arxiv.org/abs/1712.01396} {arXiv:1712.01396} \BibitemShut {NoStop}%
\bibitem [{\citenamefont {{Diego}}\ \emph {et~al.}(2019)\citenamefont
  {{Diego}}, \citenamefont {{Hannuksela}}, \citenamefont {{Kelly}},
  \citenamefont {{Broadhurst}}, \citenamefont {{Kim}}, \citenamefont {{Li}},\
  and\ \citenamefont {{Smoot}}}]{Diego2019b}%
  \BibitemOpen
  \bibfield  {author} {\bibinfo {author} {\bibfnamefont {J.~M.}\ \bibnamefont
  {{Diego}}}, \bibinfo {author} {\bibfnamefont {O.~A.}\ \bibnamefont
  {{Hannuksela}}}, \bibinfo {author} {\bibfnamefont {P.~L.}\ \bibnamefont
  {{Kelly}}}, \bibinfo {author} {\bibfnamefont {T.}~\bibnamefont
  {{Broadhurst}}}, \bibinfo {author} {\bibfnamefont {K.}~\bibnamefont {{Kim}}},
  \bibinfo {author} {\bibfnamefont {T.~G.~F.}\ \bibnamefont {{Li}}},\ and\
  \bibinfo {author} {\bibfnamefont {G.~F.}\ \bibnamefont {{Smoot}}},\
  }\bibfield  {title} {\bibinfo {title} {{Observational signatures of
  microlensing in gravitational waves at LIGO/Virgo frequencies}},\ }\href@noop
  {} {\bibfield  {journal} {\bibinfo  {journal} {arXiv e-prints}\ } (\bibinfo
  {year} {2019})},\ \Eprint {https://arxiv.org/abs/1903.04513}
  {arXiv:1903.04513} \BibitemShut {NoStop}%
\bibitem [{\citenamefont {{Diego}}\ \emph {et~al.}(2018)\citenamefont
  {{Diego}}, \citenamefont {{Kaiser}}, \citenamefont {{Broadhurst}},
  \citenamefont {{Kelly}}, \citenamefont {{Rodney}}, \citenamefont
  {{Morishita}}, \citenamefont {{Oguri}}, \citenamefont {{Ross}}, \citenamefont
  {{Zitrin}}, \citenamefont {{Jauzac}}, \citenamefont {{Richard}},
  \citenamefont {{Williams}}, \citenamefont {{Vega-Ferrero}}, \citenamefont
  {{Frye}},\ and\ \citenamefont {{Filippenko}}}]{Diego2018}%
  \BibitemOpen
  \bibfield  {author} {\bibinfo {author} {\bibfnamefont {J.~M.}\ \bibnamefont
  {{Diego}}}, \bibinfo {author} {\bibfnamefont {N.}~\bibnamefont {{Kaiser}}},
  \bibinfo {author} {\bibfnamefont {T.}~\bibnamefont {{Broadhurst}}}, \bibinfo
  {author} {\bibfnamefont {P.~L.}\ \bibnamefont {{Kelly}}}, \bibinfo {author}
  {\bibfnamefont {S.}~\bibnamefont {{Rodney}}}, \bibinfo {author}
  {\bibfnamefont {T.}~\bibnamefont {{Morishita}}}, \bibinfo {author}
  {\bibfnamefont {M.}~\bibnamefont {{Oguri}}}, \bibinfo {author} {\bibfnamefont
  {T.~W.}\ \bibnamefont {{Ross}}}, \bibinfo {author} {\bibfnamefont
  {A.}~\bibnamefont {{Zitrin}}}, \bibinfo {author} {\bibfnamefont
  {M.}~\bibnamefont {{Jauzac}}}, \bibinfo {author} {\bibfnamefont
  {J.}~\bibnamefont {{Richard}}}, \bibinfo {author} {\bibfnamefont
  {L.}~\bibnamefont {{Williams}}}, \bibinfo {author} {\bibfnamefont
  {J.}~\bibnamefont {{Vega-Ferrero}}}, \bibinfo {author} {\bibfnamefont
  {B.}~\bibnamefont {{Frye}}},\ and\ \bibinfo {author} {\bibfnamefont {A.~V.}\
  \bibnamefont {{Filippenko}}},\ }\bibfield  {title} {\bibinfo {title} {{Dark
  Matter under the Microscope: Constraining Compact Dark Matter with Caustic
  Crossing Events}},\ }\href {https://doi.org/10.3847/1538-4357/aab617}
  {\bibfield  {journal} {\bibinfo  {journal} {\apj}\ }\textbf {\bibinfo
  {volume} {857}},\ \bibinfo {eid} {25} (\bibinfo {year} {2018})},\ \Eprint
  {https://arxiv.org/abs/1706.10281} {arXiv:1706.10281} \BibitemShut {NoStop}%
\bibitem [{\citenamefont {{Dai}}\ \emph {et~al.}(2017)\citenamefont {{Dai}},
  \citenamefont {{Venumadhav}},\ and\ \citenamefont {{Sigurdson}}}]{Dai2017}%
  \BibitemOpen
  \bibfield  {author} {\bibinfo {author} {\bibfnamefont {L.}~\bibnamefont
  {{Dai}}}, \bibinfo {author} {\bibfnamefont {T.}~\bibnamefont
  {{Venumadhav}}},\ and\ \bibinfo {author} {\bibfnamefont {K.}~\bibnamefont
  {{Sigurdson}}},\ }\bibfield  {title} {\bibinfo {title} {{Effect of lensing
  magnification on the apparent distribution of black hole mergers}},\ }\href
  {https://doi.org/10.1103/PhysRevD.95.044011} {\bibfield  {journal} {\bibinfo
  {journal} {\prd}\ }\textbf {\bibinfo {volume} {95}},\ \bibinfo {eid} {044011}
  (\bibinfo {year} {2017})},\ \Eprint {https://arxiv.org/abs/1605.09398}
  {arXiv:1605.09398} \BibitemShut {NoStop}%
\bibitem [{\citenamefont {{Broadhurst}}\ \emph {et~al.}(2018)\citenamefont
  {{Broadhurst}}, \citenamefont {{Diego}},\ and\ \citenamefont
  {{Smoot}}}]{Broadhurst2018}%
  \BibitemOpen
  \bibfield  {author} {\bibinfo {author} {\bibfnamefont {T.}~\bibnamefont
  {{Broadhurst}}}, \bibinfo {author} {\bibfnamefont {J.~M.}\ \bibnamefont
  {{Diego}}},\ and\ \bibinfo {author} {\bibfnamefont {G.}~\bibnamefont
  {{Smoot}}, \bibfnamefont {III}},\ }\bibfield  {title} {\bibinfo {title}
  {{Reinterpreting Low Frequency LIGO/Virgo Events as Magnified Stellar-Mass
  Black Holes at Cosmological Distances}},\ }\href@noop {} {\bibfield
  {journal} {\bibinfo  {journal} {arXiv e-prints}\ } (\bibinfo {year}
  {2018})},\ \Eprint {https://arxiv.org/abs/1802.05273} {arXiv:1802.05273}
  \BibitemShut {NoStop}%
\bibitem [{\citenamefont {{Lai}}\ \emph {et~al.}(2018)\citenamefont {{Lai}},
  \citenamefont {{Hannuksela}}, \citenamefont {{Herrera-Mart{\'{\i}}n}},
  \citenamefont {{Diego}}, \citenamefont {{Broadhurst}},\ and\ \citenamefont
  {{Li}}}]{Lai2018}%
  \BibitemOpen
  \bibfield  {author} {\bibinfo {author} {\bibfnamefont {K.-H.}\ \bibnamefont
  {{Lai}}}, \bibinfo {author} {\bibfnamefont {O.~A.}\ \bibnamefont
  {{Hannuksela}}}, \bibinfo {author} {\bibfnamefont {A.}~\bibnamefont
  {{Herrera-Mart{\'{\i}}n}}}, \bibinfo {author} {\bibfnamefont {J.~M.}\
  \bibnamefont {{Diego}}}, \bibinfo {author} {\bibfnamefont {T.}~\bibnamefont
  {{Broadhurst}}},\ and\ \bibinfo {author} {\bibfnamefont {T.~G.~F.}\
  \bibnamefont {{Li}}},\ }\bibfield  {title} {\bibinfo {title} {{Discovering
  intermediate-mass black hole lenses through gravitational wave lensing}},\
  }\href {https://doi.org/10.1103/PhysRevD.98.083005} {\bibfield  {journal}
  {\bibinfo  {journal} {\prd}\ }\textbf {\bibinfo {volume} {98}},\ \bibinfo
  {eid} {083005} (\bibinfo {year} {2018})},\ \Eprint
  {https://arxiv.org/abs/1801.07840} {arXiv:1801.07840 [gr-qc]} \BibitemShut
  {NoStop}%
\bibitem [{\citenamefont {{Dai}}\ \emph {et~al.}(2018)\citenamefont {{Dai}},
  \citenamefont {{Li}}, \citenamefont {{Zackay}}, \citenamefont {{Mao}},\ and\
  \citenamefont {{Lu}}}]{Dai2018}%
  \BibitemOpen
  \bibfield  {author} {\bibinfo {author} {\bibfnamefont {L.}~\bibnamefont
  {{Dai}}}, \bibinfo {author} {\bibfnamefont {S.-S.}\ \bibnamefont {{Li}}},
  \bibinfo {author} {\bibfnamefont {B.}~\bibnamefont {{Zackay}}}, \bibinfo
  {author} {\bibfnamefont {S.}~\bibnamefont {{Mao}}},\ and\ \bibinfo {author}
  {\bibfnamefont {Y.}~\bibnamefont {{Lu}}},\ }\bibfield  {title} {\bibinfo
  {title} {{Detecting lensing-induced diffraction in astrophysical
  gravitational waves}},\ }\href {https://doi.org/10.1103/PhysRevD.98.104029}
  {\bibfield  {journal} {\bibinfo  {journal} {\prd}\ }\textbf {\bibinfo
  {volume} {98}},\ \bibinfo {eid} {104029} (\bibinfo {year} {2018})},\ \Eprint
  {https://arxiv.org/abs/1810.00003} {arXiv:1810.00003 [gr-qc]} \BibitemShut
  {NoStop}%
\bibitem [{\citenamefont {{Christian}}\ \emph {et~al.}(2018)\citenamefont
  {{Christian}}, \citenamefont {{Vitale}},\ and\ \citenamefont
  {{Loeb}}}]{Christian2018}%
  \BibitemOpen
  \bibfield  {author} {\bibinfo {author} {\bibfnamefont {P.}~\bibnamefont
  {{Christian}}}, \bibinfo {author} {\bibfnamefont {S.}~\bibnamefont
  {{Vitale}}},\ and\ \bibinfo {author} {\bibfnamefont {A.}~\bibnamefont
  {{Loeb}}},\ }\bibfield  {title} {\bibinfo {title} {{Detecting stellar lensing
  of gravitational waves with ground-based observatories}},\ }\href
  {https://doi.org/10.1103/PhysRevD.98.103022} {\bibfield  {journal} {\bibinfo
  {journal} {\prd}\ }\textbf {\bibinfo {volume} {98}},\ \bibinfo {eid} {103022}
  (\bibinfo {year} {2018})},\ \Eprint {https://arxiv.org/abs/1802.02586}
  {arXiv:1802.02586 [astro-ph.HE]} \BibitemShut {NoStop}%
\bibitem [{\citenamefont {{Oguri}}(2018)}]{Oguri2018b}%
  \BibitemOpen
  \bibfield  {author} {\bibinfo {author} {\bibfnamefont {M.}~\bibnamefont
  {{Oguri}}},\ }\bibfield  {title} {\bibinfo {title} {{Effect of gravitational
  lensing on the distribution of gravitational waves from distant binary black
  hole mergers}},\ }\href {https://doi.org/10.1093/mnras/sty2145} {\bibfield
  {journal} {\bibinfo  {journal} {\mnras}\ }\textbf {\bibinfo {volume} {480}},\
  \bibinfo {pages} {3842} (\bibinfo {year} {2018})},\ \Eprint
  {https://arxiv.org/abs/1807.02584} {arXiv:1807.02584} \BibitemShut {NoStop}%
\bibitem [{\citenamefont {{Li}}\ \emph {et~al.}(2018)\citenamefont {{Li}},
  \citenamefont {{Tang}}, \citenamefont {{Lin}},\ and\ \citenamefont
  {{Wang}}}]{Li2018}%
  \BibitemOpen
  \bibfield  {author} {\bibinfo {author} {\bibfnamefont {X.}~\bibnamefont
  {{Li}}}, \bibinfo {author} {\bibfnamefont {L.}~\bibnamefont {{Tang}}},
  \bibinfo {author} {\bibfnamefont {H.-N.}\ \bibnamefont {{Lin}}},\ and\
  \bibinfo {author} {\bibfnamefont {L.-L.}\ \bibnamefont {{Wang}}},\ }\bibfield
   {title} {\bibinfo {title} {{Testing the variation of the fine structure
  constant with strongly lensed gravitational waves}},\ }\href
  {https://doi.org/10.1088/1674-1137/42/9/095104} {\bibfield  {journal}
  {\bibinfo  {journal} {Chinese Physics C}\ }\textbf {\bibinfo {volume} {42}},\
  \bibinfo {eid} {095104} (\bibinfo {year} {2018})},\ \Eprint
  {https://arxiv.org/abs/1809.00474} {arXiv:1809.00474 [gr-qc]} \BibitemShut
  {NoStop}%
\bibitem [{\citenamefont {Takahashi}\ and\ \citenamefont
  {Nakamura}(2003)}]{takahashi2003gravitational}%
  \BibitemOpen
  \bibfield  {author} {\bibinfo {author} {\bibfnamefont {R.}~\bibnamefont
  {Takahashi}}\ and\ \bibinfo {author} {\bibfnamefont {T.}~\bibnamefont
  {Nakamura}},\ }\bibfield  {title} {\bibinfo {title} {Wave effects in the
  gravitational lensing of gravitational waves from chirping binaries},\
  }\href@noop {} {\bibfield  {journal} {\bibinfo  {journal} {The Astrophysical
  Journal}\ }\textbf {\bibinfo {volume} {595}},\ \bibinfo {pages} {1039}
  (\bibinfo {year} {2003})}\BibitemShut {NoStop}%
\bibitem [{\citenamefont {{Cutler}}\ and\ \citenamefont
  {{Flanagan}}(1994)}]{Cutler1994}%
  \BibitemOpen
  \bibfield  {author} {\bibinfo {author} {\bibfnamefont {C.}~\bibnamefont
  {{Cutler}}}\ and\ \bibinfo {author} {\bibfnamefont {{\'E}.~E.}\ \bibnamefont
  {{Flanagan}}},\ }\bibfield  {title} {\bibinfo {title} {{Gravitational waves
  from merging compact binaries: How accurately can one extract the binary's
  parameters from the inspiral waveform\textbackslash?}},\ }\href
  {https://doi.org/10.1103/PhysRevD.49.2658} {\bibfield  {journal} {\bibinfo
  {journal} {\prd}\ }\textbf {\bibinfo {volume} {49}},\ \bibinfo {pages} {2658}
  (\bibinfo {year} {1994})},\ \Eprint {https://arxiv.org/abs/gr-qc/9402014}
  {arXiv:gr-qc/9402014 [gr-qc]} \BibitemShut {NoStop}%
\bibitem [{\citenamefont {{Finn}}(1996)}]{Finn1996}%
  \BibitemOpen
  \bibfield  {author} {\bibinfo {author} {\bibfnamefont {L.~S.}\ \bibnamefont
  {{Finn}}},\ }\bibfield  {title} {\bibinfo {title} {{Binary inspiral,
  gravitational radiation, and cosmology}},\ }\href
  {https://doi.org/10.1103/PhysRevD.53.2878} {\bibfield  {journal} {\bibinfo
  {journal} {\prd}\ }\textbf {\bibinfo {volume} {53}},\ \bibinfo {pages} {2878}
  (\bibinfo {year} {1996})},\ \Eprint {https://arxiv.org/abs/gr-qc/9601048}
  {arXiv:gr-qc/9601048 [gr-qc]} \BibitemShut {NoStop}%
\bibitem [{Note1()}]{Note1}%
  \BibitemOpen
  \bibinfo {note} {One should note that the most likely value for $\mu $ is
  slightly below 1.0 (i.e a subtle demagnification) which, when ignored, leads
  to an overprediction of the distance and hence a systematic bias in the
  estimation of $H_o$ if not properly accounted for}\BibitemShut {NoStop}%
\bibitem [{\citenamefont {{The LIGO Scientific Collaboration }}\ and\
  \citenamefont {{ the Virgo Collaboration}}(2019)}]{LigoCatalog2019II}%
  \BibitemOpen
  \bibfield  {author} {\bibinfo {author} {\bibnamefont {{The LIGO Scientific
  Collaboration }}}\ and\ \bibinfo {author} {\bibnamefont {{ the Virgo
  Collaboration}}},\ }\bibfield  {title} {\bibinfo {title} {{Binary Black Hole
  Population Properties Inferred from the First and Second Observing Runs of
  Advanced LIGO and Advanced Virgo}},\ }\href
  {https://doi.org/10.3847/2041-8213/ab3800} {\bibfield  {journal} {\bibinfo
  {journal} {\apjl}\ }\textbf {\bibinfo {volume} {882}},\ \bibinfo {eid} {L24}
  (\bibinfo {year} {2019})}\BibitemShut {NoStop}%
\bibitem [{\citenamefont {{Kroupa}}(2001)}]{Kroupa2001}%
  \BibitemOpen
  \bibfield  {author} {\bibinfo {author} {\bibfnamefont {P.}~\bibnamefont
  {{Kroupa}}},\ }\bibfield  {title} {\bibinfo {title} {{On the variation of the
  initial mass function}},\ }\href
  {https://doi.org/10.1046/j.1365-8711.2001.04022.x} {\bibfield  {journal}
  {\bibinfo  {journal} {\mnras}\ }\textbf {\bibinfo {volume} {322}},\ \bibinfo
  {pages} {231} (\bibinfo {year} {2001})},\ \Eprint
  {https://arxiv.org/abs/astro-ph/0009005} {arXiv:astro-ph/0009005 [astro-ph]}
  \BibitemShut {NoStop}%
\bibitem [{\citenamefont {{Diego}}(2019)}]{Diego2019}%
  \BibitemOpen
  \bibfield  {author} {\bibinfo {author} {\bibfnamefont {J.~M.}\ \bibnamefont
  {{Diego}}},\ }\bibfield  {title} {\bibinfo {title} {{The Universe at extreme
  magnification}},\ }\href {https://doi.org/10.1051/0004-6361/201833670}
  {\bibfield  {journal} {\bibinfo  {journal} {\aap}\ }\textbf {\bibinfo
  {volume} {625}},\ \bibinfo {eid} {A84} (\bibinfo {year} {2019})},\ \Eprint
  {https://arxiv.org/abs/1806.04668} {arXiv:1806.04668 [astro-ph.GA]}
  \BibitemShut {NoStop}%
\bibitem [{\citenamefont {{Watson}}\ \emph {et~al.}(2014)\citenamefont
  {{Watson}}, \citenamefont {{Iliev}}, \citenamefont {{Diego}}, \citenamefont
  {{Gottl{\"o}ber}}, \citenamefont {{Knebe}}, \citenamefont
  {{Mart{\'{\i}}nez-Gonz{\'a}lez}},\ and\ \citenamefont
  {{Yepes}}}]{Watson2014}%
  \BibitemOpen
  \bibfield  {author} {\bibinfo {author} {\bibfnamefont {W.~A.}\ \bibnamefont
  {{Watson}}}, \bibinfo {author} {\bibfnamefont {I.~T.}\ \bibnamefont
  {{Iliev}}}, \bibinfo {author} {\bibfnamefont {J.~M.}\ \bibnamefont
  {{Diego}}}, \bibinfo {author} {\bibfnamefont {S.}~\bibnamefont
  {{Gottl{\"o}ber}}}, \bibinfo {author} {\bibfnamefont {A.}~\bibnamefont
  {{Knebe}}}, \bibinfo {author} {\bibfnamefont {E.}~\bibnamefont
  {{Mart{\'{\i}}nez-Gonz{\'a}lez}}},\ and\ \bibinfo {author} {\bibfnamefont
  {G.}~\bibnamefont {{Yepes}}},\ }\bibfield  {title} {\bibinfo {title}
  {{Statistics of extreme objects in the Juropa Hubble Volume simulation}},\
  }\href {https://doi.org/10.1093/mnras/stt2173} {\bibfield  {journal}
  {\bibinfo  {journal} {\mnras}\ }\textbf {\bibinfo {volume} {437}},\ \bibinfo
  {pages} {3776} (\bibinfo {year} {2014})},\ \Eprint
  {https://arxiv.org/abs/1305.1976} {arXiv:1305.1976} \BibitemShut {NoStop}%
\bibitem [{\citenamefont {{Hilbert}}\ \emph {et~al.}(2008)\citenamefont
  {{Hilbert}}, \citenamefont {{White}}, \citenamefont {{Hartlap}},\ and\
  \citenamefont {{Schneider}}}]{Hilbert2008}%
  \BibitemOpen
  \bibfield  {author} {\bibinfo {author} {\bibfnamefont {S.}~\bibnamefont
  {{Hilbert}}}, \bibinfo {author} {\bibfnamefont {S.~D.~M.}\ \bibnamefont
  {{White}}}, \bibinfo {author} {\bibfnamefont {J.}~\bibnamefont {{Hartlap}}},\
  and\ \bibinfo {author} {\bibfnamefont {P.}~\bibnamefont {{Schneider}}},\
  }\bibfield  {title} {\bibinfo {title} {{Strong-lensing optical depths in a
  {$\Lambda$}CDM universe - II. The influence of the stellar mass in
  galaxies}},\ }\href {https://doi.org/10.1111/j.1365-2966.2008.13190.x}
  {\bibfield  {journal} {\bibinfo  {journal} {\mnras}\ }\textbf {\bibinfo
  {volume} {386}},\ \bibinfo {pages} {1845} (\bibinfo {year} {2008})},\ \Eprint
  {https://arxiv.org/abs/0712.1593} {arXiv:0712.1593} \BibitemShut {NoStop}%
\bibitem [{\citenamefont {{Takahashi}}\ \emph {et~al.}(2011)\citenamefont
  {{Takahashi}}, \citenamefont {{Oguri}}, \citenamefont {{Sato}},\ and\
  \citenamefont {{Hamana}}}]{Takahashi2011}%
  \BibitemOpen
  \bibfield  {author} {\bibinfo {author} {\bibfnamefont {R.}~\bibnamefont
  {{Takahashi}}}, \bibinfo {author} {\bibfnamefont {M.}~\bibnamefont
  {{Oguri}}}, \bibinfo {author} {\bibfnamefont {M.}~\bibnamefont {{Sato}}},\
  and\ \bibinfo {author} {\bibfnamefont {T.}~\bibnamefont {{Hamana}}},\
  }\bibfield  {title} {\bibinfo {title} {{Probability Distribution Functions of
  Cosmological Lensing: Convergence, Shear, and Magnification}},\ }\href
  {https://doi.org/10.1088/0004-637X/742/1/15} {\bibfield  {journal} {\bibinfo
  {journal} {\apj}\ }\textbf {\bibinfo {volume} {742}},\ \bibinfo {eid} {15}
  (\bibinfo {year} {2011})},\ \Eprint {https://arxiv.org/abs/1106.3823}
  {arXiv:1106.3823} \BibitemShut {NoStop}%
\bibitem [{\citenamefont {{Madau}}\ and\ \citenamefont
  {{Dickinson}}(2014)}]{Madau2014}%
  \BibitemOpen
  \bibfield  {author} {\bibinfo {author} {\bibfnamefont {P.}~\bibnamefont
  {{Madau}}}\ and\ \bibinfo {author} {\bibfnamefont {M.}~\bibnamefont
  {{Dickinson}}},\ }\bibfield  {title} {\bibinfo {title} {{Cosmic
  Star-Formation History}},\ }\href
  {https://doi.org/10.1146/annurev-astro-081811-125615} {\bibfield  {journal}
  {\bibinfo  {journal} {\araa}\ }\textbf {\bibinfo {volume} {52}},\ \bibinfo
  {pages} {415} (\bibinfo {year} {2014})},\ \Eprint
  {https://arxiv.org/abs/1403.0007} {arXiv:1403.0007} \BibitemShut {NoStop}%
\bibitem [{\citenamefont {{Clesse}}\ and\ \citenamefont
  {{Garc{\'\i}a-Bellido}}(2017{\natexlab{a}})}]{Clesse2017}%
  \BibitemOpen
  \bibfield  {author} {\bibinfo {author} {\bibfnamefont {S.}~\bibnamefont
  {{Clesse}}}\ and\ \bibinfo {author} {\bibfnamefont {J.}~\bibnamefont
  {{Garc{\'\i}a-Bellido}}},\ }\bibfield  {title} {\bibinfo {title} {{The
  clustering of massive Primordial Black Holes as Dark Matter: Measuring their
  mass distribution with advanced LIGO}},\ }\href
  {https://doi.org/10.1016/j.dark.2016.10.002} {\bibfield  {journal} {\bibinfo
  {journal} {Physics of the Dark Universe}\ }\textbf {\bibinfo {volume} {15}},\
  \bibinfo {pages} {142} (\bibinfo {year} {2017}{\natexlab{a}})},\ \Eprint
  {https://arxiv.org/abs/1603.05234} {arXiv:1603.05234 [astro-ph.CO]}
  \BibitemShut {NoStop}%
\bibitem [{\citenamefont {{Portegies Zwart}}\ and\ \citenamefont
  {{McMillan}}(2000)}]{Zwart2000}%
  \BibitemOpen
  \bibfield  {author} {\bibinfo {author} {\bibfnamefont {S.~F.}\ \bibnamefont
  {{Portegies Zwart}}}\ and\ \bibinfo {author} {\bibfnamefont {S.~L.~W.}\
  \bibnamefont {{McMillan}}},\ }\bibfield  {title} {\bibinfo {title} {{Black
  Hole Mergers in the Universe}},\ }\href {https://doi.org/10.1086/312422}
  {\bibfield  {journal} {\bibinfo  {journal} {\apjl}\ }\textbf {\bibinfo
  {volume} {528}},\ \bibinfo {pages} {L17} (\bibinfo {year} {2000})},\ \Eprint
  {https://arxiv.org/abs/astro-ph/9910061} {arXiv:astro-ph/9910061 [astro-ph]}
  \BibitemShut {NoStop}%
\bibitem [{\citenamefont {{Banerjee}}\ \emph {et~al.}(2010)\citenamefont
  {{Banerjee}}, \citenamefont {{Baumgardt}},\ and\ \citenamefont
  {{Kroupa}}}]{Banerjee2010}%
  \BibitemOpen
  \bibfield  {author} {\bibinfo {author} {\bibfnamefont {S.}~\bibnamefont
  {{Banerjee}}}, \bibinfo {author} {\bibfnamefont {H.}~\bibnamefont
  {{Baumgardt}}},\ and\ \bibinfo {author} {\bibfnamefont {P.}~\bibnamefont
  {{Kroupa}}},\ }\bibfield  {title} {\bibinfo {title} {{Stellar-mass black
  holes in star clusters: implications for gravitational wave radiation}},\
  }\href {https://doi.org/10.1111/j.1365-2966.2009.15880.x} {\bibfield
  {journal} {\bibinfo  {journal} {\mnras}\ }\textbf {\bibinfo {volume} {402}},\
  \bibinfo {pages} {371} (\bibinfo {year} {2010})},\ \Eprint
  {https://arxiv.org/abs/0910.3954} {arXiv:0910.3954 [astro-ph.SR]}
  \BibitemShut {NoStop}%
\bibitem [{\citenamefont {{Rodriguez}}\ \emph {et~al.}(2015)\citenamefont
  {{Rodriguez}}, \citenamefont {{Morscher}}, \citenamefont {{Pattabiraman}},
  \citenamefont {{Chatterjee}}, \citenamefont {{Haster}},\ and\ \citenamefont
  {{Rasio}}}]{Rodriguez2015}%
  \BibitemOpen
  \bibfield  {author} {\bibinfo {author} {\bibfnamefont {C.~L.}\ \bibnamefont
  {{Rodriguez}}}, \bibinfo {author} {\bibfnamefont {M.}~\bibnamefont
  {{Morscher}}}, \bibinfo {author} {\bibfnamefont {B.}~\bibnamefont
  {{Pattabiraman}}}, \bibinfo {author} {\bibfnamefont {S.}~\bibnamefont
  {{Chatterjee}}}, \bibinfo {author} {\bibfnamefont {C.-J.}\ \bibnamefont
  {{Haster}}},\ and\ \bibinfo {author} {\bibfnamefont {F.~A.}\ \bibnamefont
  {{Rasio}}},\ }\bibfield  {title} {\bibinfo {title} {{Binary Black Hole
  Mergers from Globular Clusters: Implications for Advanced LIGO}},\ }\href
  {https://doi.org/10.1103/PhysRevLett.115.051101} {\bibfield  {journal}
  {\bibinfo  {journal} {\prl}\ }\textbf {\bibinfo {volume} {115}},\ \bibinfo
  {eid} {051101} (\bibinfo {year} {2015})},\ \Eprint
  {https://arxiv.org/abs/1505.00792} {arXiv:1505.00792 [astro-ph.HE]}
  \BibitemShut {NoStop}%
\bibitem [{\citenamefont {{Ziosi}}\ \emph {et~al.}(2014)\citenamefont
  {{Ziosi}}, \citenamefont {{Mapelli}}, \citenamefont {{Branchesi}},\ and\
  \citenamefont {{Tormen}}}]{Ziosi2014}%
  \BibitemOpen
  \bibfield  {author} {\bibinfo {author} {\bibfnamefont {B.~M.}\ \bibnamefont
  {{Ziosi}}}, \bibinfo {author} {\bibfnamefont {M.}~\bibnamefont {{Mapelli}}},
  \bibinfo {author} {\bibfnamefont {M.}~\bibnamefont {{Branchesi}}},\ and\
  \bibinfo {author} {\bibfnamefont {G.}~\bibnamefont {{Tormen}}},\ }\bibfield
  {title} {\bibinfo {title} {{Dynamics of stellar black holes in young star
  clusters with different metallicities - II. Black hole-black hole
  binaries}},\ }\href {https://doi.org/10.1093/mnras/stu824} {\bibfield
  {journal} {\bibinfo  {journal} {\mnras}\ }\textbf {\bibinfo {volume} {441}},\
  \bibinfo {pages} {3703} (\bibinfo {year} {2014})},\ \Eprint
  {https://arxiv.org/abs/1404.7147} {arXiv:1404.7147 [astro-ph.GA]}
  \BibitemShut {NoStop}%
\bibitem [{\citenamefont {{Postnov}}\ and\ \citenamefont
  {{Kuranov}}(2019)}]{Postnov2019}%
  \BibitemOpen
  \bibfield  {author} {\bibinfo {author} {\bibfnamefont {K.~A.}\ \bibnamefont
  {{Postnov}}}\ and\ \bibinfo {author} {\bibfnamefont {A.~G.}\ \bibnamefont
  {{Kuranov}}},\ }\bibfield  {title} {\bibinfo {title} {{Black hole spins in
  coalescing binary black holes}},\ }\href
  {https://doi.org/10.1093/mnras/sty3313} {\bibfield  {journal} {\bibinfo
  {journal} {\mnras}\ }\textbf {\bibinfo {volume} {483}},\ \bibinfo {pages}
  {3288} (\bibinfo {year} {2019})},\ \Eprint {https://arxiv.org/abs/1706.00369}
  {arXiv:1706.00369 [astro-ph.HE]} \BibitemShut {NoStop}%
\bibitem [{\citenamefont {{Clesse}}\ and\ \citenamefont
  {{Garc{\'\i}a-Bellido}}(2017{\natexlab{b}})}]{Clesse2017B}%
  \BibitemOpen
  \bibfield  {author} {\bibinfo {author} {\bibfnamefont {S.}~\bibnamefont
  {{Clesse}}}\ and\ \bibinfo {author} {\bibfnamefont {J.}~\bibnamefont
  {{Garc{\'\i}a-Bellido}}},\ }\bibfield  {title} {\bibinfo {title} {{Detecting
  the gravitational wave background from primordial black hole dark matter}},\
  }\href {https://doi.org/10.1016/j.dark.2017.10.001} {\bibfield  {journal}
  {\bibinfo  {journal} {Physics of the Dark Universe}\ }\textbf {\bibinfo
  {volume} {18}},\ \bibinfo {pages} {105} (\bibinfo {year}
  {2017}{\natexlab{b}})},\ \Eprint {https://arxiv.org/abs/1610.08479}
  {arXiv:1610.08479 [astro-ph.CO]} \BibitemShut {NoStop}%
\bibitem [{\citenamefont {{The LIGO Scientific Collaboration}}\ and\
  \citenamefont {{the Virgo Collaboration}}(2019)}]{LIGO2019}%
  \BibitemOpen
  \bibfield  {author} {\bibinfo {author} {\bibnamefont {{The LIGO Scientific
  Collaboration}}}\ and\ \bibinfo {author} {\bibnamefont {{the Virgo
  Collaboration}}},\ }\bibfield  {title} {\bibinfo {title} {{Search for the
  isotropic stochastic background using data from Advanced LIGO's second
  observing run}},\ }\href@noop {} {\bibfield  {journal} {\bibinfo  {journal}
  {arXiv e-prints}\ ,\ \bibinfo {eid} {arXiv:1903.02886}} (\bibinfo {year}
  {2019})},\ \Eprint {https://arxiv.org/abs/1903.02886} {arXiv:1903.02886
  [gr-qc]} \BibitemShut {NoStop}%
\bibitem [{\citenamefont {{Chang}}\ and\ \citenamefont
  {{Refsdal}}(1979)}]{Chang1979}%
  \BibitemOpen
  \bibfield  {author} {\bibinfo {author} {\bibfnamefont {K.}~\bibnamefont
  {{Chang}}}\ and\ \bibinfo {author} {\bibfnamefont {S.}~\bibnamefont
  {{Refsdal}}},\ }\bibfield  {title} {\bibinfo {title} {{Flux variations of QSO
  0957+561 A, B and image splitting by stars near the light path}},\ }\href
  {https://doi.org/10.1038/282561a0} {\bibfield  {journal} {\bibinfo  {journal}
  {\nat}\ }\textbf {\bibinfo {volume} {282}},\ \bibinfo {pages} {561} (\bibinfo
  {year} {1979})}\BibitemShut {NoStop}%
\bibitem [{\citenamefont {{Chang}}\ and\ \citenamefont
  {{Refsdal}}(1984)}]{Chang1984}%
  \BibitemOpen
  \bibfield  {author} {\bibinfo {author} {\bibfnamefont {K.}~\bibnamefont
  {{Chang}}}\ and\ \bibinfo {author} {\bibfnamefont {S.}~\bibnamefont
  {{Refsdal}}},\ }\bibfield  {title} {\bibinfo {title} {{Star disturbances in
  gravitational lens galaxies}},\ }\href@noop {} {\bibfield  {journal}
  {\bibinfo  {journal} {\aap}\ }\textbf {\bibinfo {volume} {132}},\ \bibinfo
  {pages} {168} (\bibinfo {year} {1984})}\BibitemShut {NoStop}%
\bibitem [{\citenamefont {{Kayser}}\ \emph {et~al.}(1986)\citenamefont
  {{Kayser}}, \citenamefont {{Refsdal}},\ and\ \citenamefont
  {{Stabell}}}]{Kayser1986}%
  \BibitemOpen
  \bibfield  {author} {\bibinfo {author} {\bibfnamefont {R.}~\bibnamefont
  {{Kayser}}}, \bibinfo {author} {\bibfnamefont {S.}~\bibnamefont
  {{Refsdal}}},\ and\ \bibinfo {author} {\bibfnamefont {R.}~\bibnamefont
  {{Stabell}}},\ }\bibfield  {title} {\bibinfo {title} {{Astrophysical
  applications of gravitational micro-lensing}},\ }\href@noop {} {\bibfield
  {journal} {\bibinfo  {journal} {\aap}\ }\textbf {\bibinfo {volume} {166}},\
  \bibinfo {pages} {36} (\bibinfo {year} {1986})}\BibitemShut {NoStop}%
\bibitem [{\citenamefont {{Paczynski}}(1986)}]{Paczynski1986}%
  \BibitemOpen
  \bibfield  {author} {\bibinfo {author} {\bibfnamefont {B.}~\bibnamefont
  {{Paczynski}}},\ }\bibfield  {title} {\bibinfo {title} {{Gravitational
  microlensing at large optical depth}},\ }\href
  {https://doi.org/10.1086/163919} {\bibfield  {journal} {\bibinfo  {journal}
  {\apj}\ }\textbf {\bibinfo {volume} {301}},\ \bibinfo {pages} {503} (\bibinfo
  {year} {1986})}\BibitemShut {NoStop}%
\bibitem [{\citenamefont {{Kelly}}\ \emph {et~al.}(2018)\citenamefont
  {{Kelly}}, \citenamefont {{Diego}}, \citenamefont {{Rodney}}, \citenamefont
  {{Kaiser}}, \citenamefont {{Broadhurst}}, \citenamefont {{Zitrin}},
  \citenamefont {{Treu}}, \citenamefont {{P{\'e}rez-Gonz{\'a}lez}},
  \citenamefont {{Morishita}}, \citenamefont {{Jauzac}}, \citenamefont
  {{Selsing}}, \citenamefont {{Oguri}}, \citenamefont {{Pueyo}}, \citenamefont
  {{Ross}}, \citenamefont {{Filippenko}}, \citenamefont {{Smith}},
  \citenamefont {{Hjorth}}, \citenamefont {{Cenko}}, \citenamefont {{Wang}},
  \citenamefont {{Howell}}, \citenamefont {{Richard}}, \citenamefont {{Frye}},
  \citenamefont {{Jha}}, \citenamefont {{Foley}}, \citenamefont {{Norman}},
  \citenamefont {{Bradac}}, \citenamefont {{Zheng}}, \citenamefont {{Brammer}},
  \citenamefont {{Benito}}, \citenamefont {{Cava}}, \citenamefont
  {{Christensen}}, \citenamefont {{de Mink}}, \citenamefont {{Graur}},
  \citenamefont {{Grillo}}, \citenamefont {{Kawamata}}, \citenamefont
  {{Kneib}}, \citenamefont {{Matheson}}, \citenamefont {{McCully}},
  \citenamefont {{Nonino}}, \citenamefont {{P{\'e}rez-Fournon}}, \citenamefont
  {{Riess}}, \citenamefont {{Rosati}}, \citenamefont {{Schmidt}}, \citenamefont
  {{Sharon}},\ and\ \citenamefont {{Weiner}}}]{Kelly2018}%
  \BibitemOpen
  \bibfield  {author} {\bibinfo {author} {\bibfnamefont {P.~L.}\ \bibnamefont
  {{Kelly}}}, \bibinfo {author} {\bibfnamefont {J.~M.}\ \bibnamefont
  {{Diego}}}, \bibinfo {author} {\bibfnamefont {S.}~\bibnamefont {{Rodney}}},
  \bibinfo {author} {\bibfnamefont {N.}~\bibnamefont {{Kaiser}}}, \bibinfo
  {author} {\bibfnamefont {T.}~\bibnamefont {{Broadhurst}}}, \bibinfo {author}
  {\bibfnamefont {A.}~\bibnamefont {{Zitrin}}}, \bibinfo {author}
  {\bibfnamefont {T.}~\bibnamefont {{Treu}}}, \bibinfo {author} {\bibfnamefont
  {P.~G.}\ \bibnamefont {{P{\'e}rez-Gonz{\'a}lez}}}, \bibinfo {author}
  {\bibfnamefont {T.}~\bibnamefont {{Morishita}}}, \bibinfo {author}
  {\bibfnamefont {M.}~\bibnamefont {{Jauzac}}}, \bibinfo {author}
  {\bibfnamefont {J.}~\bibnamefont {{Selsing}}}, \bibinfo {author}
  {\bibfnamefont {M.}~\bibnamefont {{Oguri}}}, \bibinfo {author} {\bibfnamefont
  {L.}~\bibnamefont {{Pueyo}}}, \bibinfo {author} {\bibfnamefont {T.~W.}\
  \bibnamefont {{Ross}}}, \bibinfo {author} {\bibfnamefont {A.~V.}\
  \bibnamefont {{Filippenko}}}, \bibinfo {author} {\bibfnamefont
  {N.}~\bibnamefont {{Smith}}}, \bibinfo {author} {\bibfnamefont
  {J.}~\bibnamefont {{Hjorth}}}, \bibinfo {author} {\bibfnamefont {S.~B.}\
  \bibnamefont {{Cenko}}}, \bibinfo {author} {\bibfnamefont {X.}~\bibnamefont
  {{Wang}}}, \bibinfo {author} {\bibfnamefont {D.~A.}\ \bibnamefont
  {{Howell}}}, \bibinfo {author} {\bibfnamefont {J.}~\bibnamefont {{Richard}}},
  \bibinfo {author} {\bibfnamefont {B.~L.}\ \bibnamefont {{Frye}}}, \bibinfo
  {author} {\bibfnamefont {S.~W.}\ \bibnamefont {{Jha}}}, \bibinfo {author}
  {\bibfnamefont {R.~J.}\ \bibnamefont {{Foley}}}, \bibinfo {author}
  {\bibfnamefont {C.}~\bibnamefont {{Norman}}}, \bibinfo {author}
  {\bibfnamefont {M.}~\bibnamefont {{Bradac}}}, \bibinfo {author}
  {\bibfnamefont {W.}~\bibnamefont {{Zheng}}}, \bibinfo {author} {\bibfnamefont
  {G.}~\bibnamefont {{Brammer}}}, \bibinfo {author} {\bibfnamefont {A.~M.}\
  \bibnamefont {{Benito}}}, \bibinfo {author} {\bibfnamefont {A.}~\bibnamefont
  {{Cava}}}, \bibinfo {author} {\bibfnamefont {L.}~\bibnamefont
  {{Christensen}}}, \bibinfo {author} {\bibfnamefont {S.~E.}\ \bibnamefont {{de
  Mink}}}, \bibinfo {author} {\bibfnamefont {O.}~\bibnamefont {{Graur}}},
  \bibinfo {author} {\bibfnamefont {C.}~\bibnamefont {{Grillo}}}, \bibinfo
  {author} {\bibfnamefont {R.}~\bibnamefont {{Kawamata}}}, \bibinfo {author}
  {\bibfnamefont {J.-P.}\ \bibnamefont {{Kneib}}}, \bibinfo {author}
  {\bibfnamefont {T.}~\bibnamefont {{Matheson}}}, \bibinfo {author}
  {\bibfnamefont {C.}~\bibnamefont {{McCully}}}, \bibinfo {author}
  {\bibfnamefont {M.}~\bibnamefont {{Nonino}}}, \bibinfo {author}
  {\bibfnamefont {I.}~\bibnamefont {{P{\'e}rez-Fournon}}}, \bibinfo {author}
  {\bibfnamefont {A.~G.}\ \bibnamefont {{Riess}}}, \bibinfo {author}
  {\bibfnamefont {P.}~\bibnamefont {{Rosati}}}, \bibinfo {author}
  {\bibfnamefont {K.~B.}\ \bibnamefont {{Schmidt}}}, \bibinfo {author}
  {\bibfnamefont {K.}~\bibnamefont {{Sharon}}},\ and\ \bibinfo {author}
  {\bibfnamefont {B.~J.}\ \bibnamefont {{Weiner}}},\ }\bibfield  {title}
  {\bibinfo {title} {{Extreme magnification of an individual star at redshift
  1.5 by a galaxy-cluster lens}},\ }\href
  {https://doi.org/10.1038/s41550-018-0430-3} {\bibfield  {journal} {\bibinfo
  {journal} {Nature Astronomy}\ }\textbf {\bibinfo {volume} {2}},\ \bibinfo
  {pages} {334} (\bibinfo {year} {2018})},\ \Eprint
  {https://arxiv.org/abs/1706.10279} {arXiv:1706.10279 [astro-ph.GA]}
  \BibitemShut {NoStop}%
\bibitem [{\citenamefont {{Chen}}\ \emph {et~al.}(2019)\citenamefont {{Chen}},
  \citenamefont {{Kelly}}, \citenamefont {{Diego}}, \citenamefont {{Oguri}},
  \citenamefont {{Williams}}, \citenamefont {{Zitrin}}, \citenamefont {{Treu}},
  \citenamefont {{Smith}}, \citenamefont {{Broadhurst}}, \citenamefont
  {{Kaiser}}, \citenamefont {{Foley}}, \citenamefont {{Filippenko}},
  \citenamefont {{Salo}}, \citenamefont {{Hjorth}},\ and\ \citenamefont
  {{Selsing}}}]{Chen2019}%
  \BibitemOpen
  \bibfield  {author} {\bibinfo {author} {\bibfnamefont {W.}~\bibnamefont
  {{Chen}}}, \bibinfo {author} {\bibfnamefont {P.~L.}\ \bibnamefont {{Kelly}}},
  \bibinfo {author} {\bibfnamefont {J.~M.}\ \bibnamefont {{Diego}}}, \bibinfo
  {author} {\bibfnamefont {M.}~\bibnamefont {{Oguri}}}, \bibinfo {author}
  {\bibfnamefont {L.~L.~R.}\ \bibnamefont {{Williams}}}, \bibinfo {author}
  {\bibfnamefont {A.}~\bibnamefont {{Zitrin}}}, \bibinfo {author}
  {\bibfnamefont {T.~L.}\ \bibnamefont {{Treu}}}, \bibinfo {author}
  {\bibfnamefont {N.}~\bibnamefont {{Smith}}}, \bibinfo {author} {\bibfnamefont
  {T.~J.}\ \bibnamefont {{Broadhurst}}}, \bibinfo {author} {\bibfnamefont
  {N.}~\bibnamefont {{Kaiser}}}, \bibinfo {author} {\bibfnamefont {R.~J.}\
  \bibnamefont {{Foley}}}, \bibinfo {author} {\bibfnamefont {A.~V.}\
  \bibnamefont {{Filippenko}}}, \bibinfo {author} {\bibfnamefont
  {L.}~\bibnamefont {{Salo}}}, \bibinfo {author} {\bibfnamefont
  {J.}~\bibnamefont {{Hjorth}}},\ and\ \bibinfo {author} {\bibfnamefont
  {J.}~\bibnamefont {{Selsing}}},\ }\bibfield  {title} {\bibinfo {title}
  {{Searching for Highly Magnified Stars at Cosmological Distances: Discovery
  of a Redshift 0.94 Blue Supergiant in Archival Images of the Galaxy Cluster
  MACS J0416.1-2403}},\ }\href@noop {} {\bibfield  {journal} {\bibinfo
  {journal} {arXiv e-prints}\ } (\bibinfo {year} {2019})},\ \Eprint
  {https://arxiv.org/abs/1902.05510} {arXiv:1902.05510} \BibitemShut {NoStop}%
\bibitem [{\citenamefont {{Schneider}}\ \emph {et~al.}(1992)\citenamefont
  {{Schneider}}, \citenamefont {{Ehlers}},\ and\ \citenamefont
  {{Falco}}}]{SchneiderBook1992}%
  \BibitemOpen
  \bibfield  {author} {\bibinfo {author} {\bibfnamefont {P.}~\bibnamefont
  {{Schneider}}}, \bibinfo {author} {\bibfnamefont {J.}~\bibnamefont
  {{Ehlers}}},\ and\ \bibinfo {author} {\bibfnamefont {E.~E.}\ \bibnamefont
  {{Falco}}},\ }\href {https://doi.org/10.1007/978-3-662-03758-4} {\emph
  {\bibinfo {title} {Gravitational Lenses}}}\ (\bibinfo  {publisher}
  {Springer-Verlag Berlin Heidelberg New York},\ \bibinfo {year} {1992})\ p.\
  \bibinfo {pages} {112}\BibitemShut {NoStop}%
\bibitem [{\citenamefont {{Ulmer}}\ and\ \citenamefont
  {{Goodman}}(1995)}]{UlmerGoodman1995}%
  \BibitemOpen
  \bibfield  {author} {\bibinfo {author} {\bibfnamefont {A.}~\bibnamefont
  {{Ulmer}}}\ and\ \bibinfo {author} {\bibfnamefont {J.}~\bibnamefont
  {{Goodman}}},\ }\bibfield  {title} {\bibinfo {title} {{Femtolensing: Beyond
  the semiclassical approximation}},\ }\href {https://doi.org/10.1086/175422}
  {\bibfield  {journal} {\bibinfo  {journal} {\apj}\ }\textbf {\bibinfo
  {volume} {442}},\ \bibinfo {pages} {67} (\bibinfo {year} {1995})},\ \Eprint
  {https://arxiv.org/abs/astro-ph/9406042} {astro-ph/9406042} \BibitemShut
  {NoStop}%
\bibitem [{\citenamefont {Hannuksela}\ \emph {et~al.}(2019)\citenamefont
  {Hannuksela}, \citenamefont {Haris}, \citenamefont {Ng}, \citenamefont
  {Kumar}, \citenamefont {Mehta}, \citenamefont {Keitel}, \citenamefont {Li},\
  and\ \citenamefont {Ajith}}]{hannuksela2019search}%
  \BibitemOpen
  \bibfield  {author} {\bibinfo {author} {\bibfnamefont {O.}~\bibnamefont
  {Hannuksela}}, \bibinfo {author} {\bibfnamefont {K.}~\bibnamefont {Haris}},
  \bibinfo {author} {\bibfnamefont {K.}~\bibnamefont {Ng}}, \bibinfo {author}
  {\bibfnamefont {S.}~\bibnamefont {Kumar}}, \bibinfo {author} {\bibfnamefont
  {A.}~\bibnamefont {Mehta}}, \bibinfo {author} {\bibfnamefont
  {D.}~\bibnamefont {Keitel}}, \bibinfo {author} {\bibfnamefont
  {T.}~\bibnamefont {Li}},\ and\ \bibinfo {author} {\bibfnamefont
  {P.}~\bibnamefont {Ajith}},\ }\bibfield  {title} {\bibinfo {title} {Search
  for gravitational lensing signatures in ligo-virgo binary black hole
  events},\ }\href@noop {} {\bibfield  {journal} {\bibinfo  {journal} {arXiv
  preprint arXiv:1901.02674}\ } (\bibinfo {year} {2019})}\BibitemShut {NoStop}%
\bibitem [{Note2()}]{Note2}%
  \BibitemOpen
  \bibinfo {note} {See however the discussion above about the reduced
  probability of observing the counterimage with negative parity}\BibitemShut
  {NoStop}%
\bibitem [{\citenamefont {{Broadhurst}}\ \emph {et~al.}(2019)\citenamefont
  {{Broadhurst}}, \citenamefont {{Diego}},\ and\ \citenamefont
  {{Smoot}}}]{Broadhurst2019}%
  \BibitemOpen
  \bibfield  {author} {\bibinfo {author} {\bibfnamefont {T.}~\bibnamefont
  {{Broadhurst}}}, \bibinfo {author} {\bibfnamefont {J.~M.}\ \bibnamefont
  {{Diego}}},\ and\ \bibinfo {author} {\bibfnamefont {G.~F.}\ \bibnamefont
  {{Smoot}}, \bibfnamefont {III}},\ }\bibfield  {title} {\bibinfo {title}
  {{Twin LIGO/Virgo Detections of a Viable Gravitationally-Lensed Black Hole
  Merger}},\ }\href@noop {} {\bibfield  {journal} {\bibinfo  {journal} {arXiv
  e-prints}\ } (\bibinfo {year} {2019})},\ \Eprint
  {https://arxiv.org/abs/1901.03190} {arXiv:1901.03190} \BibitemShut {NoStop}%
\bibitem [{\citenamefont {{Singer}}\ \emph {et~al.}(2019)\citenamefont
  {{Singer}}, \citenamefont {{Goldstein}},\ and\ \citenamefont
  {{Bloom}}}]{Singer2019}%
  \BibitemOpen
  \bibfield  {author} {\bibinfo {author} {\bibfnamefont {L.~P.}\ \bibnamefont
  {{Singer}}}, \bibinfo {author} {\bibfnamefont {D.~A.}\ \bibnamefont
  {{Goldstein}}},\ and\ \bibinfo {author} {\bibfnamefont {J.~S.}\ \bibnamefont
  {{Bloom}}},\ }\bibfield  {title} {\bibinfo {title} {{The Two LIGO/Virgo
  Binary Black Hole Mergers on 2019 August 28 Were Not Strongly Lensed}},\
  }\href@noop {} {\bibfield  {journal} {\bibinfo  {journal} {arXiv e-prints}\
  ,\ \bibinfo {eid} {arXiv:1910.03601}} (\bibinfo {year} {2019})},\ \Eprint
  {https://arxiv.org/abs/1910.03601} {arXiv:1910.03601 [astro-ph.CO]}
  \BibitemShut {NoStop}%
\bibitem [{\citenamefont {{Morishita}}\ \emph {et~al.}(2017)\citenamefont
  {{Morishita}}, \citenamefont {{Abramson}}, \citenamefont {{Treu}},
  \citenamefont {{Schmidt}}, \citenamefont {{Vulcani}},\ and\ \citenamefont
  {{Wang}}}]{Morishita2017}%
  \BibitemOpen
  \bibfield  {author} {\bibinfo {author} {\bibfnamefont {T.}~\bibnamefont
  {{Morishita}}}, \bibinfo {author} {\bibfnamefont {L.~E.}\ \bibnamefont
  {{Abramson}}}, \bibinfo {author} {\bibfnamefont {T.}~\bibnamefont {{Treu}}},
  \bibinfo {author} {\bibfnamefont {K.~B.}\ \bibnamefont {{Schmidt}}}, \bibinfo
  {author} {\bibfnamefont {B.}~\bibnamefont {{Vulcani}}},\ and\ \bibinfo
  {author} {\bibfnamefont {X.}~\bibnamefont {{Wang}}},\ }\bibfield  {title}
  {\bibinfo {title} {{Characterizing Intracluster Light in the Hubble Frontier
  Fields}},\ }\href {https://doi.org/10.3847/1538-4357/aa8403} {\bibfield
  {journal} {\bibinfo  {journal} {\apj}\ }\textbf {\bibinfo {volume} {846}},\
  \bibinfo {eid} {139} (\bibinfo {year} {2017})},\ \Eprint
  {https://arxiv.org/abs/1610.08503} {arXiv:1610.08503} \BibitemShut {NoStop}%
\bibitem [{\citenamefont {{Liu}}\ \emph {et~al.}(2018)\citenamefont {{Liu}},
  \citenamefont {{Guo}},\ and\ \citenamefont {{Cai}}}]{Liu2018}%
  \BibitemOpen
  \bibfield  {author} {\bibinfo {author} {\bibfnamefont {L.}~\bibnamefont
  {{Liu}}}, \bibinfo {author} {\bibfnamefont {Z.-K.}\ \bibnamefont {{Guo}}},\
  and\ \bibinfo {author} {\bibfnamefont {R.-G.}\ \bibnamefont {{Cai}}},\
  }\bibfield  {title} {\bibinfo {title} {{Effects of the surrounding primordial
  black holes on the merger rate of primordial black hole binaries}},\
  }\href@noop {} {\bibfield  {journal} {\bibinfo  {journal} {arXiv e-prints}\ }
  (\bibinfo {year} {2018})},\ \Eprint {https://arxiv.org/abs/1812.05376}
  {arXiv:1812.05376} \BibitemShut {NoStop}%
\bibitem [{\citenamefont {{Lehmann}}\ \emph {et~al.}(2018)\citenamefont
  {{Lehmann}}, \citenamefont {{Profumo}},\ and\ \citenamefont
  {{Yant}}}]{Lehman2018}%
  \BibitemOpen
  \bibfield  {author} {\bibinfo {author} {\bibfnamefont {B.~V.}\ \bibnamefont
  {{Lehmann}}}, \bibinfo {author} {\bibfnamefont {S.}~\bibnamefont
  {{Profumo}}},\ and\ \bibinfo {author} {\bibfnamefont {J.}~\bibnamefont
  {{Yant}}},\ }\bibfield  {title} {\bibinfo {title} {{The maximal-density mass
  function for primordial black hole dark matter}},\ }\href
  {https://doi.org/10.1088/1475-7516/2018/04/007} {\bibfield  {journal}
  {\bibinfo  {journal} {\jcap}\ }\textbf {\bibinfo {volume} {2018}},\ \bibinfo
  {eid} {007} (\bibinfo {year} {2018})},\ \Eprint
  {https://arxiv.org/abs/1801.00808} {arXiv:1801.00808 [astro-ph.CO]}
  \BibitemShut {NoStop}%
\bibitem [{\citenamefont {{Niikura}}\ \emph {et~al.}(2019)\citenamefont
  {{Niikura}}, \citenamefont {{Takada}}, \citenamefont {{Yasuda}},
  \citenamefont {{Lupton}}, \citenamefont {{Sumi}}, \citenamefont {{More}},
  \citenamefont {{Kurita}}, \citenamefont {{Sugiyama}}, \citenamefont {{More}},
  \citenamefont {{Oguri}},\ and\ \citenamefont {{Chiba}}}]{Niikura2019}%
  \BibitemOpen
  \bibfield  {author} {\bibinfo {author} {\bibfnamefont {H.}~\bibnamefont
  {{Niikura}}}, \bibinfo {author} {\bibfnamefont {M.}~\bibnamefont {{Takada}}},
  \bibinfo {author} {\bibfnamefont {N.}~\bibnamefont {{Yasuda}}}, \bibinfo
  {author} {\bibfnamefont {R.~H.}\ \bibnamefont {{Lupton}}}, \bibinfo {author}
  {\bibfnamefont {T.}~\bibnamefont {{Sumi}}}, \bibinfo {author} {\bibfnamefont
  {S.}~\bibnamefont {{More}}}, \bibinfo {author} {\bibfnamefont
  {T.}~\bibnamefont {{Kurita}}}, \bibinfo {author} {\bibfnamefont
  {S.}~\bibnamefont {{Sugiyama}}}, \bibinfo {author} {\bibfnamefont
  {A.}~\bibnamefont {{More}}}, \bibinfo {author} {\bibfnamefont
  {M.}~\bibnamefont {{Oguri}}},\ and\ \bibinfo {author} {\bibfnamefont
  {M.}~\bibnamefont {{Chiba}}},\ }\bibfield  {title} {\bibinfo {title}
  {{Microlensing constraints on primordial black holes with Subaru/HSC
  Andromeda observations}},\ }\href {https://doi.org/10.1038/s41550-019-0723-1}
  {\bibfield  {journal} {\bibinfo  {journal} {Nature Astronomy}\ }\textbf
  {\bibinfo {volume} {3}},\ \bibinfo {pages} {524} (\bibinfo {year} {2019})},\
  \Eprint {https://arxiv.org/abs/1701.02151} {arXiv:1701.02151 [astro-ph.CO]}
  \BibitemShut {NoStop}%
\bibitem [{\citenamefont {{Smyth}}\ \emph {et~al.}(2019)\citenamefont
  {{Smyth}}, \citenamefont {{Profumo}}, \citenamefont {{English}},
  \citenamefont {{Jeltema}}, \citenamefont {{McKinnon}},\ and\ \citenamefont
  {{Guhathakurta}}}]{Smyth2019}%
  \BibitemOpen
  \bibfield  {author} {\bibinfo {author} {\bibfnamefont {N.}~\bibnamefont
  {{Smyth}}}, \bibinfo {author} {\bibfnamefont {S.}~\bibnamefont {{Profumo}}},
  \bibinfo {author} {\bibfnamefont {S.}~\bibnamefont {{English}}}, \bibinfo
  {author} {\bibfnamefont {T.}~\bibnamefont {{Jeltema}}}, \bibinfo {author}
  {\bibfnamefont {K.}~\bibnamefont {{McKinnon}}},\ and\ \bibinfo {author}
  {\bibfnamefont {P.}~\bibnamefont {{Guhathakurta}}},\ }\bibfield  {title}
  {\bibinfo {title} {{Updated Constraints on Asteroid-Mass Primordial Black
  Holes as Dark Matter}},\ }\href@noop {} {\bibfield  {journal} {\bibinfo
  {journal} {arXiv e-prints}\ ,\ \bibinfo {eid} {arXiv:1910.01285}} (\bibinfo
  {year} {2019})},\ \Eprint {https://arxiv.org/abs/1910.01285}
  {arXiv:1910.01285 [astro-ph.CO]} \BibitemShut {NoStop}%
\bibitem [{\citenamefont {{Carr}}\ \emph {et~al.}(2017)\citenamefont {{Carr}},
  \citenamefont {{Raidal}}, \citenamefont {{Tenkanen}}, \citenamefont
  {{Vaskonen}},\ and\ \citenamefont {{Veerm{\"a}e}}}]{Carr2017}%
  \BibitemOpen
  \bibfield  {author} {\bibinfo {author} {\bibfnamefont {B.}~\bibnamefont
  {{Carr}}}, \bibinfo {author} {\bibfnamefont {M.}~\bibnamefont {{Raidal}}},
  \bibinfo {author} {\bibfnamefont {T.}~\bibnamefont {{Tenkanen}}}, \bibinfo
  {author} {\bibfnamefont {V.}~\bibnamefont {{Vaskonen}}},\ and\ \bibinfo
  {author} {\bibfnamefont {H.}~\bibnamefont {{Veerm{\"a}e}}},\ }\bibfield
  {title} {\bibinfo {title} {{Primordial black hole constraints for extended
  mass functions}},\ }\href {https://doi.org/10.1103/PhysRevD.96.023514}
  {\bibfield  {journal} {\bibinfo  {journal} {\prd}\ }\textbf {\bibinfo
  {volume} {96}},\ \bibinfo {eid} {023514} (\bibinfo {year} {2017})},\ \Eprint
  {https://arxiv.org/abs/1705.05567} {arXiv:1705.05567 [astro-ph.CO]}
  \BibitemShut {NoStop}%
\bibitem [{\citenamefont {{Carr}}\ \emph {et~al.}(2019)\citenamefont {{Carr}},
  \citenamefont {{Clesse}}, \citenamefont {{Garcia-Bellido}},\ and\
  \citenamefont {{Kuhnel}}}]{Carr2019}%
  \BibitemOpen
  \bibfield  {author} {\bibinfo {author} {\bibfnamefont {B.}~\bibnamefont
  {{Carr}}}, \bibinfo {author} {\bibfnamefont {S.}~\bibnamefont {{Clesse}}},
  \bibinfo {author} {\bibfnamefont {J.}~\bibnamefont {{Garcia-Bellido}}},\ and\
  \bibinfo {author} {\bibfnamefont {F.}~\bibnamefont {{Kuhnel}}},\ }\bibfield
  {title} {\bibinfo {title} {{Cosmic Conundra Explained by Thermal History and
  Primordial Black Holes}},\ }\href@noop {} {\bibfield  {journal} {\bibinfo
  {journal} {arXiv e-prints}\ ,\ \bibinfo {eid} {arXiv:1906.08217}} (\bibinfo
  {year} {2019})},\ \Eprint {https://arxiv.org/abs/1906.08217}
  {arXiv:1906.08217 [astro-ph.CO]} \BibitemShut {NoStop}%
\bibitem [{\citenamefont {{Raidal}}\ \emph {et~al.}(2019)\citenamefont
  {{Raidal}}, \citenamefont {{Spethmann}}, \citenamefont {{Vaskonen}},\ and\
  \citenamefont {{Veerm{\"a}e}}}]{Raidal2019}%
  \BibitemOpen
  \bibfield  {author} {\bibinfo {author} {\bibfnamefont {M.}~\bibnamefont
  {{Raidal}}}, \bibinfo {author} {\bibfnamefont {C.}~\bibnamefont
  {{Spethmann}}}, \bibinfo {author} {\bibfnamefont {V.}~\bibnamefont
  {{Vaskonen}}},\ and\ \bibinfo {author} {\bibfnamefont {H.}~\bibnamefont
  {{Veerm{\"a}e}}},\ }\bibfield  {title} {\bibinfo {title} {{Formation and
  evolution of primordial black hole binaries in the early universe}},\ }\href
  {https://doi.org/10.1088/1475-7516/2019/02/018} {\bibfield  {journal}
  {\bibinfo  {journal} {\jcap}\ }\textbf {\bibinfo {volume} {2}},\ \bibinfo
  {eid} {018} (\bibinfo {year} {2019})},\ \Eprint
  {https://arxiv.org/abs/1812.01930} {arXiv:1812.01930} \BibitemShut {NoStop}%
\bibitem [{\citenamefont {{Fryer}}\ \emph {et~al.}(2012)\citenamefont
  {{Fryer}}, \citenamefont {{Belczynski}}, \citenamefont {{Wiktorowicz}},
  \citenamefont {{Dominik}}, \citenamefont {{Kalogera}},\ and\ \citenamefont
  {{Holz}}}]{Fryer2012}%
  \BibitemOpen
  \bibfield  {author} {\bibinfo {author} {\bibfnamefont {C.~L.}\ \bibnamefont
  {{Fryer}}}, \bibinfo {author} {\bibfnamefont {K.}~\bibnamefont
  {{Belczynski}}}, \bibinfo {author} {\bibfnamefont {G.}~\bibnamefont
  {{Wiktorowicz}}}, \bibinfo {author} {\bibfnamefont {M.}~\bibnamefont
  {{Dominik}}}, \bibinfo {author} {\bibfnamefont {V.}~\bibnamefont
  {{Kalogera}}},\ and\ \bibinfo {author} {\bibfnamefont {D.~E.}\ \bibnamefont
  {{Holz}}},\ }\bibfield  {title} {\bibinfo {title} {{Compact Remnant Mass
  Function: Dependence on the Explosion Mechanism and Metallicity}},\ }\href
  {https://doi.org/10.1088/0004-637X/749/1/91} {\bibfield  {journal} {\bibinfo
  {journal} {\apj}\ }\textbf {\bibinfo {volume} {749}},\ \bibinfo {eid} {91}
  (\bibinfo {year} {2012})},\ \Eprint {https://arxiv.org/abs/1110.1726}
  {arXiv:1110.1726 [astro-ph.SR]} \BibitemShut {NoStop}%
\bibitem [{\citenamefont {{Virgo Collaboration}}(2019)}]{LigoCatalog2019I}%
  \BibitemOpen
  \bibfield  {author} {\bibinfo {author} {\bibnamefont {{Virgo
  Collaboration}}},\ }\bibfield  {title} {\bibinfo {title} {{GWTC-1: A
  Gravitational-Wave Transient Catalog of Compact Binary Mergers Observed by
  LIGO and Virgo during the First and Second Observing Runs}},\ }\href
  {https://doi.org/10.1103/PhysRevX.9.031040} {\bibfield  {journal} {\bibinfo
  {journal} {Physical Review X}\ }\textbf {\bibinfo {volume} {9}},\ \bibinfo
  {eid} {031040} (\bibinfo {year} {2019})}\BibitemShut {NoStop}%
\bibitem [{\citenamefont {{Wei}}\ and\ \citenamefont {{Wu}}(2017)}]{Wei2017}%
  \BibitemOpen
  \bibfield  {author} {\bibinfo {author} {\bibfnamefont {J.-J.}\ \bibnamefont
  {{Wei}}}\ and\ \bibinfo {author} {\bibfnamefont {X.-F.}\ \bibnamefont
  {{Wu}}},\ }\bibfield  {title} {\bibinfo {title} {{Strongly lensed
  gravitational waves and electromagnetic signals as powerful cosmic rulers}},\
  }\href {https://doi.org/10.1093/mnras/stx2210} {\bibfield  {journal}
  {\bibinfo  {journal} {\mnras}\ }\textbf {\bibinfo {volume} {472}},\ \bibinfo
  {pages} {2906} (\bibinfo {year} {2017})},\ \Eprint
  {https://arxiv.org/abs/1707.04152} {arXiv:1707.04152} \BibitemShut {NoStop}%
\bibitem [{\citenamefont {{Chen}}\ and\ \citenamefont
  {{Huang}}(2018)}]{Chen2018}%
  \BibitemOpen
  \bibfield  {author} {\bibinfo {author} {\bibfnamefont {Z.}~\bibnamefont
  {{Chen}}}\ and\ \bibinfo {author} {\bibfnamefont {Q.}~\bibnamefont
  {{Huang}}},\ }\bibfield  {title} {\bibinfo {title} {{Merger Rate Distribution
  of Primordial Black Hole Binaries}},\ }\href
  {https://doi.org/10.3847/1538-4357/aad6e2} {\bibfield  {journal} {\bibinfo
  {journal} {\apj}\ }\textbf {\bibinfo {volume} {864}},\ \bibinfo {eid} {61}
  (\bibinfo {year} {2018})},\ \Eprint {https://arxiv.org/abs/1801.10327}
  {arXiv:1801.10327} \BibitemShut {NoStop}%
\bibitem [{\citenamefont {{Jackson}}\ \emph {et~al.}(2019)\citenamefont
  {{Jackson}}, \citenamefont {{Liu}},\ and\ \citenamefont
  {{Naselsky}}}]{Jackson2019}%
  \BibitemOpen
  \bibfield  {author} {\bibinfo {author} {\bibfnamefont {A.~D.}\ \bibnamefont
  {{Jackson}}}, \bibinfo {author} {\bibfnamefont {H.}~\bibnamefont {{Liu}}},\
  and\ \bibinfo {author} {\bibfnamefont {P.}~\bibnamefont {{Naselsky}}},\
  }\bibfield  {title} {\bibinfo {title} {{Noise residuals for GW150914 using
  maximum likelihood and numerical relativity templates}},\ }\href
  {https://doi.org/10.1088/1475-7516/2019/05/014} {\bibfield  {journal}
  {\bibinfo  {journal} {\jcap}\ }\textbf {\bibinfo {volume} {2019}},\ \bibinfo
  {eid} {014} (\bibinfo {year} {2019})},\ \Eprint
  {https://arxiv.org/abs/1903.02401} {arXiv:1903.02401 [astro-ph.IM]}
  \BibitemShut {NoStop}%
\bibitem [{\citenamefont {{The LIGO Scientific Collaboration}}\ \emph
  {et~al.}(2018)\citenamefont {{The LIGO Scientific Collaboration}},
  \citenamefont {{the Virgo Collaboration}}, \citenamefont {{Abbott}},
  \citenamefont {{Abbott}}, \citenamefont {{Abbott}}, \citenamefont
  {{Abraham}}, \citenamefont {{Acernese}}, \citenamefont {{Ackley}},
  \citenamefont {{Adams}}, \citenamefont {{Adhikari}},\ and\ \citenamefont
  {et~al.}}]{LIGO2018}%
  \BibitemOpen
  \bibfield  {author} {\bibinfo {author} {\bibnamefont {{The LIGO Scientific
  Collaboration}}}, \bibinfo {author} {\bibnamefont {{the Virgo
  Collaboration}}}, \bibinfo {author} {\bibfnamefont {B.~P.}\ \bibnamefont
  {{Abbott}}}, \bibinfo {author} {\bibfnamefont {R.}~\bibnamefont {{Abbott}}},
  \bibinfo {author} {\bibfnamefont {T.~D.}\ \bibnamefont {{Abbott}}}, \bibinfo
  {author} {\bibfnamefont {S.}~\bibnamefont {{Abraham}}}, \bibinfo {author}
  {\bibfnamefont {F.}~\bibnamefont {{Acernese}}}, \bibinfo {author}
  {\bibfnamefont {K.}~\bibnamefont {{Ackley}}}, \bibinfo {author}
  {\bibfnamefont {C.}~\bibnamefont {{Adams}}}, \bibinfo {author} {\bibfnamefont
  {R.~X.}\ \bibnamefont {{Adhikari}}},\ and\ \bibinfo {author} {\bibnamefont
  {et~al.}},\ }\bibfield  {title} {\bibinfo {title} {{GWTC-1: A
  Gravitational-Wave Transient Catalog of Compact Binary Mergers Observed by
  LIGO and Virgo during the First and Second Observing Runs}},\ }\href@noop {}
  {\bibfield  {journal} {\bibinfo  {journal} {arXiv e-prints}\ } (\bibinfo
  {year} {2018})},\ \Eprint {https://arxiv.org/abs/1811.12907}
  {arXiv:1811.12907 [astro-ph.HE]} \BibitemShut {NoStop}%
\bibitem [{\citenamefont {{Abbott}}\ \emph {et~al.}(2016)\citenamefont
  {{Abbott}}, \citenamefont {{Abbott}}, \citenamefont {{Abbott}}, \citenamefont
  {{Abernathy}}, \citenamefont {{Acernese}}, \citenamefont {{Ackley}},
  \citenamefont {{Adams}}, \citenamefont {{Adams}}, \citenamefont {{Addesso}},
  \citenamefont {{Adhikari}},\ and\ \citenamefont
  {et~al.}}]{2016PhRvD..93l2004A}%
  \BibitemOpen
  \bibfield  {author} {\bibinfo {author} {\bibfnamefont {B.~P.}\ \bibnamefont
  {{Abbott}}}, \bibinfo {author} {\bibfnamefont {R.}~\bibnamefont {{Abbott}}},
  \bibinfo {author} {\bibfnamefont {T.~D.}\ \bibnamefont {{Abbott}}}, \bibinfo
  {author} {\bibfnamefont {M.~R.}\ \bibnamefont {{Abernathy}}}, \bibinfo
  {author} {\bibfnamefont {F.}~\bibnamefont {{Acernese}}}, \bibinfo {author}
  {\bibfnamefont {K.}~\bibnamefont {{Ackley}}}, \bibinfo {author}
  {\bibfnamefont {C.}~\bibnamefont {{Adams}}}, \bibinfo {author} {\bibfnamefont
  {T.}~\bibnamefont {{Adams}}}, \bibinfo {author} {\bibfnamefont
  {P.}~\bibnamefont {{Addesso}}}, \bibinfo {author} {\bibfnamefont {R.~X.}\
  \bibnamefont {{Adhikari}}},\ and\ \bibinfo {author} {\bibnamefont {et~al.}},\
  }\bibfield  {title} {\bibinfo {title} {{Observing gravitational-wave
  transient GW150914 with minimal assumptions}},\ }\href
  {https://doi.org/10.1103/PhysRevD.93.122004} {\bibfield  {journal} {\bibinfo
  {journal} {\prd}\ }\textbf {\bibinfo {volume} {93}},\ \bibinfo {eid} {122004}
  (\bibinfo {year} {2016})},\ \Eprint {https://arxiv.org/abs/1602.03843}
  {arXiv:1602.03843 [gr-qc]} \BibitemShut {NoStop}%
\bibitem [{\citenamefont {{Abbott}}\ \emph {et~al.}(2017)\citenamefont
  {{Abbott}}, \citenamefont {{Abbott}}, \citenamefont {{Abbott}}, \citenamefont
  {{Abernathy}}, \citenamefont {{Acernese}}, \citenamefont {{Ackley}},
  \citenamefont {{Adams}}, \citenamefont {{Adams}}, \citenamefont {{Addesso}},
  \citenamefont {{Adhikari}},\ and\ \citenamefont
  {et~al.}}]{2017PhRvD..95d2003A}%
  \BibitemOpen
  \bibfield  {author} {\bibinfo {author} {\bibfnamefont {B.~P.}\ \bibnamefont
  {{Abbott}}}, \bibinfo {author} {\bibfnamefont {R.}~\bibnamefont {{Abbott}}},
  \bibinfo {author} {\bibfnamefont {T.~D.}\ \bibnamefont {{Abbott}}}, \bibinfo
  {author} {\bibfnamefont {M.~R.}\ \bibnamefont {{Abernathy}}}, \bibinfo
  {author} {\bibfnamefont {F.}~\bibnamefont {{Acernese}}}, \bibinfo {author}
  {\bibfnamefont {K.}~\bibnamefont {{Ackley}}}, \bibinfo {author}
  {\bibfnamefont {C.}~\bibnamefont {{Adams}}}, \bibinfo {author} {\bibfnamefont
  {T.}~\bibnamefont {{Adams}}}, \bibinfo {author} {\bibfnamefont
  {P.}~\bibnamefont {{Addesso}}}, \bibinfo {author} {\bibfnamefont {R.~X.}\
  \bibnamefont {{Adhikari}}},\ and\ \bibinfo {author} {\bibnamefont {et~al.}},\
  }\bibfield  {title} {\bibinfo {title} {{All-sky search for short
  gravitational-wave bursts in the first Advanced LIGO run}},\ }\href
  {https://doi.org/10.1103/PhysRevD.95.042003} {\bibfield  {journal} {\bibinfo
  {journal} {\prd}\ }\textbf {\bibinfo {volume} {95}},\ \bibinfo {eid} {042003}
  (\bibinfo {year} {2017})},\ \Eprint {https://arxiv.org/abs/1611.02972}
  {arXiv:1611.02972 [gr-qc]} \BibitemShut {NoStop}%
\bibitem [{\citenamefont {Maggiore}(2008)}]{maggiore}%
  \BibitemOpen
  \bibfield  {author} {\bibinfo {author} {\bibfnamefont {M.}~\bibnamefont
  {Maggiore}},\ }\href@noop {} {\emph {\bibinfo {title} {Gravitational Waves:
  Volume 1: Theory and Experiments}}},\ Vol.~\bibinfo {volume} {1}\ (\bibinfo
  {publisher} {Oxford university press},\ \bibinfo {year} {2008})\BibitemShut
  {NoStop}%
\bibitem [{\citenamefont {Punturo}\ \emph {et~al.}(2010)\citenamefont
  {Punturo}, \citenamefont {Abernathy}, \citenamefont {Acernese}, \citenamefont
  {Allen}, \citenamefont {Andersson}, \citenamefont {Arun}, \citenamefont
  {Barone}, \citenamefont {Barr}, \citenamefont {Barsuglia}, \citenamefont
  {Beker} \emph {et~al.}}]{3gdetectors1}%
  \BibitemOpen
  \bibfield  {author} {\bibinfo {author} {\bibfnamefont {M.}~\bibnamefont
  {Punturo}}, \bibinfo {author} {\bibfnamefont {M.}~\bibnamefont {Abernathy}},
  \bibinfo {author} {\bibfnamefont {F.}~\bibnamefont {Acernese}}, \bibinfo
  {author} {\bibfnamefont {B.}~\bibnamefont {Allen}}, \bibinfo {author}
  {\bibfnamefont {N.}~\bibnamefont {Andersson}}, \bibinfo {author}
  {\bibfnamefont {K.}~\bibnamefont {Arun}}, \bibinfo {author} {\bibfnamefont
  {F.}~\bibnamefont {Barone}}, \bibinfo {author} {\bibfnamefont
  {B.}~\bibnamefont {Barr}}, \bibinfo {author} {\bibfnamefont {M.}~\bibnamefont
  {Barsuglia}}, \bibinfo {author} {\bibfnamefont {M.}~\bibnamefont {Beker}},
  \emph {et~al.},\ }\bibfield  {title} {\bibinfo {title} {The einstein
  telescope: a third-generation gravitational wave observatory},\ }\href@noop
  {} {\bibfield  {journal} {\bibinfo  {journal} {Classical and Quantum
  Gravity}\ }\textbf {\bibinfo {volume} {27}},\ \bibinfo {pages} {194002}
  (\bibinfo {year} {2010})}\BibitemShut {NoStop}%
\bibitem [{\citenamefont {Abbott}\ \emph {et~al.}(2017)\citenamefont {Abbott},
  \citenamefont {Abbott}, \citenamefont {Abbott}, \citenamefont {Abernathy},
  \citenamefont {Ackley}, \citenamefont {Adams}, \citenamefont {Addesso},
  \citenamefont {Adhikari}, \citenamefont {Adya}, \citenamefont {Affeldt} \emph
  {et~al.}}]{3gdetectors2}%
  \BibitemOpen
  \bibfield  {author} {\bibinfo {author} {\bibfnamefont {B.~P.}\ \bibnamefont
  {Abbott}}, \bibinfo {author} {\bibfnamefont {R.}~\bibnamefont {Abbott}},
  \bibinfo {author} {\bibfnamefont {T.}~\bibnamefont {Abbott}}, \bibinfo
  {author} {\bibfnamefont {M.}~\bibnamefont {Abernathy}}, \bibinfo {author}
  {\bibfnamefont {K.}~\bibnamefont {Ackley}}, \bibinfo {author} {\bibfnamefont
  {C.}~\bibnamefont {Adams}}, \bibinfo {author} {\bibfnamefont
  {P.}~\bibnamefont {Addesso}}, \bibinfo {author} {\bibfnamefont
  {R.}~\bibnamefont {Adhikari}}, \bibinfo {author} {\bibfnamefont
  {V.}~\bibnamefont {Adya}}, \bibinfo {author} {\bibfnamefont {C.}~\bibnamefont
  {Affeldt}}, \emph {et~al.},\ }\bibfield  {title} {\bibinfo {title} {Exploring
  the sensitivity of next generation gravitational wave detectors},\
  }\href@noop {} {\bibfield  {journal} {\bibinfo  {journal} {Classical and
  Quantum Gravity}\ }\textbf {\bibinfo {volume} {34}},\ \bibinfo {pages}
  {044001} (\bibinfo {year} {2017})}\BibitemShut {NoStop}%
\bibitem [{\citenamefont {Dwyer}\ \emph {et~al.}(2015)\citenamefont {Dwyer},
  \citenamefont {Sigg}, \citenamefont {Ballmer}, \citenamefont {Barsotti},
  \citenamefont {Mavalvala},\ and\ \citenamefont {Evans}}]{3gdetectors3}%
  \BibitemOpen
  \bibfield  {author} {\bibinfo {author} {\bibfnamefont {S.}~\bibnamefont
  {Dwyer}}, \bibinfo {author} {\bibfnamefont {D.}~\bibnamefont {Sigg}},
  \bibinfo {author} {\bibfnamefont {S.~W.}\ \bibnamefont {Ballmer}}, \bibinfo
  {author} {\bibfnamefont {L.}~\bibnamefont {Barsotti}}, \bibinfo {author}
  {\bibfnamefont {N.}~\bibnamefont {Mavalvala}},\ and\ \bibinfo {author}
  {\bibfnamefont {M.}~\bibnamefont {Evans}},\ }\bibfield  {title} {\bibinfo
  {title} {Gravitational wave detector with cosmological reach},\ }\href@noop
  {} {\bibfield  {journal} {\bibinfo  {journal} {Physical Review D}\ }\textbf
  {\bibinfo {volume} {91}},\ \bibinfo {pages} {082001} (\bibinfo {year}
  {2015})}\BibitemShut {NoStop}%
\bibitem [{\citenamefont {Abernathy}\ \emph {et~al.}(2011)\citenamefont
  {Abernathy}, \citenamefont {Acernese}, \citenamefont {Ajith}, \citenamefont
  {Allen}, \citenamefont {Amaro-Seoane} \emph {et~al.}}]{3gdetectors4}%
  \BibitemOpen
  \bibfield  {author} {\bibinfo {author} {\bibfnamefont {M.}~\bibnamefont
  {Abernathy}}, \bibinfo {author} {\bibfnamefont {F.}~\bibnamefont {Acernese}},
  \bibinfo {author} {\bibfnamefont {P.}~\bibnamefont {Ajith}}, \bibinfo
  {author} {\bibfnamefont {B.}~\bibnamefont {Allen}}, \bibinfo {author}
  {\bibfnamefont {P.}~\bibnamefont {Amaro-Seoane}}, \emph {et~al.},\ }\bibfield
   {title} {\bibinfo {title} {Einstein gravitational wave telescope conceptual
  design study},\ }\href@noop {} {\bibfield  {journal} {\bibinfo  {journal}
  {available from European Gravitational Observatory, document number
  ET-0106A-10}\ } (\bibinfo {year} {2011})}\BibitemShut {NoStop}%
\bibitem [{\citenamefont {Li}\ \emph {et~al.}(2018)\citenamefont {Li},
  \citenamefont {Mao}, \citenamefont {Zhao},\ and\ \citenamefont
  {Lu}}]{li2018gravitational}%
  \BibitemOpen
  \bibfield  {author} {\bibinfo {author} {\bibfnamefont {S.-S.}\ \bibnamefont
  {Li}}, \bibinfo {author} {\bibfnamefont {S.}~\bibnamefont {Mao}}, \bibinfo
  {author} {\bibfnamefont {Y.}~\bibnamefont {Zhao}},\ and\ \bibinfo {author}
  {\bibfnamefont {Y.}~\bibnamefont {Lu}},\ }\bibfield  {title} {\bibinfo
  {title} {Gravitational lensing of gravitational waves: A statistical
  perspective},\ }\href@noop {} {\bibfield  {journal} {\bibinfo  {journal}
  {Monthly Notices of the Royal Astronomical Society}\ }\textbf {\bibinfo
  {volume} {476}},\ \bibinfo {pages} {2220} (\bibinfo {year}
  {2018})}\BibitemShut {NoStop}%
\end{thebibliography}%

\end{document}